\documentclass[aps,twocolumn,floatfix,showpacs,amsmath,amsfonts]{revtex4}

\usepackage{graphicx}
\usepackage{bm}
\usepackage{units}
\usepackage{color}
\usepackage{upgreek}
\usepackage{gensymb}

\begin{document}

\title{Exciton-polaritons in cuprous oxide: Theory and comparison with experiment}

\author{Frank Schweiner}
\author{Jan Ertl}
\author{J\"org Main}
\author{G\"unter Wunner}
\affiliation{Institut f\"ur Theoretische Physik 1, Universit\"at Stuttgart,
  70550 Stuttgart, Germany}
\author{Christoph Uihlein}
\affiliation{Experimentelle Physik 2, Technische Universit\"at Dortmund, 44221 Dortmund, Germany}
\date{\today}

\begin{abstract}
The observation of giant Rydberg excitons in cuprous oxide $\left(\mathrm{Cu_{2}O}\right)$
up to a principal quantum number of $n=25$ by T.~Kazimierczuk \emph{et al.}
[Nature \textbf{514}, 343, (2014)] inevitably raises the question whether these quasi-particles must be described within a multi-polariton framework since excitons and photons
are always coupled in the solid.
In this paper we present the theory of exciton-polaritons in $\mathrm{Cu_{2}O}$.
To this end we extend the Hamiltonian 
which includes the complete valence band structure, the exchange interaction,
and the central-cell corrections effects, and which has been recently 
deduced by F.~Schweiner \emph{et al.} [Phys.~Rev.~B \textbf{95}, 195201, (2017)],
for finite values of the exciton momentum $\hbar K$.
We derive formulas to calculate not only dipole
but also quadrupole oscillator strengths when using the complete basis of
F.~Schweiner \emph{et al.}, which has recently been 
proven as a powerful tool to calculate exciton spectra.
Very complex polariton spectra for the three
orientations of $\boldsymbol{K}$ along the axes 
$[001]$, $[110]$, and $[111]$ of high symmetry are obtained
and a strong mixing of exciton states is reported.
The main focus is on the $1S$ ortho exciton-polariton,
for which pronounced polariton
effects have been measured in experiments.
We set up a $5\times 5$ matrix model,
which accounts for both the polariton effect and the $K$-dependent splitting,
and which allows treating the anisotropic polariton dispersion for any direction of $\boldsymbol{K}$.
We especially discuss the dispersions for $\boldsymbol{K}$ being oriented
in the planes perpendicular to $[1\bar{1}0]$ and $[111]$, 
for which experimental transmission spectra have been measured.
Furthermore, we compare our results with experimental values of the
$K$-dependent splitting, the group velocity,
and the oscillator strengths of this exciton-polariton.
The results are in good agreement. This proves the 
validity of the $5\times 5$ matrix model as a useful theoretical
model for further investigations on the $1S$ ortho exciton-polariton.
\end{abstract}


\maketitle

\section{Introduction}

Excitons are Coulomb-bound pairs of a positively
charged hole in the valence band and a negatively charged
electron in the conduction band of a semiconductor.
Hence, these elementary excitations of a semiconductor
are often regarded as the hydrogen analog of the solid state.
It is now more than 80 years since Frenkel~\cite{TOE_1,TOE_2,TOE_4},
Peierls~\cite{TOE_3}, and Wannier~\cite{TOE_5} formulated the concept of 
excitons. After the experimental discovery of these quasi-particles in cuprous oxide
$\left(\mathrm{Cu_{2}O}\right)$ by Gross and Karryjew in 1952~\cite{GRE_4},
excitons in bulk semiconductors became an important topic in solid-state physics
from the late 1950s to the 1970s (see, e.g., Refs.~\cite{TOE,12,79_7,79_8,79_6,79_9} 
and further references therein).

Very recently, new attention has been drawn to the field of excitons by an experimental 
observation of the yellow exciton series in $\mathrm{Cu_{2}O}$ 
up to a large principal quantum number of $n=25$~\cite{GRE}.
This discovery has opened up the research field of giant Rydberg excitons
and led to a variety of new experimental and theoretical 
investigations~\cite{GRE,QC,75,76,50,28,80,100,125,175,78,79,150,74,77,200,225,275,70,94,95,96,97,98}.
Furthermore, it may pave the way, e.g.,  to a deeper
understanding of inter-particle interactions in the solid~\cite{GRE}
and to applications in quantum
information technology~\cite{77}.

Even though the spectrum of Rydberg excitons in $\mathrm{Cu_{2}O}$ 
can be described quite well in a first approximation by the hydrogen-like model
of Wannier, one must keep in mind that excitons are complex many-body
states of the solid and, hence, that there are significant limitations to the
hydrogen-like model and to the atom-like description of these quasi-particles~\cite{87}.

Some essential corrections to the hydrogen-like model
comprise, e.g., the inclusion of the complete 
cubic valence band structure~\cite{17_17_18,17_17_26,7_11,17_17,17_15,44_12},
which leads to a complicated fine-structure splitting, the
central-cell corrections, which account for deviations from the hydrogen-like
model in the limit of a small exciton radius~\cite{HP8,HP9,HP1,TOE,SO,1,PAE,IPEP},
and the exchange interaction~\cite{E2,E3,TOE,1,SO}.
Furthermore, interactions with other quasi-particles, like, e.g.,
phonons have to be considered~\cite{AP3,M1_7,2}.
All of these effects have already been discussed recently for the
excitons in $\mathrm{Cu_{2}O}$~\cite{75,28,100,125,150,200}.

Above all, there is another fundamental difference between atoms and excitons
as regards their interaction with light. By analogy with the interaction
of atoms with light one may suppose that absorption of light in a crystal
can be described as the excitation of an exciton with the simultaneous disappearance
of a photon~\cite{TOE}. Indeed, in the weak-coupling limit, the incident
light acts only as a perturbation on the different energy states or
excitations of the solid like, e.g., excitons~\cite{SO}.
However, since the excited states in the solid are connected with
a polarization and since an oscillating polarization emits
again an electromagnetic wave acting back onto the incident wave,
there is an interplay between light and matter. If the frequency of
light is within the range of the resonance frequency of an excitation,
the coupling is strong and thus anomalous dispersion can be observed~\cite{TOE}. Due
to this coupling, excitons and photons cannot be treated as independent entities or
good eigenstates, but new quasi-particles must be introduced, which are called
polaritons and which represent the quanta of the mixed state of polarization
and electromagnetic wave~\cite{P2,TOE_46a,TOE_206,Haug_H58,SOK4_63H1,SOK5_75C1,CO,SOK5_85H1,
SOK5_98K1}.

In this regard the $1S$ ortho exciton of $\mathrm{Cu_{2}O}$, in particular, is of
interest as for this state characteristic polariton effects
like propagation beats and the conservation 
of coherence over macroscopic distances
have already been demonstrated experimentally~\cite{GRE_26}.
Furthermore, the threefold degeneracy of this state is lifted for 
finite momentum of the center of mass $\hbar K$,
and its oscillator strengths~\cite{P3_107_12} as well as its dispersion
are anisotropic~\cite{9_1,8,9,100}.
All of these effects and the number of experimental
results for, e.g., the oscillator strength,
the $\boldsymbol{K}$-dependent spectra and the group velocity~\cite{P3_107_12},
with which theoretical results can be compared, make
the $1S$ ortho exciton an ideal candidate for theoretical investigations.

In this paper we present the theory of exciton-polaritons in
$\mathrm{Cu_{2}O}$. We extend the Hamiltonian of Ref.~\cite{200},
which accounts for the complete valence band structure, the exchange interaction,
and the central-cell corrections, for a finite momentum $\hbar K$ of the center of mass.
The corresponding Schr\"odinger equation can then be solved using a complete
basis. This method allows us not only to calculate 
dipole and quadrupole oscillator strengths but also the size of the nonanalytic
exchange interaction. As the splitting due to the nonanalytic exchange interaction at $K=0$
is identical to the longitudinal-transverse 
splitting (LT-splitting) when treating polaritons, this interaction
needs to be considered for a correct treatment of the complete problem.
We show how to calculate the polariton dispersion for the complete exciton spectrum.

Due to the presence of the complete valence band structure, the exchange interaction,
and the central-cell corrections, we obtain very complex polariton dispersions
for the excitons in $\mathrm{Cu_{2}O}$, which by far exceed previous
investigations on the polariton dispersion using a 
simple hydrogen-like model~\cite{74}.
%
The main emphasis in this paper is on
the only exciton state for which a pronounced polariton effect
has been observed in several experiments: the $1S$ ortho exciton
state.
At first, we discuss the effects of a finite momentum 
$\hbar\boldsymbol{K}$ of the center of
mass on the spectrum. The $K$-dependent splitting,
which has been discussed in Refs.~\cite{9_1,8,9,100}, 
is now treated using the $K$-dependent Hamiltonian in explicit form
and including the central-cell corrections.
The calculated splittings are in the same order of magnitude as 
the splittings observed experimentally~\cite{9_1,8,9}.

Previous publications on the $1S$ ortho exciton 
polariton~\cite{74,GRE_26,P3_107_12,GRE_28} were confining 
the polariton problem to situations where the $K$-vector is parallel to one of the 
principal symmetry axes of the crystal. This ensures that the polarization of the 
exciton is always parallel to the electric field vector and perpendicular to the 
wave vector. What is still missing is a proper treatment of the general polariton 
problem with an \emph{arbitrary} orientation of the $K$-vector, which considers
the $K$-dependent splitting and the polariton effect simultaneously. The challenge is the anisotropy 
of the exciton-photon interaction, leading to a polarization that is neither parallel 
to the electric field nor orthogonal to the wave vector. Solving this very complex problem 
is a significant step ahead in a proper treatment 
of the exciton polariton problem. We are able to show that the general polariton problem 
can be reduced to solving an eigenvalue problem involving two photon states and three exciton 
states; thus giving rise to five distinct polariton branches. The general case leads to 
polariton states with transverse and longitudinal polarization components. These hybrid 
polaritons states can therefore be regarded as a mixture between a transverse polariton 
and a longitudinal exciton.  

Using the $5\times 5$ matrix model for the anisotropic dispersion of the 
$1S$ ortho exciton-polariton we are able to calculate
the group velocity, the oscillator strengths,
and spectra for different orientations of $\boldsymbol{K}$
and to compare them with experimental results~\cite{P3_107_12,GRE_28,GRE_26,8}.
We especially discuss the two cases of the polariton spectra
in the planes perpendicular to
$[1\bar{1}0]$ and $[111]$. In the first case the vector 
$\boldsymbol{K}$ is oriented in one of the six mirror planes 
of the cubic group $O_{\mathrm{h}}$ so that
the polariton states can be classified according to two different irreducible
representations of the group $C_{\mathrm{s}}$. As regards the second case, there
is no such symmetry left and five polariton branches are obtained
independently from the polarization of light. We are able to
compare our results with experimental transmission spectra.
The good agreement between theory and experiment not only
emphasizes the special nature of the $1S$ ortho exciton polariton
but also proves the validity of the $5\times 5$ matrix model 
as a useful theoretical model for further investigations 
on this polariton.

Our investigations with respect to the exciton-polariton problem 
in $\mathrm{Cu_{2}O}$ are gaining 
additional importance by recent second harmonic generation (SHG) 
measurements performed by J.~Mund~\emph{et~al.}~\cite{93}. The experiments were carried 
out with a laser generating 
ultra short pulses having a band width covering several meV. The resonant excitation of 
the $1S$ ortho exciton by means of a two-photon absorption process generates exciton 
polaritons that are giving rise to SHG radiation. The polariton nature of the quadrupole 
exciton explains why it is possible to generate SHG in a crystal with inversion symmetry. 

The paper is organized as follows:
In Sec.~\ref{sec:Theory} we present the
Hamiltonian of excitons in $\mathrm{Cu_{2}O}$
when considering a finite momentum $\hbar K$ 
of the center of mass and
show how to solve the corresponding Schr\"odinger equation 
in a complete basis.
In Sec.~\ref{sec:Oscillator} we present formulas to calculate
dipole and quadrupole oscillator oscillator strengths.
Having introduced the multi-polariton concept of exciton-polartions
in Sec.~\ref{sub:polaritons-without}, we discuss the rotating wave approximation, the nonanalytic
exchange interaction and criteria for the observability of polariton effects 
in Secs.~\ref{sub:rwa}, \ref{sub:NAexchange},
and~\ref{sub:criteria}, respectively.
In Sec.~\ref{sec:complete} we then use the Hamiltonian, 
which accounts for the complete valence band
structure, the exchange interaction and the central-cell corrections
to calculate the complex dispersion
of the polaritons connected with the excitons states of $2\leq n\leq 4$.
We also discuss the observability of polariton effects.
The $K$ dependent splitting and the dispersion of the $1S$ 
ortho exciton polariton are treated separately from the other polaritons
in Sec.~\ref{sec:complete1S}. After the discussion of the 
splitting in Sec.~\ref{sec:1S1}, we introduce the $5\times 5$
matrix model in Sec.~\ref{sec:1S}. The dispersion of the
ortho exciton polariton, the group velocity and the oscillator
strengths are compared with experimental results in Sec.~\ref{sec:1S2}.
We finally give a summary and outlook in Sec.~\ref{sec:Summary}.

\section{Hamiltonian \label{sec:Theory}}

In this section we shortly present the theory of
excitons with a finite momentum of the center of mass
in $\mathrm{Cu_{2}O}$, where the cubic valence band structure,
the exchange interaction and the central-cell corrections 
need to be considered.

The Hamiltonian of the exciton is given by
\begin{eqnarray}
H & = & E_{\mathrm{g}}+V\left(\boldsymbol{r}_{e}-\boldsymbol{r}_{h}\right)+H_{\mathrm{e}}\left(\boldsymbol{p}_{\mathrm{e}}\right)+H_{\mathrm{h}}\left(\boldsymbol{p}_{\mathrm{\mathrm{h}}}\right)\nonumber \\
 & + & H_{\mathrm{CCC}}\left(\boldsymbol{r}_{e}-\boldsymbol{r}_{h}\right)+H_{\mathrm{exch}}\left(\boldsymbol{r}_{e}-\boldsymbol{r}_{h}\right)\label{eq:Hpeph}
\end{eqnarray}
with the energy $E_{\mathrm{g}}$ of the band gap.
The screened Coulomb interaction $V$,
the central cell corrections
$H_{\mathrm{CCC}}$, the analytic exchange exchange interaction
$H_{\mathrm{exch}}$ and the kinetic energies $H_{\mathrm{e}}\left(\boldsymbol{p}_{\mathrm{e}}\right)$ 
and $H_{\mathrm{h}}\left(\boldsymbol{p}_{\mathrm{h}}\right)$ of the electron and the
hole are given in Ref.~\cite{200}.

We now introduce relative 
and center-of-mass coordinates
\begin{subequations}
\begin{align}
\boldsymbol{r} = &\: \boldsymbol{r}_{\mathrm{e}}-\boldsymbol{r}_{\mathrm{h}},\\
\displaybreak[1]
\boldsymbol{R} = &\: \boldsymbol{\alpha}\boldsymbol{r}_{\mathrm{e}}+\boldsymbol{\gamma}\boldsymbol{r}_{\mathrm{h}}.
\end{align}
\label{eq:rRa}%
\end{subequations}
The factors $\boldsymbol{\alpha}$ and $\boldsymbol{\gamma}$ are in general $3\times 3$ matrices
with $\left|\det\left(\boldsymbol{\alpha}+\boldsymbol{\gamma}\right)\right|=1$~\cite{17_15}.
As the Hamiltonian~(\ref{eq:Hpeph}) depends
only on the relative coordinate $\boldsymbol{r}=\boldsymbol{r}_{\mathrm{e}}-\boldsymbol{r}_{\mathrm{h}}$
of electron and hole, the momentum of the center of
mass 
\begin{equation}
\boldsymbol{P}=\boldsymbol{p}_{\mathrm{e}}+\boldsymbol{p}_{\mathrm{h}}=-i\hbar\left(\boldsymbol{\alpha}+\boldsymbol{\gamma}\right)\nabla_{\boldsymbol{R}}
\end{equation}
with
\begin{subequations}
\begin{align}
\nabla_{\boldsymbol{r}_{\mathrm{e}}} = &\: \boldsymbol{\alpha}\nabla_{\boldsymbol{R}}+\nabla_{\boldsymbol{r}},\\
\displaybreak[1]
\nabla_{\boldsymbol{r}_{\mathrm{h}}} = &\: \boldsymbol{\gamma}\nabla_{\boldsymbol{R}}-\nabla_{\boldsymbol{r}},
\end{align}
\label{eq:pPa}%
\end{subequations}
is a constant of motion, i.e., we can set $\boldsymbol{P}=\hbar\boldsymbol{K}$~\cite{24,17_15}.
According to Ref.~\cite{17_15}, the matrices $\boldsymbol{\alpha}$ and $\boldsymbol{\gamma}$
can be chosen arbitrarily, as long as $\left|\det\left(\boldsymbol{\alpha}+\boldsymbol{\gamma}\right)\right|=1$ holds, 
but should be adapted to the problem.
Note that if we insert Eqs.~(\ref{eq:rRa}) and~(\ref{eq:pPa}) into the Hamiltonian~(\ref{eq:Hpeph})
in general a coupling term between the relative motion and the motion
of center of mass appears.
Only for a specific choice of $\boldsymbol{\alpha}$ and $\boldsymbol{\gamma}$ this coupling term vanishes.
However, the correct values of 
$\boldsymbol{\alpha}$ and $\boldsymbol{\gamma}$ are difficult to find~\cite{24,100}.
In particular, the generalized relative
and center of mass coordinate transformation of Ref.~\cite{100} holds only if the parameters
$\eta_i$ in the kinetic energy of the hole are set to zero (cf.~Ref.~\cite{100}).
When assuming isotropic coefficients $\boldsymbol{\alpha}=\alpha\boldsymbol{1}$ and
$\boldsymbol{\gamma}=(1-\alpha)\boldsymbol{1}$ it is not possible to find a constant
value of $\alpha$ for which the relative motion and the motion of the center of mass are
decoupled. This is connected to the fact that the exciton mass in $\mathrm{Cu_{2}O}$
is not isotropic~\cite{100}. Hence, the more complicated generalized transformation with $3\times 3$
matrices would be needed.

Nevertheless, the results of the Hamiltonian are independent of the choice
of $\boldsymbol{\alpha}$ and $\boldsymbol{\gamma}$. As we extend the theory
of excitons in semiconductors with a cubic valence band structure
for polariton effects, we simply
use the coordinates and momenta 
of relative and center of mass motion 
with $\boldsymbol{\alpha}=m_{\mathrm{e}}/\left(m_{\mathrm{e}}+m_{\mathrm{h}}\right)\boldsymbol{1}$ and 
$\boldsymbol{\gamma}=m_{\mathrm{h}}/\left(m_{\mathrm{e}}+m_{\mathrm{h}}\right)\boldsymbol{1}$
known from the hydrogen atom in the following:
\begin{subequations}
\begin{align}
\boldsymbol{r} = &\: \boldsymbol{r}_{\mathrm{e}}-\boldsymbol{r}_{\mathrm{h}},\\
\displaybreak[1]
\boldsymbol{R} = &\: \left(m_{\mathrm{e}}\boldsymbol{r}_{\mathrm{e}}+m_{\mathrm{h}}\boldsymbol{r}_{\mathrm{h}}\right)/\left(m_{\mathrm{e}}+m_{\mathrm{h}}\right),\\
\displaybreak[1]
\boldsymbol{p} = &\: \left(m_{\mathrm{h}}\boldsymbol{p}_{\mathrm{e}}-m_{\mathrm{e}}\boldsymbol{p}_{\mathrm{h}}\right)/\left(m_{\mathrm{e}}+m_{\mathrm{h}}\right),\\
\displaybreak[1]
\boldsymbol{P} = &\: \boldsymbol{p}_{\mathrm{e}}+\boldsymbol{p}_{\mathrm{h}}=\hbar\boldsymbol{K}.
\end{align}
\label{eq:rRpPa}%
\end{subequations}
We can then write the Hamiltonian in the form
\begin{equation}
H=H_{0}+\left(\hbar K\right)H_{1}+\left(\hbar K\right)^{2}H_{2}.\label{eq:Hges}
\end{equation}
The first part $H_0$ is exactly the Hamiltonian
of relative motion presented and discussed in Ref.~\cite{200}, while
the last part $H_2$ describes the motion of the center of mass in the degenerate
band case. The term $H_{1}$ depends on the relative momentum $\boldsymbol{p}$
and thus couples the relative motion and the motion of the center of mass.

\begin{table}[t]

\protect\caption{Material parameters used in the calculations.
\label{tab:1}}

\begin{centering}
\begin{tabular}{lll}
\hline 
band gap energy & $E_{\mathrm{g}}=2.17208\,\mathrm{eV}$ & \cite{GRE}\tabularnewline
electron mass & $m_{\mathrm{e}}=0.99\, m_{0}$ & \cite{M2}\tabularnewline
spin-orbit coupling & $\Delta=0.131\,\mathrm{eV}$ & \cite{80}\tabularnewline
valence band parameters & $\gamma_{1}=1.76$ & \cite{80,100}\tabularnewline
 & $\gamma_2=0.7532$ & \cite{80,100}\tabularnewline
 & $\gamma_3=-0.3668$ & \cite{80,100}\tabularnewline
 & $\eta_1=-0.020$ & \cite{80,100}\tabularnewline 
 & $\eta_2=-0.0037$ & \cite{80,100}\tabularnewline
 & $\eta_3=-0.0337$ & \cite{80,100}\tabularnewline
lattice constant & $a=0.42696\,\mathrm{nm}$ & \cite{20_24}\tabularnewline
dielectric constants & $\varepsilon_{\mathrm{s}1}=7.5$ & \cite{SOK1_82L1}\tabularnewline
 & $\varepsilon_{\mathrm{b}1}=\varepsilon_{\mathrm{s}2}=7.11$ & \cite{SOK1_82L1}\tabularnewline
 & $\varepsilon_{\mathrm{b}2}=6.46$ & \cite{SOK1_82L1}\tabularnewline
energy of $\Gamma_{4}^{-}$-LO phonons & $\hbar\omega_{\mathrm{LO1}}=18.7\,\mathrm{meV}$ & \cite{1}\tabularnewline
 & $\hbar\omega_{\mathrm{LO2}}=87\,\mathrm{meV}$ & \cite{1}\tabularnewline
exchange energy & $J_0=0.792\,\mathrm{eV}$ & \cite{200}\tabularnewline
central-cell correction & $V_0=0.539\,\mathrm{eV}$ & \cite{200}\tabularnewline
\hline 
\end{tabular}
\par\end{centering}

\end{table}

For the case that the wave vector $\boldsymbol{K}$ is oriented along one of the
directions of high symmetry, i.e., along $[001]$, $[110]$
or $[111]$,
one can rotate the coordinate system to make
the quantization axis coincide with the direction of $\boldsymbol{K}$
and then express the Hamiltonian~(\ref{eq:Hges}) in terms of irreducible tensors~\cite{ED,7,7_11,44}.
Explicit expressions for $H_0$, $H_1$, and $H_2$ for these special orientations of the 
wave vector $\boldsymbol{K}$ are given in 
the Supplemental Material~\cite{300_SM}.

The Schr\"odinger equation corresponding to the Hamiltonian~(\ref{eq:Hges})
is solved using the method presented in Refs.~\cite{100,125,200}
with a complete basis.
The ansatz for the exciton wave function now reads
\begin{subequations}
\begin{eqnarray}
\left|\Psi_{\nu\boldsymbol{K}}\right\rangle  & = & \sum_{NLJFF_{t}M_{F_{t}}}c_{NLJFF_{t}M_{F_{t}}}^{\nu\boldsymbol{K}}\left|\Pi\right\rangle,\\
\nonumber \\
\left|\Pi\right\rangle  & = & \left|N,\, L;\,\left(I,\, S_{\mathrm{h}}\right)\, J;\, F,\, S_{\mathrm{e}};\, F_{t},\, M_{F_{t}}\right\rangle\label{eq:basis}
\end{eqnarray}\label{eq:ansatz}%
\end{subequations}
with complex coefficients $c$ and the quantum numbers explained in Refs.~\cite{100,125,200}. 
Note that the coefficients here depend on the wave vector $\boldsymbol{K}$, which enters
the Hamiltonian~(\ref{eq:Hges}). 
The index $\nu$ is a number to distinguish the different exciton states.

Inserting the ansatz~(\ref{eq:ansatz}) in the Schr\"odinger
equation $H\Psi=E\Psi$ and multiplying from the left with
another basis state $\left\langle \Pi'\right|$,
we obtain a matrix representation 
of the Schr\"odinger equation of the form
\begin{equation}
\boldsymbol{D}\boldsymbol{c}=E\boldsymbol{M}\boldsymbol{c}.\label{eq:gev}
\end{equation}
The vector $\boldsymbol{c}$ contains the coefficients of the ansatz~(\ref{eq:ansatz}).
All matrix elements of $H_1$ and $H_2$, which enter the 
symmetric matrices $\boldsymbol{D}$ and
$\boldsymbol{M}$ are given in the Supplemental Material~\cite{300_SM}.
The matrix elements of $H_0$ are already given in the Appendices of Refs.~\cite{100} and~\cite{200}.
The generalized eigenvalue problem~(\ref{eq:gev})
can finally be solved using an appropriate LAPACK routine~\cite{Lapack}.

All material parameters used in the calculations are listed in Table~\ref{tab:1}.

\section{Oscillator strengths \label{sec:Oscillator}}

Having solved the generalized eigenvalue problem~(\ref{eq:gev}), one can directly
calculate the relative oscillator strengths for the transitions 
from the ground state of the solid to the exciton states.
In this section we will derive the according formulas for dipole and
quadrupole transitions.
The oscillator strength is strongly connected to the
interaction between the excitons and photons. 
According to Ref.~\cite{TOE} the probability per unit time
for a transition from the ground state $\Phi_0$ of the semiconductor
to an exciton state $\Psi_{vc,\,\nu\boldsymbol{K}}^{\sigma\tau}$ is proportional to $\left|M\right|^2$
with
\begin{equation}
M=\int\Psi_{vc,\,\nu\boldsymbol{K}}^{\sigma\tau *}\left[\frac{-e}{m_0}\boldsymbol{A}_{0}\left(\boldsymbol{\kappa},\xi\right)\sum_{l=1}^{N}e^{i\boldsymbol{\kappa}\boldsymbol{r}_{l}}\boldsymbol{p}_{l}\right]\Phi_{0}\mathrm{d}\boldsymbol{r}_{1}\cdots\mathrm{d}\boldsymbol{r}_{N}.\label{eq:matelemtrans}
\end{equation}
Here $\boldsymbol{A}_{0}\left(\boldsymbol{\kappa},\xi\right)$ denotes the amplitude of the 
vector potential of the radiation field with the wave vector $\boldsymbol{\kappa}$ and the
polarization $\xi$. $N$ denotes the number of electrons with the coordinates $\boldsymbol{r}_{l}$.

Within the scope of the simple band model the wave function
of an exciton is given by
\begin{eqnarray}
\Psi_{vc,\,\nu\boldsymbol{K}}^{\sigma\tau} & = & \sum_{\boldsymbol{q}}f_{vc\,\nu}\left(\boldsymbol{q}\right)\Phi_{vc}^{\sigma\tau}\left(\boldsymbol{q}-\gamma\boldsymbol{K},\,\boldsymbol{q}+\alpha\boldsymbol{K}\right),\label{eq:Psiexc}
\end{eqnarray}
where $\tau$ and $-\sigma$ denote the spins of the electron and the hole, respectively.
As we will discuss below, the spin orbit splitting in $\mathrm{Cu_{2}O}$ does not have a
perceptible effect on the oscillator strength. Hence, we will keep the derivation of the
formula for the oscillator strength more simple by assuming a threefold degenerate
$\Gamma_5^+$ valence band and treating the electron spin and the hole spin separately.

The envelope function $f_{vc\,\nu}\left(\boldsymbol{q}\right)$ in Eq.~(\ref{eq:Psiexc}) is the Fourier transform of the 
solution $F_{vc\,\nu}\left(\boldsymbol{\beta}\right)$ of the Wannier equation~\cite{TOE_5,TOE},
\begin{equation}
f_{vc\,\nu}\left(\boldsymbol{q}\right)=\frac{1}{\sqrt{N}}\sum_{\boldsymbol{\beta}}F_{vc\,\nu}\left(\boldsymbol{\beta}\right)e^{-i\boldsymbol{q}\boldsymbol{\beta}},
\end{equation}
with $\nu$ denoting the different exciton states. 
Note that the coordinate
$\boldsymbol{\beta}$ is a lattice vector which in general takes only discrete
values~\cite{1}. 
This coordinate is identical to the relative coordinate $\boldsymbol{r}$
used in Sec.~\ref{sec:Theory} in the continuum approximation.

The wave function~(\ref{eq:Psiexc}) contains
a Slater determinant of Bloch functions with one electron being
in a Bloch state of the conduction band and $N-1$ electrons in Bloch
states of the valence bands:
\begin{eqnarray}
& & \Phi_{vc}^{\sigma\tau}\left(\boldsymbol{k}_{\mathrm{h}},\,\boldsymbol{k}_{\mathrm{e}}\right) = \nonumber\\
& & \mathcal{A}\psi_{v\boldsymbol{k}_{1}\alpha}\psi_{v\boldsymbol{k}_{1}\beta}\cdots\psi_{v\boldsymbol{k}_{\mathrm{h}}\sigma}\psi_{c\boldsymbol{k}_{\mathrm{e}}\tau}\cdots\psi_{v\boldsymbol{k}_{N}\beta}.\label{eq:slater}
\end{eqnarray}
Here $\mathcal{A}$ denotes the antisymmetrization operator.
With the ground state of the semiconductor
\begin{equation}
\Phi_{0} = \mathcal{A}\psi_{v\boldsymbol{k}_{1}\alpha}\psi_{v\boldsymbol{k}_{1}\beta}\cdots\psi_{v\boldsymbol{k}_{i}\alpha}\psi_{v\boldsymbol{k}_{i}\beta}\cdots\psi_{v\boldsymbol{k}_{N}\beta}\label{eq:slater2}
\end{equation}
we can express the exciton state~(\ref{eq:Psiexc}) as
\begin{equation}
\left|\Psi_{vc,\,\nu\boldsymbol{K}}^{\sigma\tau}\right\rangle = \sum_{\boldsymbol{q}}f_{vc\,\nu}\left(\boldsymbol{q}\right)c_{c\left(\boldsymbol{q}+\alpha\boldsymbol{K}\right)\tau}^{\dagger}c_{v\left(\boldsymbol{q}-\gamma\boldsymbol{K}\right)\sigma}^{\phantom{\dagger}}\left|\Phi_{0}\right\rangle,
\end{equation}
using creation and annihilation operators.
The operator in square brackets in Eq.~(\ref{eq:matelemtrans}) can be written in second quantization as
\begin{eqnarray}
& & \sum_{nn'}\sum_{\boldsymbol{k}\boldsymbol{k}'}\sum_{\sigma'\tau'}\frac{-e}{m_0}\boldsymbol{A}_{0}\left(\boldsymbol{\kappa},\xi\right)\nonumber\\
&  & \qquad\quad\times\left\langle \psi_{n'\boldsymbol{k}'\tau'}\left|e^{i\boldsymbol{\kappa}\boldsymbol{r}}\boldsymbol{p}\right|\psi_{n\boldsymbol{k}\sigma'}\right\rangle c_{n'\boldsymbol{k}'\tau'}^{\dagger}c_{n\boldsymbol{k}\sigma'}.
\end{eqnarray}

After some transformations, using $\boldsymbol{A}_{0}\left(\boldsymbol{\kappa},\xi\right)=A_{0}\left(\boldsymbol{\kappa},\xi\right)\hat{\boldsymbol{e}}_{\xi\boldsymbol{\kappa}}$ and neglecting Umklapp processes, we arrive at 
\begin{eqnarray}
M & = & \frac{-e\hbar}{m_0}A_{0}\left(\boldsymbol{\kappa},\xi\right)N\delta_{\tau\sigma}\delta_{\boldsymbol{\kappa},\boldsymbol{K}}\sum_{\boldsymbol{q}}f_{vc\,\nu}^{*}\left(\boldsymbol{q}\right)\int_{\mathrm{WSC}}\!\!\!\mathrm{d}\boldsymbol{r}\nonumber\\
& \times & u_{c\left(\boldsymbol{q}+\alpha\boldsymbol{K}\right)}^{*}\left(\boldsymbol{r}\right)\hat{\boldsymbol{e}}_{\xi\boldsymbol{\kappa}}
\left[\left(\boldsymbol{q}-\gamma\boldsymbol{K}\right)u_{v\left(\boldsymbol{q}-\gamma\boldsymbol{K}\right)}\left(\boldsymbol{r}\right)\right.\nonumber\\
& & \qquad\qquad\qquad\qquad\qquad-\left.i\nabla u_{v\left(\boldsymbol{q}-\gamma\boldsymbol{K}\right)}\left(\boldsymbol{r}\right)\right]
\label{eq:matfourier2}
\end{eqnarray}
with an integral over the Wigner-Seitz cell (WSC)~\cite{SST}.

To obtain expressions for the dipole and quadrupole oscillator
strength, we use $\boldsymbol{k}\cdot\boldsymbol{p}$ perturbation theory
and keep all terms up to first order in $\boldsymbol{q}$ and $\boldsymbol{K}$.
It is~\cite{SST,1}
\begin{equation}
u_{m\boldsymbol{k}}\left(\boldsymbol{r}\right) \approx u_{m\boldsymbol{0}}\left(\boldsymbol{r}\right)+\frac{\hbar}{m_{0}}\sum_{n\neq m}\frac{\boldsymbol{k}\cdot\boldsymbol{p}_{nm}}{\left(E_{m}-E_{n}\right)}u_{n\boldsymbol{0}}\left(\boldsymbol{r}\right)+\ldots\label{eq:kppert}
\end{equation}
with $\boldsymbol{p}_{mn}=\left\langle u_{m\boldsymbol{0}}\left|\boldsymbol{p}\right|u_{n\boldsymbol{0}}\right\rangle$
and the energy $E_{n}=E_{n}\left(\boldsymbol{k}=\boldsymbol{0}\right)$ of the band
$n$ at the $\Gamma$ point.
Due to the orthogonality of the Bloch functions,
the first summand in the integral of Eq.~(\ref{eq:matfourier2}) vanishes
up to first order in $\boldsymbol{q}$ and $\boldsymbol{K}$.
We obtain 
\begin{align}
M = & -\frac{e\hbar}{m_{0}^{2}}A_{0}\left(\boldsymbol{\kappa},\xi\right)\delta_{\tau\sigma}\delta_{\boldsymbol{\kappa},\boldsymbol{K}}\sum_{\boldsymbol{q}}f_{vc\,\nu}^{*}\left(\boldsymbol{q}\right)\nonumber \\
\times & \left[\left\langle u_{c\boldsymbol{0}}\right|\hat{\boldsymbol{e}}_{\xi\boldsymbol{\kappa}}\cdot\boldsymbol{p}\left|u_{v\boldsymbol{0}}\right\rangle\right.
\nonumber \\
+ & \left.\left\langle u_{c\boldsymbol{0}}\right|\left(\hat{\boldsymbol{e}}_{\xi\boldsymbol{\kappa}}\cdot\boldsymbol{p}\right)M_v\left(\boldsymbol{p}\cdot\left(\boldsymbol{q}-\gamma\boldsymbol{K}\right)\right)\left|u_{v\boldsymbol{0}}\right\rangle \right.
\nonumber \\
+ & \left.\left\langle u_{c\boldsymbol{0}}\right|\left(\left(\boldsymbol{q}+\alpha\boldsymbol{K}\right)\cdot\boldsymbol{p}\right)M_c\left(\boldsymbol{p}\cdot\hat{\boldsymbol{e}}_{\xi\boldsymbol{\kappa}}\right)\left|u_{v\boldsymbol{0}}\right\rangle \right],\label{eq:transPP}
\end{align}
where we defined 
\begin{equation}
M_m=\sum_{n\neq m}\frac{\left|u_{n\boldsymbol{0}}\right\rangle \left\langle u_{n\boldsymbol{0}}\right|}{\left(E_{m}-E_{n}\right)}.\label{eq:Mm}
\end{equation}
The sum over $\boldsymbol{q}$ can be evaluated using 
\begin{equation}
\frac{1}{\sqrt{N}}\sum_{\boldsymbol{q}}\, q_{i}^{\chi}f_{vc\,\nu}\left(\boldsymbol{q}\right)=\left.(-i)^{\chi}\frac{\partial^{\chi}}{\partial\beta_{i}^{\chi}}F_{vc\,\nu}\left(\boldsymbol{\beta}\right)\right|_{\boldsymbol{\beta}=\boldsymbol{0}}\label{eq:Fourierq}
\end{equation}
and we arrive at
\begin{eqnarray}
M & = & -\frac{e\hbar}{m_{0}^{2}}A_{0}\left(\boldsymbol{\kappa},\xi\right)\sqrt{N}\delta_{\tau\sigma}\delta_{\boldsymbol{\kappa},\boldsymbol{K}}\nonumber\\
& \times & \lim_{\boldsymbol{r}\rightarrow\boldsymbol{0}}\left[\hat{\boldsymbol{e}}_{\xi\boldsymbol{\kappa}}\cdot\boldsymbol{p}_{cv}\,F_{vc\,\nu}^{*}\left(\boldsymbol{r}\right)\phantom{\int}\right.\nonumber\\
& + & \hat{\boldsymbol{e}}_{\xi\boldsymbol{\kappa}}\cdot(\tilde{\boldsymbol{M}}_{v}+\tilde{\boldsymbol{M}}_{c})\cdot(i\nabla_{\boldsymbol{\beta}}F_{vc\,\nu}^{*}\left(\boldsymbol{r}\right))\nonumber\\
& + & \left.\hat{\boldsymbol{e}}_{\xi\boldsymbol{\kappa}}\cdot(-\gamma\tilde{\boldsymbol{M}}_{v}+\alpha\tilde{\boldsymbol{M}}_{c})\cdot(F_{vc\,\nu}^{*}\left(\boldsymbol{r}\right)\boldsymbol{K})\right],
\label{eq:Mfinal}
\end{eqnarray}
where we replaced $\boldsymbol{\beta}$ with $\boldsymbol{r}$ and defined the matrices
$\tilde{\boldsymbol{M}}_{v}$, $\tilde{\boldsymbol{M}}_{c}$ with the components
\begin{subequations}
\begin{align}
(\tilde{\boldsymbol{M}}_{v})_{ij} &=\left\langle u_{c\boldsymbol{0}}\right|p_i M_{v}p_j \left|u_{v\boldsymbol{0}}\right\rangle,\\
(\tilde{\boldsymbol{M}}_{c})_{ij} &=\left\langle u_{c\boldsymbol{0}}\right|p_i M_{c}p_j \left|u_{v\boldsymbol{0}}\right\rangle.
\end{align}
\label{eq:Mtilde}%
\end{subequations}
In $\mathrm{Cu_{2}O}$ the first term or $\boldsymbol{p}_{cv}$ vanishes
since valence and conduction band have the same parity.
The term $\delta_{\tau\sigma}$
has to be replaced by $\sqrt{2}\delta_{S,0}$ when using the eigenstates
of the total spin $S=S_{\mathrm{e}}+S_{\mathrm{h}}=\tau-\sigma$~\cite{TO}.

The operators $M_c$ and $M_v$ are projection operators. 
For reasons of symmetry these operators 
transform according to the irreducible representation $\Gamma_{1}^{+}$ of $O_{\mathrm{h}}$.
On the other hand, the operator $\boldsymbol{p}$ transforms according
to $\Gamma_{4}^{-}$. The symmetry of the operator between the Bloch
functions in Eq.~(\ref{eq:Mtilde}) is therefore
\begin{equation}
\Gamma_{4}^{-}\otimes\Gamma_{1}^{+}\otimes\Gamma_{4}^{-}=\Gamma_{1}^{+}\oplus\Gamma_{3}^{+}\oplus\Gamma_{4}^{+}\oplus\Gamma_{5}^{+}.
\end{equation}
The symmetry of the Bloch functions (without spin) is 
\begin{equation}
\Gamma_{v}\otimes\Gamma_{c}=\Gamma_{5}^{+}\otimes\Gamma_{1}^{+}=\Gamma_{5}^{+}.
\end{equation}
Hence, the expression requires that
the operator has the symmetry $\Gamma_{5}^{+}$~\cite{G1}. We
can then consider the coupling coefficients for the case $\Gamma_{4}^{-}\otimes\Gamma_{4}^{-}\rightarrow\Gamma_{5}^{+}$.
With the basis functions $\left|X\right\rangle $, $\left|Y\right\rangle $,
$\left|Z\right\rangle $ of $\Gamma_{4}^{-}$ and the basis functions
$|\tilde{X}\rangle=\left|YZ\right\rangle $,
$|\tilde{Y}\rangle=\left|ZX\right\rangle $, and $|\tilde{Z}\rangle=\left|XY\right\rangle $
of $\Gamma_{5}^{+}$, we see that, e.g., the $\Gamma_{5}^{+}$ like part of the products
$\left|X\right\rangle _{1}\left|Y\right\rangle _{2}$ and $\left|Y\right\rangle _{1}\left|X\right\rangle _{2}$
transforms as $|\tilde{Z}\rangle /\sqrt{2}$. The 
other expressions are obtained via cyclic permutation~\cite{G3}.


In Sec.~\ref{sec:Theory} we have introduced the quasi spin $I=1$.
If we compare the states $\left|I,\, M_{I}\right\rangle $ with the three
functions $\left|u_{v\boldsymbol{0}}^{xy}\right\rangle$, $\left|u_{v\boldsymbol{0}}^{yz}\right\rangle$, and $\left|u_{v\boldsymbol{0}}^{zx}\right\rangle$
transforming according to $\Gamma_5^+$, it is~\cite{8}
\begin{subequations}
\begin{align}
\left|1,+1\right\rangle _{I} = &\: -\left(\left|u_{v\boldsymbol{0}}^{yz}\right\rangle+i \left|u_{v\boldsymbol{0}}^{zx}\right\rangle\right)/\sqrt{2},\\
\displaybreak[1]
\left|1,0\right\rangle _{I} = &\: \left|u_{v\boldsymbol{0}}^{xy}\right\rangle,\\
\displaybreak[1]
\left|1,-1\right\rangle _{I} = &\: +\left(\left|u_{v\boldsymbol{0}}^{yz}\right\rangle-i \left|u_{v\boldsymbol{0}}^{zx}\right\rangle\right)/\sqrt{2}.
\end{align}
\end{subequations}

In the envelope function of the exciton the
angular dependence is given by the spherical harmonics $Y_{LM}$.
We know that in Eq.~(\ref{eq:Mfinal}) the gradient of the envelope function 
at $\boldsymbol{r}=\boldsymbol{0}$ is non-zero only if $L=1$ holds. Furthermore, 
the envelope function itself vanishes at $\boldsymbol{r}=\boldsymbol{0}$ if $L\neq 0$ holds.

Let us assume that the light is polarized in $x$-direction, i.e., $\hat{\boldsymbol{e}}_{\xi\boldsymbol{\kappa}}=\hat{\boldsymbol{e}}_{x}$.
Let us furthermore write the function $\left|\Psi_{\nu\boldsymbol{K}}\right\rangle$ of the exciton
in the form of Eq.~(\ref{eq:ansatz}), where the spins, the envelope with the angular momentum
$L$ and the function of the hole with the quasi-spin $I$ enter.
The dipole-term in Eq.~(\ref{eq:Mfinal}) is then proportional to
\begin{widetext}
\begin{align}
 & \lim_{\boldsymbol{r}\rightarrow\boldsymbol{0}}\left\langle S=0,\,M_S=0\right|\left(\left\langle u_{v\boldsymbol{0}}^{xy}\right|\frac{\partial}{\partial y}
+ \left\langle u_{v\boldsymbol{0}}^{zx}\right|\frac{\partial}{\partial z}\right)\left|\Psi_{\nu\boldsymbol{K}}\right\rangle\nonumber \\
\displaybreak[1]
 = & \lim_{r\rightarrow 0}\left\langle S=0,\,M_S=0\right|\frac{\partial}{\partial r}\left(\left\langle I=1,\,M_I=0\right|(-i)\sqrt{\frac{3}{8\pi}}\left(\left\langle L=1,\,M_L=1\right|+\left\langle L=1,\,M_L=-1\right|\right)\right.\nonumber\\
 & \left.\qquad\qquad\qquad\qquad\qquad-\frac{i}{\sqrt{2}}\left(\left\langle I=1,\,M_I=-1\right|+\left\langle I=1,\,M_I=1\right|\right)\sqrt{\frac{3}{4\pi}}\left\langle L=1,\,M_L=0\right|\right)\left|\Psi_{\nu\boldsymbol{K}}\right\rangle\nonumber \\
 \displaybreak[1]
 = & -i\sqrt{\frac{3}{4\pi}}\lim_{r\rightarrow 0}\frac{\partial}{\partial r}\left( _D\left\langle 2,\,1\right|+ _D\left\langle 2,\,-1\right|\right)\left|\Psi_{\nu\boldsymbol{K}}\right\rangle
\end{align}
Here the state $\left|F_t,\,M_{F_t}\right\rangle_D$ 
for the dipole transition ($D$) is a short notation for
\begin{eqnarray}
\left|\left(S_{\mathrm{e}},\,S_{\mathrm{h}}\right)\,S,\,I;\,I+S,\,L;\,F_t,\,M_{F_t}\right\rangle=\left|\left(1/2,\,1/2\right)\,0,\,1;\,1,\,1;\,F_t,\,M_{F_t}\right\rangle,\label{eq:FMD}
\end{eqnarray}
in which the coupling scheme of the spins and angular momenta is different from the one of Eq.~(\ref{eq:basis})
due to the requirement that $S$ must be a good quantum number:
$S_{\mathrm{e}}+S_{\mathrm{h}}=S\rightarrow(I+S)+L=F_t$.
As the quantization axis we choose the $z$-axis, which is parallel
to one of the principal axes of the crystal lattice.
In an analogous way, the quadrupole term can be written as
\begin{align}
 & \lim_{\boldsymbol{r}\rightarrow\boldsymbol{0}}\left\langle S=0,\,M_S=0\right|\left(\left\langle u_{v\boldsymbol{0}}^{xy}\right|K_y
+ \left\langle u_{v\boldsymbol{0}}^{zx}\right|K_z\right)\left|\Psi_{\nu\boldsymbol{K}}\right\rangle\nonumber \\
\displaybreak[1]
 = & \lim_{r\rightarrow 0}\left\langle S=0,\,M_S=0\right|\frac{1}{\sqrt{4\pi}}\left\langle L=0,\,M_L=0\right|
\left(\left\langle I=1,\,M_I=0\right|K_y\right.\nonumber\\
 & \left.\qquad\qquad\qquad\qquad\qquad-\frac{i}{\sqrt{2}}\left(\left\langle I=1,\,M_I=-1\right|+\left\langle I=1,\,M_I=1\right|\right)K_z\right)\left|\Psi_{\nu\boldsymbol{K}}\right\rangle \nonumber\\
\displaybreak[1]
 = & \lim_{r\rightarrow 0}\frac{1}{\sqrt{4\pi}}\left( _Q\left\langle 1,\,0\right|K_y-\frac{i}{\sqrt{2}}\left( _Q\left\langle 1,\,-1\right|+ _Q\left\langle 1,\,1\right|\right)K_z\right)\left|\Psi_{\nu\boldsymbol{K}}\right\rangle
\end{align}
with the state $\left|F_t,\,M_{F_t}\right\rangle_Q$ 
for the quadrupole transition ($Q$) being a short notation for
\begin{eqnarray}
\left|\left(S_{\mathrm{e}},\,S_{\mathrm{h}}\right)\,S,\,I;\,I+S,\,L;\,F_t,\,M_{F_t}\right\rangle=\left|\left(1/2,\,1/2\right)\,0,\,1;\,1,\,0;\,F_t,\,M_{F_t}\right\rangle.
\end{eqnarray}
Note that this state is similar to the one of Eq.~(\ref{eq:FMD}) but only $L$ is set to zero.
We finally arrive at the following expression for the relative oscillator strength:
\begin{eqnarray}
f^{\mathrm{rel}}_{\xi\nu\boldsymbol{K}} = \left| \lim_{r\rightarrow 0}\left[-i\left(\tilde{M}_v^*+\tilde{M}_c^*\right)\frac{\partial}{\partial r}\langle T^D_{\xi\boldsymbol{K}}|\Psi_{\nu\boldsymbol{K}}\rangle+\left(-\gamma\tilde{M}_v^*+\alpha\tilde{M}_c^*\right)\frac{K}{\sqrt{6}}\langle T^Q_{\xi\boldsymbol{K}}|\Psi_{\nu\boldsymbol{K}}\rangle\right]\right|^2.\phantom{i}\label{eq:frelres}
\end{eqnarray}
\end{widetext}

For the dipole term in Eq.~(\ref{eq:frelres}) the
two states $|T^D_{\xi\boldsymbol{K}}\rangle$ are given by
\begin{equation}
|T^D_{\xi\boldsymbol{K}}\rangle = \sum_{i=1}^{3} \hat{e}_{\xi\boldsymbol{K}i}|\pi_i^D\rangle,\qquad\xi=1,\,2
\end{equation}
with the three components of the polarization vector $\hat{\boldsymbol{e}}_{\xi\boldsymbol{K}}$
and three states which transform according 
to $\Gamma_{4}^{-}$~\cite{125}:
\begin{subequations}
\begin{align}
|\pi_x^D\rangle= &\; \frac{i}{\sqrt{2}}\left[\left|2,\,-1\right\rangle_D+\left|2,\,1\right\rangle_D\right],\\
\displaybreak[1]
|\pi_y^D\rangle= &\; \frac{1}{\sqrt{2}}\left[\left|2,\,-1\right\rangle_D-\left|2,\,1\right\rangle_D\right],\\
\displaybreak[1]
|\pi_z^D\rangle= &\; \frac{i}{\sqrt{2}}\left[\left|2,\,-2\right\rangle_D-\left|2,\,2\right\rangle_D\right].
\end{align}
\label{eq:Dxyz}%
\end{subequations}

The state $|T^Q_{\xi\boldsymbol{K}}\rangle$ in the quadrupole term of 
Eq.~(\ref{eq:frelres}) reads
\begin{equation}
|T^Q_{\xi\boldsymbol{K}}\rangle = \sum_{i=1}^{3} \hat{e}_{\xi\boldsymbol{K}i}|\pi_i^Q\rangle,\qquad\xi=1,\,2\label{eq:TQ}
\end{equation}
with the three states which transform according 
to $\Gamma_{5}^{+}$~\cite{125}:
\begin{subequations}
\begin{align}
\left|\pi_x^Q\right\rangle= &\; \hat{K}_y \left|1,\,0\right\rangle_Q\nonumber\\
&\; \qquad+\hat{K}_z \frac{i}{\sqrt{2}}\left[\left|1,\,-1\right\rangle_Q+\left|1,\,1\right\rangle_Q\right],\\
\displaybreak[1]
\left|\pi_y^Q\right\rangle= &\; \hat{K}_x \left|1,\,0\right\rangle_Q\nonumber\\
&\; \qquad+\hat{K}_z \frac{1}{\sqrt{2}}\left[\left|1,\,-1\right\rangle_Q-\left|1,\,1\right\rangle_Q\right],\\
\displaybreak[1]
\left|\pi_z^Q\right\rangle= &\; \hat{K}_y \frac{1}{\sqrt{2}}\left[\left|1,\,-1\right\rangle_Q-\left|1,\,1\right\rangle_Q\right]\nonumber\\
&\; \qquad+\hat{K}_x \frac{i}{\sqrt{2}}\left[\left|1,\,-1\right\rangle_Q+\left|1,\,1\right\rangle_Q\right].
\end{align}
\label{eq:Qxyz}%
\end{subequations}
If we now set $K=0$ in Eq.~(\ref{eq:frelres}), we see that we have derived the expression
for the relative oscillator strength, which has already been used in Refs.~\cite{100,125}.

We can finally make an assumption as regards the size of the parameters
$\tilde{M}_v$ and $\tilde{M}_c$. Since in $\mathrm{Cu_{2}O}$ the uppermost valence bands as well as the lowermost conduction
band have positive parity, we see 
from Eq.~(\ref{eq:Mtilde})
that only bands with negative parity will contribute to the sums in Eq.~(\ref{eq:Mm}).
In $\mathrm{Cu_{2}O}$ there are only two bands of negative parity, which
lie $449\,\mathrm{meV}$ above the lowest conduction band and
$5.6\,\mathrm{eV}$ below the highest conduction band~\cite{1}.
Hence, as regards the denominators of the form $\left(E_m-E_n\right)$ in $M_c$ and $M_v$
the energy difference between the $\Gamma_7^+$ and the $\Gamma_8^+$
valence band due to the spin-orbit coupling
is negligible small in comparion to the energy difference between these bands
and the bands of negative parity.
Thus, the spin-orbit coupling does not have a sizeable effect
on the oscillator strength.
Furthermore, the denominator $\left(E_c-E_n\right)$ in $M_c$ is much smaller than
the denominator $\left(E_v-E_n\right)$ in $M_v$ and it is $M_c\gg M_v$. 
We therefore neglect $M_v$ in this paper.
As a consequence, to obtain absolute oscillator strengths, only one parameter
$\eta$ in the final expression
\begin{equation}
f_{\xi\nu\boldsymbol{K}}=\eta\left| \lim_{r\rightarrow 0}\left[-i\frac{\partial}{\partial r}\langle T^D_{\xi\boldsymbol{K}} |\Psi_{\nu\boldsymbol{K}}\rangle +\frac{\alpha K}{\sqrt{6}}\langle T^Q_{\xi\boldsymbol{K}} |\Psi_{\nu\boldsymbol{K}}\rangle\right]\right|^2\label{eq:fgesres}
\end{equation}
has to be determined via a comparison with experimental values.
Note the specific form of Eq.~(\ref{eq:fgesres}), where the sum
of dipole and quadrupole matrix element is squared. 
In contrast to the hydrogen-like
model, inversion symmetry is broken when considering the
more precise model of $\mathrm{Cu_{2}O}$. Hence, the exciton 
states are mixed-parity states and, thus, an interference term occurs.
Especially for the $1S$ ortho exciton both terms in Eq.~(\ref{eq:fgesres})
are of the same magnitude. The dipole term becomes equally large
as the quadrupole term for $K\neq 0$ due to a $K$ dependent admixture
of $P$ excitons to the $1S$ ortho exciton.
Even though the parameter $\alpha$ occurs in Eq.~(\ref{eq:fgesres}),
the whole expression is independent of its value. This is related to the fact that
the wave function of the exciton also depends on $\alpha$.

In this paper we choose $\boldsymbol{K}$ to be oriented in
$[001]$, $[110]$, or $[111]$ direction and, as in Ref.~\cite{125},
we also rotate the coordinate system to make the
$z$ axis of the new coordinate system coincide 
with the direction of $\boldsymbol{K}$.
The formulas for the oscillator 
strengths for the three orientations of $\boldsymbol{K}$~\cite{300_SO} can be derived 
by analogy with Ref.~\cite{125}.
The consideration of these three directions of high
symmetry is sufficient to determine those parameters in Sec.~\ref{sec:complete1S},
which are needed in the $5\times 5$ matrix model to calculate the
polariton dispersion for any other direction of $\boldsymbol{K}$.

\section{Exciton-polaritons \label{sec:exciton-polaritons}}

In this section we recapitulate the
quantum mechanical theory of exciton-polaritons in Sec.~\ref{sub:polaritons-without}
and discuss the rotating-wave approximation in Sec.~\ref{sub:rwa}.
To obtain the correct treatment of the $K$-dependent problem, we consider
the nonanalytic exchange interaction in Sec.~\ref{sub:NAexchange}.
In Sec.~\ref{sub:criteria} we shortly present the criteria for the observability of
polariton effects.

\subsection{Polariton transformation\label{sub:polaritons-without}}

The quantum mechanical theory of polaritons was first developed by Hopfield, 
Fano, and Agranovich~\cite{P2,TOE_46a,TOE_206}. According to Refs.~\cite{83,P1,P2},
the second-quantized Hamiltonian for the interaction of excitons and photons
\begin{eqnarray}
H & = & \sum_{\xi\boldsymbol{K}}\left[\hbar\omega_{\xi\boldsymbol{K}}\left(a_{\xi\boldsymbol{K}}^{\dagger}a_{\xi\boldsymbol{K}}+\frac{1}{2}\right)\right.\nonumber\\
\nonumber\\
 & + & \left.\sum_{\nu}E_{\nu\boldsymbol{K}}\left(B_{\nu\boldsymbol{K}}^{\dagger}B_{\nu\boldsymbol{K}}+\frac{1}{2}\right)\right.\nonumber \\
\nonumber \\
 & + & \left.i\sum_{\nu}C_{\xi\,\nu\boldsymbol{K}}\left(a_{\xi\boldsymbol{K}}^{\dagger}+a_{\xi-\boldsymbol{K}}\right)\left(B_{\nu\boldsymbol{K}}-B_{\nu-\boldsymbol{K}}^{\dagger}\right)\right.\nonumber\\
\nonumber\\
 & + & \left.\sum_{\nu}D_{\xi\,\nu\boldsymbol{K}}\left(a_{\xi\boldsymbol{K}}^{\dagger}+a_{\xi-\boldsymbol{K}}\right)\left(a_{\xi\boldsymbol{K}}+a_{\xi-\boldsymbol{K}}^{\dagger}\right)\right]\label{eq:polarH}
\end{eqnarray}
can be derived either from a microscopic model of excitons with the Hamiltonian describing the interaction
between radiation and matter or from the equation of motion for the exciton polarization.
In the Hamiltonian~(\ref{eq:polarH}) the operators $B_{\nu\boldsymbol{K}}^{\dagger}$ and
$B_{\nu\boldsymbol{K}}^{\phantom{\dagger}}$ create and annihilate an exciton with
energy $E_{\nu\boldsymbol{K}}$, respectivley,
and obey Bose commutation rules in the following. Likewise, the operator $a_{\xi\boldsymbol{K}}^{\dagger}$ 
($a_{\xi\boldsymbol{K}}^{\phantom{\dagger}}$) creates (annihilates) a photon with polarization $\xi$ and
energy $\hbar\omega_{\xi\boldsymbol{K}}=\hbar cK/\sqrt{\varepsilon_{\mathrm{b2}}}$.
The coupling coefficients in the exciton-photon and the photon-photon
interaction terms of Eq.~(\ref{eq:polarH}) are given by
\begin{equation}
C_{\xi\,\nu\boldsymbol{K}}=\left[\frac{\kappa_{\mathrm{SI}}\pi\beta_{\xi\nu\boldsymbol{K}}E_{\nu\boldsymbol{K}}^{3}}{\varepsilon_{\mathrm{b2}}\hbar\omega_{\xi\boldsymbol{K}}}\right]^{\frac{1}{2}}
\end{equation}
with $\kappa_{\mathrm{SI}}=1/4\pi\varepsilon_0$ and
\begin{equation}
D_{\xi\,\nu\boldsymbol{K}}=C_{\xi\,\nu\boldsymbol{K}}^{2}/E_{\nu\boldsymbol{K}}.
\end{equation}
The polarizability $\beta_{\xi\nu\boldsymbol{K}}$ is proportional to the oscillator strength
of the exciton state. With our definition of the oscillator strength 
$f_{\xi\nu\boldsymbol{K}}$ (cf.~Refs.~\cite{GRE_26,GRE_28,P3_107_12}) this proportionality is given by~\cite{P3_x,P3,83}
\begin{equation}
\beta_{\xi\nu\boldsymbol{K}}=\varepsilon_{0}\varepsilon_{\mathrm{b2}} f_{\xi\nu\boldsymbol{K}}.
\end{equation}

The Hamiltonian~(\ref{eq:polarH}) can be diagonalized by the Hopfield transformation~\cite{TOE_46a,P2},
which is similar to the Bogolyubov's $uv$ transformation~\cite{83_36,P1_6}:
New creation and annihilation operators $p_{\mu\xi\boldsymbol{K}}^{\dagger}$ and $p_{\mu\xi\boldsymbol{K}}$ are introduced via
\begin{subequations}
\begin{eqnarray}
a_{\xi\boldsymbol{K}} & = & \sum_{\mu}\left[u_{\mu\xi\boldsymbol{K}}p_{\mu\xi\boldsymbol{K}}+v_{\mu\xi-\boldsymbol{K}}^{*}p_{\mu\xi-\boldsymbol{K}}^{\dagger}\right],\\
\nonumber \\
B_{\nu\boldsymbol{K}} & = & \sum_{\mu}\left[u_{\mu\xi\nu\boldsymbol{K}}p_{\mu\xi\boldsymbol{K}}+v_{\mu\xi\nu-\boldsymbol{K}}^{*}p_{\mu\xi-\boldsymbol{K}}^{\dagger}\right],
\end{eqnarray}
\end{subequations}
to obtain the polariton Hamiltonian
\begin{equation}
H=\sum_{\mu\xi\boldsymbol{K}}E_{\mu\xi\boldsymbol{K}}p_{\mu\xi\boldsymbol{K}}^{\dagger}p_{\mu\xi\boldsymbol{K}}+\mathrm{const}.
\end{equation}
with $\mu$ and $E_{\mu\xi\boldsymbol{K}}$ denoting the polariton branches
and the polariton energies, respectively.
The new operators must obey Bose commutation relations
and the Hamiltonian shall be diagonal, i.e., 
\begin{equation}
\left[p_{\mu\xi\boldsymbol{K}},H\right]=E_{\mu\xi\boldsymbol{K}}p_{\mu\xi\boldsymbol{K}}\label{eq:pHk}
\end{equation}
must hold.
This provides the following conditional equation for the polariton energies~\cite{P1,TOE_206,P1_7}:
\begin{equation}
\frac{\hbar^{2}c^{2}K^{2}}{E_{\mu\xi\boldsymbol{K}}^{2}}=\varepsilon_{\mathrm{b2}}+\sum_{\nu}\frac{4\pi\kappa_{\mathrm{SI}}\beta_{\xi\nu\boldsymbol{K}}}{1-\left(E_{\mu\xi\boldsymbol{K}}/E_{\nu\boldsymbol{K}}\right)^{2}}.\label{eq:polarener}
\end{equation}

Using the phase convention of Hopfield~\cite{P2}, the soulitons
for the coefficients $u$ and $v$ of the polariton transformation can be obtained
(see Ref.~\cite{P1}).
The polariton operators can also be expressed in terms of exciton and photon operators:
\begin{eqnarray}
p_{\mu\xi\boldsymbol{K}} & = & w_{\mu\xi\boldsymbol{K}}^{(1)}a_{\xi\boldsymbol{K}}+w_{\mu\xi-\boldsymbol{K}}^{(2)*}a_{\xi-\boldsymbol{K}}^{\dagger}\nonumber \\
\nonumber \\
 & + & \sum_{\nu}\left[z_{\mu\xi\nu\boldsymbol{K}}^{(1)}B_{\nu\boldsymbol{K}}+z_{\mu\xi\nu-\boldsymbol{K}}^{(2)*}B_{\nu-\boldsymbol{K}}^{\dagger}\right].
\end{eqnarray}
Since all creation and annihilation operators of the three (quasi-) particles obey Bose commutation
relations, we can determine the coefficients $w$ and $z$ by evaluating
\begin{subequations}
\begin{alignat}{2}
\left[p_{\mu\xi\boldsymbol{K}},\, a_{\xi\boldsymbol{K}}^{\dagger}\right] & = +w_{\mu\xi\boldsymbol{K}}^{(1)} = u_{\mu\xi\boldsymbol{K}}^{*},\\
\left[p_{\mu\xi\boldsymbol{K}},\, a_{\xi-\boldsymbol{K}}^{\phantom{\dagger}}\right] & = -w_{\mu\xi\boldsymbol{K}}^{(2)*} = v_{\mu\xi\boldsymbol{K}}^{*}
\end{alignat}
\end{subequations}
and
\begin{subequations}
\begin{alignat}{2}
\left[p_{\mu\xi\boldsymbol{K}},\, B_{\nu-\boldsymbol{K}}^{\dagger}\right] & = +z_{\mu\xi\nu\boldsymbol{K}}^{(1)} = u_{\mu\xi\nu\boldsymbol{K}}^{*},\\
\left[p_{\mu\xi\boldsymbol{K}},\, B_{\nu\boldsymbol{K}}^{\phantom{\dagger}}\right] & = -z_{\mu\xi\nu\boldsymbol{K}}^{(2)*} = v_{\mu\xi\nu\boldsymbol{K}}^{*}.
\end{alignat}
\end{subequations}
The coefficients $w_{\mu\xi\boldsymbol{K}}^{(i)}$ or the sum
\begin{equation}
W_{\mu\xi\boldsymbol{K}} = \sum_{i=1}^{2}|w_{\mu\xi\boldsymbol{K}}^{(i)}|^2\label{eq:photlike}
\end{equation}
then allow one to determine
whether the polariton is more photon-like $(W_{\mu\xi\boldsymbol{K}}\rightarrow 1)$
or more exciton-like $(W_{\mu\xi\boldsymbol{K}}\rightarrow 0)$.

\subsection{Rotating-wave approximation\label{sub:rwa}}

In the literature polaritons are often treated within 
the so-called rotating-wave approximation~\cite{83}.
In this case
the term with the coefficient $D$ and the anti-resonant terms
of the form $aB$ and $a^{\dagger}B^{\dagger}$ are neglected in the Hamiltonian~(\ref{eq:polarH}).
The resulting Hamiltonian
\begin{eqnarray}
H & = & \sum_{\xi\boldsymbol{K}}\left[\hbar\omega_{\xi\boldsymbol{K}}a_{\xi\boldsymbol{K}}^{\dagger}a_{\xi\boldsymbol{K}}+\sum_{\nu}E_{\nu\boldsymbol{K}}B_{\nu\boldsymbol{K}}^{\dagger}B_{\nu\boldsymbol{K}}\right.\nonumber\\
 & & \qquad+\left.\sum_{\nu}C_{\xi\,\nu\boldsymbol{K}}\left(a_{\xi\boldsymbol{K}}^{\dagger}B_{\nu\boldsymbol{K}}+a_{\xi\boldsymbol{K}}B_{\nu\boldsymbol{K}}^{\dagger}\right)\right]\label{eq:polarJC}
\end{eqnarray}
is then called the Jaynes-Cummings Hamiltonian~\cite{P3_32},
where the vacuum energy of the photons has also been neglected~\cite{83} and where the operators 
$a_{\xi\boldsymbol{K}}$ have been replaced with $ia_{\xi\boldsymbol{K}}$.
Note that this replacement does not change the physics of the problem since it only adds global phases to the
occupation-number states. The occupation-number operator and the commutation relations remain
unchanged.

The coefficient $C_{\xi\,\nu\boldsymbol{K}}$
can be written as~\cite{83,P3}
\begin{equation}
C_{\xi\,\nu\boldsymbol{K}}=\left[\frac{\kappa_{\mathrm{SI}}\pi\beta_{\xi\nu\boldsymbol{K}}E_{\nu\boldsymbol{K}}^{3}}{\varepsilon_{\mathrm{b2}}\hbar\omega_{\xi\boldsymbol{K}}}\right]^{\frac{1}{2}}\approx \frac{1}{2}\left(\frac{K_0}{K}\right)^{\frac{1}{2}}\hbar\Omega_{\mathrm{R}}\label{eq:CJC}
\end{equation}
with the wave vector at the exciton-photon resonance $K_0=E_{\nu\boldsymbol{K}_0}\sqrt{\varepsilon_{\mathrm{b2}}}/\hbar c$
and the Rabi frequency 
\begin{equation}
\Omega_{\mathrm{R}}=E_{\nu\boldsymbol{K}_0}\sqrt{\frac{4\pi\kappa_{\mathrm{SI}}\beta_{\xi\nu\boldsymbol{K}}}{\varepsilon_{\mathrm{b2}}\hbar^2}}.\label{eq:Omegar}
\end{equation}
The rotating wave approximation is generally valid if $\hbar\Omega_{\mathrm{R}}\ll E_{\nu\boldsymbol{K}_0}$ holds.
This is, e.g., the case for anorganic semiconductors and especially for $\mathrm{Cu_{2}O}$~\cite{P3}.

Close to the resonance $\left(K\approx K_0\right)$ one can assume $C_{\xi\,\nu\boldsymbol{K}}\approx\hbar\Omega_{\mathrm{R}}/2$.
Note that for $K\rightarrow 0$ the coupling constant~(\ref{eq:CJC}) diverges, which is \emph{a manifestation of the infrared catastrophe in 
quantum electrodynamics}~\cite{83}. Hence, the simplifications made above are valid only in the vicinity
of the exciton-photon resonance. Otherwise, the full Hamiltonian~(\ref{eq:polarH}) has to be diagonalized.

In the rotating-wave approximation the polariton transformation
is more simple as there is no interaction between states with different
values of $\boldsymbol{K}$. Using the ansatz
\begin{eqnarray}
p_{\mu\xi\boldsymbol{K}} & = & w_{\mu\xi\boldsymbol{K}}a_{\xi\boldsymbol{K}}+\sum_{i}z_{\mu\xi\nu_{i}\boldsymbol{K}}B_{\nu_{i}\boldsymbol{K}},
\end{eqnarray}
the Bose commutation relations of the creation and annihilation operators,
and the condition~(\ref{eq:pHk})
for the polariton operator, one ends up with
the eigenvalue problem
\begin{equation}
\boldsymbol{P}_{\xi\{\nu\}\boldsymbol{K}}\boldsymbol{z}_{\mu\xi\{\nu\}\boldsymbol{K}}=E_{\mu\xi\boldsymbol{K}}\boldsymbol{z}_{\mu\xi\{\nu\}\boldsymbol{K}}\label{eq:rwaev}
\end{equation}
with
\begin{widetext}
\begin{equation}
\boldsymbol{P}_{\xi\{\nu\}\boldsymbol{K}}=\left(\begin{array}{cccccc}
\hbar\omega_{\xi\boldsymbol{K}} & \frac{1}{2}\hbar\Omega_{\mathrm{R},\,\nu_{1}} & \frac{1}{2}\hbar\Omega_{\mathrm{R},\,\nu_{2}} & \cdots & \frac{1}{2}\hbar\Omega_{\mathrm{R},\,\nu_{n}} & \cdots\\
\frac{1}{2}\hbar\Omega_{\mathrm{R},\,\nu_{1}} & E_{\nu_{1}\boldsymbol{K}} & 0 & \cdots & 0 & \cdots\\
\frac{1}{2}\hbar\Omega_{\mathrm{R},\,\nu_{2}} & 0 & E_{\nu_{2}\boldsymbol{K}} &  & \vdots\\
\vdots & \vdots &  & \ddots & 0 & \cdots\\
\frac{1}{2}\hbar\Omega_{\mathrm{R},\,\nu_{n}} & 0 & \cdots & 0 & E_{\nu_{n}\boldsymbol{K}}\\
\vdots & \vdots &  & \vdots &  & \ddots
\end{array}\right)
\qquad\mathrm{and}\qquad
\boldsymbol{z}_{\mu\xi\{\nu\}\boldsymbol{K}}=\left(\begin{array}{c}
w_{\mu\xi\boldsymbol{K}}\\
z_{\mu\xi\nu_{1}\boldsymbol{K}}\\
z_{\mu\xi\nu_{2}\boldsymbol{K}}\\
\vdots\\
z_{\mu\xi\nu_{n}\boldsymbol{K}}\\
\vdots
\end{array}\right).\label{eq:rwaev2}
\end{equation}
\end{widetext}
Knowing the energies $E_{\nu\boldsymbol{K}}$ and the Rabi frequencies
$\Omega_{\mathrm{R},\, \nu}$ of the exciton states, one can
directly obtain the corresponding polariton energies
by determining the eigenvalues of Eq.~(\ref{eq:rwaev}).

Finally, as the polariton is a mixed
state of a photon and excitons, one can again 
determine the photon-like part
\begin{equation}
W_{\mu\xi\boldsymbol{K}} = |w_{\mu\xi\boldsymbol{K}}|^2\label{eq:photlike2}
\end{equation}
of the polariton
or the contribution 
\begin{equation}
Z_{\mu\xi\nu\boldsymbol{K}} = |z_{\mu\xi\nu\boldsymbol{K}}|^2\label{eq:exclike}
\end{equation}
of the exciton with the energy 
$E_{\nu\boldsymbol{K}}$ to the polariton.

\subsection{Nonanalytic exchange interaction\label{sub:NAexchange}}

There is another interaction affecting the exciton
states: the nonanalytic (NA) exchange interaction.  
It is well known that the splitting caused by $H_{\mathrm{exch}}^{\mathrm{NA}}$ 
is identical to the longitudinal-transverse splitting (LT-splitting) when
treating polaritons~\cite{TOE_50}. Hence, it is indispensable
to include the nonanalytic exchange interaction in the theory to
obtain a correct treatment of the complete problem.

In this section we will derive an expression for the nonanalytic exchange interaction.
We start with the formula of Ref.~\cite{150}
for the nonanalytic exchange energy between two exciton 
states $\Psi_{vc,\,\nu\boldsymbol{K}}^{\sigma\tau}$
and $\Psi_{vc,\,\nu'\boldsymbol{K}'}^{\sigma'\tau'}$
in second quantization
\begin{eqnarray}
H_{\mathrm{exch}}^{\mathrm{NA}} & = & \sum_{\nu\nu'\boldsymbol{K}}\frac{m_{\nu\boldsymbol{K}}^{*}m_{\nu'\boldsymbol{K}}^{\phantom{*}}}{\varepsilon_{0}\varepsilon_{\mathrm{b2}} V_{\mathrm{uc}}K^{2}}B_{\nu\boldsymbol{K}}^{\dagger}B_{\nu'\boldsymbol{K}}\label{eq:Hexch}
\end{eqnarray}
with the volume $V_{\mathrm{uc}}$ of one unit cell and
\begin{widetext}
\begin{align}
m_{\nu\boldsymbol{K}} = &\: \delta_{\sigma\tau}\frac{e}{\sqrt{N}}\sum_{\boldsymbol{q}}\, f_{vc\,\nu}\left(\boldsymbol{q}\right)\left\{ -\frac{\hbar}{m_{0}}\frac{\boldsymbol{K}\cdot\boldsymbol{p}_{vc}}{E_{v}-E_{c}}\right.+\frac{\hbar^2}{m_{0}^2}\sum_{n\neq v,c}\left[\frac{\left[\left(\boldsymbol{q}-\gamma\boldsymbol{K}\right)\cdot\boldsymbol{p}_{vn}\right]\,\left[\left(\boldsymbol{q}+\alpha\boldsymbol{K}\right)\cdot\boldsymbol{p}_{nc}\right]}{\left(E_{v}-E_{n}\right)\left(E_{c}-E_{n}\right)}\right.\nonumber\\
  & \nonumber\\
  & \qquad+\frac{\left[\left(\boldsymbol{q}+\alpha\boldsymbol{K}\right)\cdot\boldsymbol{p}_{vn}\right]\,\left[\left(\boldsymbol{q}+\alpha\boldsymbol{K}\right)\cdot\boldsymbol{p}_{nc}\right]}{\left(E_{c}-E_{v}\right)\left(E_{c}-E_{n}\right)}+\left.\left.\frac{\left[\left(\boldsymbol{q}-\gamma\boldsymbol{K}\right)\cdot\boldsymbol{p}_{nc}\right]\,\left[\left(\boldsymbol{q}-\gamma\boldsymbol{K}\right)\cdot\boldsymbol{p}_{vn}\right]}{\left(E_{v}-E_{c}\right)\left(E_{v}-E_{n}\right)}\right]\right\}.\label{eq:mvcnu}
\end{align}
\end{widetext}
Here $m_{\nu\boldsymbol{K}}$ is a short notation for the 
function $m_{vc\,\nu}\left(\boldsymbol{K},\,\boldsymbol{0}\right)$ of Ref.~\cite{150}.
For the definitions of $\Psi_{vc,\,\nu\boldsymbol{K}}^{\sigma\tau}$, $f_{vc\,\nu}\left(\boldsymbol{q}\right)$,
and $\boldsymbol{p}_{mn}$ see Sec.~\ref{sec:Oscillator}.
The exchange energy includes the term $\delta_{\sigma\tau}\delta_{\sigma'\tau'}$.
Introducing the total spin $S=S_{\mathrm{e}}+S_{\mathrm{h}}=\tau-\sigma$
of electron and hole, this term can be written
for singlet and triplet states as $2\delta_{S,0}$~\cite{TO}. 

Using Eq.~(\ref{eq:Fourierq}) and
rearranging the different terms in Eq.~(\ref{eq:mvcnu}) yields
\begin{eqnarray}
m_{\nu\boldsymbol{K}} & = & \delta_{\sigma\tau}\frac{e\hbar^2}{m_{0}^{2}}\frac{K}{\left(E_c-E_v\right)}\nonumber \\
\nonumber \\
& \times & \lim_{\boldsymbol{r}\rightarrow\boldsymbol{0}}\left[\left(\tilde{\boldsymbol{N}}_{v}+\tilde{\boldsymbol{N}}_{c}\right)\cdot\left(-i\nabla_{\boldsymbol{r}}F_{vc\,\nu}\left(\boldsymbol{r}\right)\right)\right.\nonumber \\
\nonumber \\
& + & \left.\left(-\gamma\tilde{\boldsymbol{N}}_{v}+\alpha\tilde{\boldsymbol{N}}_{c}\right)\cdot\left(F_{vc\,\nu}\left(\boldsymbol{r}\right)\boldsymbol{K}\right)\right]\label{eq:mkfinal}
\end{eqnarray}
with the matrices 
\begin{subequations}
\begin{align}
\tilde{\boldsymbol{N}}_{v}&=\left\langle u_{v\boldsymbol{0}}\right|\boldsymbol{p}M_{v}(\hat{\boldsymbol{K}}\cdot\boldsymbol{p})\left|u_{c\boldsymbol{0}}\right\rangle,\\
\tilde{\boldsymbol{N}}_{c}&=\left\langle u_{v\boldsymbol{0}}\right|(\hat{\boldsymbol{K}}\cdot\boldsymbol{p})M_{c}\boldsymbol{p}\left|u_{c\boldsymbol{0}}\right\rangle.
\end{align}
\label{eq:Ntilde}
\end{subequations}
and $\hat{\boldsymbol{K}}=\boldsymbol{K}/K$.

Due to the similarity between Eq.~(\ref{eq:Mfinal}) and Eq.~(\ref{eq:mkfinal}), we
can perform the same calculation as in Sec.~\ref{sec:Oscillator} to obtain
\begin{eqnarray}
m_{\nu\boldsymbol{K}} & \sim & K \lim_{r\rightarrow 0}\left[-i\left(\tilde{M}_v+\tilde{M}_c\right)\frac{\partial}{\partial r}\langle L^D_{\boldsymbol{K}}|\Psi_{\nu\boldsymbol{K}}\rangle \right.\nonumber\\
& & \qquad+\left.\left(-\gamma\tilde{M}_v+\alpha\tilde{M}_c\right)\frac{K}{\sqrt{6}}\langle L^Q_{\boldsymbol{K}}|\Psi_{\nu\boldsymbol{K}}\rangle\right]\;\phantom{i}\label{eq:m0res2}
\end{eqnarray}
with the states
\begin{equation}
|L^D_{\boldsymbol{K}}\rangle = \sum_{i=1}^{3} \hat{K}_{i}|\pi_i^D\rangle, \quad |L^Q_{\boldsymbol{K}}\rangle = \sum_{i=1}^{3} \hat{K}_{i}|\pi_i^Q\rangle,\label{eq:long}
\end{equation}
where we have introduced $\hat{\boldsymbol{K}}=\boldsymbol{K}/K$.
As in Sec.~\ref{sec:Oscillator} we will assume $M_c\gg M_v$ so that we can finally state that $m_{\nu\boldsymbol{K}}$
is proportional to
\begin{eqnarray}
K\lim_{r\rightarrow 0}\left[-i\frac{\partial}{\partial r}\langle L^D_{\boldsymbol{K}}|\Psi_{\nu\boldsymbol{K}}\rangle +\frac{\alpha K}{\sqrt{6}}\langle L^Q_{\boldsymbol{K}}|\Psi_{\nu\boldsymbol{K}}\rangle\right].\label{eq:m0res3}
\end{eqnarray}

We can see from Eq.~(\ref{eq:Hexch}) that there is no interaction between states with different
values of $\boldsymbol{K}$, which is the same case as for the Hamiltonian~(\ref{eq:polarJC})
of the polariton interaction in the rotating-wave approximation. 
Knowing the exciton energies $E_{\nu\boldsymbol{K}}$
and the corresponding wave functions $\left|\Psi_{\nu\boldsymbol{K}}\right\rangle$,
we can simultaneously diagonalize the polariton Hamiltonian and the NA-exchange Hamiltonian
by solving the eigenvalue problem
\begin{equation}
\left(\boldsymbol{P}_{\xi\{\nu\}\boldsymbol{K}}+\boldsymbol{N}_{\{\nu\}\boldsymbol{K}}\right)\boldsymbol{z}_{\mu\xi\{\nu\}\boldsymbol{K}}=E_{\mu\xi\boldsymbol{K}}\boldsymbol{z}_{\mu\xi\{\nu\}\boldsymbol{K}}\label{eq:PNAeig}
\end{equation}
with the matrix
\begin{widetext}
\begin{equation}
\boldsymbol{N}_{\{\nu\}\boldsymbol{K}}=\frac{\zeta}{\varepsilon_{\mathrm{b2}} K^{2}}\left(\begin{array}{cccccc}
0 & 0 & 0 & \cdots & 0 & \cdots\\
0 & m_{\nu_{1}\boldsymbol{K}}^{*}m_{\nu_{1}\boldsymbol{K}}^{\phantom{*}} & m_{\nu_{1}\boldsymbol{K}}^{*}m_{\nu_{2}\boldsymbol{K}}^{\phantom{*}} & \cdots & m_{\nu_{1}\boldsymbol{K}}^{*}m_{\nu_{n}\boldsymbol{K}}^{\phantom{*}} & \cdots\\
0 & m_{\nu_{2}\boldsymbol{K}}^{*}m_{\nu_{1}\boldsymbol{K}}^{\phantom{*}} & m_{\nu_{2}\boldsymbol{K}}^{*}m_{\nu_{2}\boldsymbol{K}}^{\phantom{*}} &  & \vdots\\
\vdots & \vdots &  & \ddots &  & \cdots\\
0 & m_{\nu_{n}\boldsymbol{K}}^{*}m_{\nu_{1}\boldsymbol{K}}^{\phantom{*}} & \cdots &  & m_{\nu_{n}\boldsymbol{K}}^{*}m_{\nu_{n}\boldsymbol{K}}^{\phantom{*}}\\
\vdots & \vdots &  & \vdots &  & \ddots
\end{array}\right),\label{eq:zetamat}
\end{equation}
\end{widetext}
and the matrix $\boldsymbol{P}_{\xi\{\nu\}\boldsymbol{K}}$ and vector $\boldsymbol{z}_{\mu\xi\{\nu\}\boldsymbol{K}}$ defined in Eq.~(\ref{eq:rwaev2}).
The constant parameter $\zeta$ can be determined by the fact that the 
splitting caused by $H_{\mathrm{exch}}^{\mathrm{NA}}$ 
is identical to the LT-splitting.

\subsection{Observability of polariton effects\label{sub:criteria}}

We now shortly discuss the criteria for the observability
of polariton effects, which were derived by Tait in Ref.~\cite{DB_12}.
To obtain these criteria, he included a damping term $\Gamma$ in the model of polaritons since damping is
always present in the solid due to the interaction
between excitons and phonons or the leakage of photons out of the solid~\cite{P3}:
\begin{equation}
\frac{\hbar^{2}c^{2}K^{2}}{E_{\mu\xi\boldsymbol{K}}^{2}}=\varepsilon_{\mathrm{b2}}+\sum_{\nu}\frac{4\pi\kappa_{\mathrm{SI}}\beta_{\xi\nu\boldsymbol{K}}E_{\nu\boldsymbol{K}}^2}{E_{\nu\boldsymbol{K}}^2-E_{\mu\xi\boldsymbol{K}}^2-i\Gamma E_{\mu\xi\boldsymbol{K}}}.\label{eq:polarenerGamma}
\end{equation}
This equation can either be solved for a fixed wave vector 
$K$ or for a fixed frequency $\omega=E_{\mu\xi\boldsymbol{K}}/\hbar$.

The first case corresponds to nonlinear optical experiments like, e.g., 
two-photon absorption. For this case a criterion
of temporal coherence between the photon and the exciton can be derived~\cite{DB_12,83}.
As long as
\begin{equation}
\hbar\Gamma < \hbar\sqrt{\kappa_{\mathrm{SI}}\frac{\pi}{\varepsilon_{\mathrm{b2}}}\beta_{\xi\nu\boldsymbol{K}}E_{\nu\boldsymbol{K}_0}^2}=\frac{\hbar}{2}\Omega_{\mathrm{R}}\label{eq:tc}
\end{equation}
holds, where $\hbar\Gamma$ is the broadening
of the linewidth due to damping, the polariton splitting is observable. 
This criterion can be interpreted in terms of Rabi oscillations, i.e.,
polariton effects are observable if a coherent energy transfer 
between an exciton and a photon is possible at least once~\cite{P3}.
Spatial coherence is already provided here by keeping $K$ fixed~\cite{83}.

The second case corresponds to reflectivity or absorption experiments.
Here the coupling between the photon and the exciton must remain
coherent during the propagation of the polariton through the solid
in the presence of damping~\cite{83}. The criterion of spatial
coherence reads~\cite{83,P3_44,P3}
\begin{equation}
\hbar\Gamma < \sqrt{\frac{16\pi}{Mc^2}\kappa_{\mathrm{SI}}E_{\nu\boldsymbol{K}_0}^3\beta_{\xi\nu\boldsymbol{K}}}=\frac{\hbar}{2}\Omega_{\mathrm{R}}\sqrt{\frac{16\varepsilon_{\mathrm{b2}}}{Mc^2}E_{\nu\boldsymbol{K}_0}},\label{eq:sc}
\end{equation}
and it is generally more difficult to satisfy than Eq.~(\ref{eq:tc})~\cite{P3,P3_45}
since $\sqrt{16\varepsilon_{\mathrm{b2}} E_{\nu\boldsymbol{K}_0}/Mc^2}\ll 1$ holds for
an exciton mass $M$ on the order of $m_0$ and an 
exciton energy on the order of a few $\mathrm{eV}$.
The criterion~(\ref{eq:sc}) is equivalent to $l\gg\lambda$ with $\lambda$ denoting the light wavelength
and $l=v_{g}/\Gamma$ the mean free path of the exciton~\cite{83}.
Hence, polariton effects can hardly be observed in semiconductors
with very shallow excitons, e.g., in GaAs~\cite{83}, when using
linear optical techniques. Therefore, polariton effects are often investigated
using nonlinear optical spectroscopic techniques due to the much less
stringent criterion~(\ref{eq:tc})~\cite{P3_46,P8}.

\section{Results including VB structure and central-cell corrections\label{sec:complete}}

In this section we will treat the exciton-polaritons with $2\leq n\leq 4$
in $\mathrm{Cu_{2}O}$ using the Hamiltonian which accounts for
the valence band structure, the exchange interaction, the central-cell corrections,
and the finite momentum $\hbar K$ of the center of mass. 
For the three orientations $[001]$, $[110]$, and $[111]$ of $\boldsymbol{K}$
considered here the cubic symmetry $O_{\mathrm{h}}$ is reduced to 
$C_{\mathrm{4v}}$, $C_{\mathrm{2v}}$, and $C_{\mathrm{3v}}$, respectively~\cite{G1}.

\begin{figure*}[t]
\begin{centering}
\includegraphics[width=2.0\columnwidth]{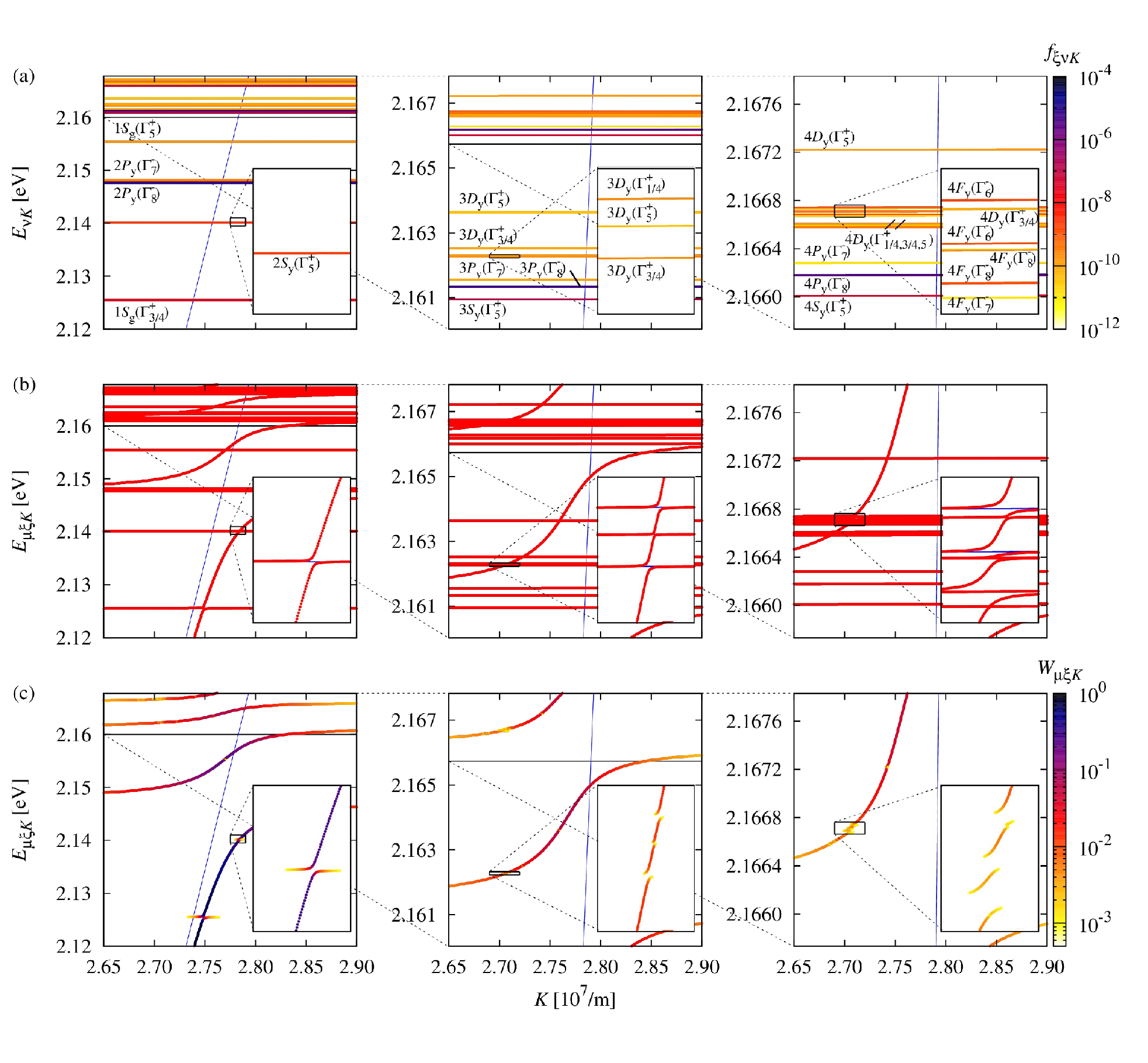}
\par\end{centering}

\protect\caption{(a)~The exciton energies $E_{\nu\boldsymbol{K}}$ in dependence on
$K=\left|\boldsymbol{K}\right|$ for $\boldsymbol{K}\parallel[001]$. 
Due to the inclusion of the complete valence band structure and the central cell corrections,
the spectrum is much more complicated than in the hydrogen-like case (cf.~Refs.~\cite{74,BA_JaV}).
The colorbar shows the oscillator strengths for $\xi=\sigma_z^{\pm}$ polarized light.
For both polarizations the spectrum is identical.
We denote from which states at $K=0$ the exciton states originate (cf.~Ref.~\cite{200}).
For reasons of space we introduce the abbreviated notation $\Gamma_{i/j}^{\pm}$
to replace $\Gamma_{i}^{\pm},\,\Gamma_{j}^{\pm}$.
The blue solid line gives the photon dispersion 
$\hbar\omega_{\xi\boldsymbol{K}}=\hbar Kc/\sqrt{\varepsilon_{\mathrm{b2}}}$.
(b)~Polariton dispersion obtained by solving the eigenvalue problem~(\ref{eq:PNAeig}).
Since the $P$ excitons have a large oscillator strength, 
large avoided crossings can be observed.
They clearly shift the position of the smaller 
avoided crossings for the $2S$, $3D$ and $4F$ excitons away from the position
where the dispersion of light and the dispersion of these exciton states cross. 
(c)~Photon-like part $W_{\mu\xi\boldsymbol{K}}$ of the polariton states. 
In all cases the mixing of excitons and photons is strongest
in the vicinity of an avoided crossing. Only for the $P$ excitons an admixture
of photons far away from the resonance occurs.~\label{fig:Figg1}}
\end{figure*}

Excitons of even and odd parity behave differently in dependence
on the momentum $\hbar K$. 
On the one hand, states of the odd series having the symmetry $\Gamma_4^-$
and a component with $L=1$
show a finite dipole oscillator strength at $K=0$. 
Since the angular momentum $L$ is not a 
good quantum number in $\mathrm{Cu_{2}O}$
and $P$-like exciton states are admixed to other states of odd parity at $K=0$~\cite{200},
$P$, $F$, $H$, $\ldots$ states of symmetry $\Gamma_4^-$
have a finite dipole oscillator strength.
Hence, for these states already a
splitting at $K=0$ occurs due to the nonanalytic exchange interaction
and the exciton-photon coupling, which affect the states of this symmetry. 

On the other hand, the even exciton states do not have an oscillator strength at $K=0$
and, therefore, no splitting occurs.
For those even states which have the symmetry $\Gamma_5^+$
and an $L=0$-component at $K=0$ the coupling to light via a finite
$K$ vector is possible. Since $S$-like exciton states are admixed 
to other states of even parity at $K=0$~\cite{200},
$S$, $D$, $G$, $\ldots$ states of symmetry $\Gamma_5^+$
obtain a finite quadrupole oscillator strength being proportional to $K^2$.

At first, we will compare our numerical results for the relative oscillator
strengths of the $1S$ excitons with the absolute value from the experiment.
This will allow us to calculate absolute oscillator strengths
for all exciton states.
Having determined the correct size of the nonanalytic exchange interaction
for the $nP$ excitons and, hence, for all other exciton states,
we can then investigate the dispersion 
of exciton-polaritons for the three
orientations of $\boldsymbol{K}$ along the axes of high symmetry.

Since the formula derived in Sec.~\ref{sec:Oscillator}
allows us only to calculate relative oscillator strengths
$f^{\mathrm{rel}}_{\xi\nu\boldsymbol{K}}$ but not absolute
oscillator strengths for the different polarizations $\xi$ and exciton states $\nu\boldsymbol{K}$,
we determine the scaling factor $\eta$ in
\begin{equation}
f_{\xi\nu\boldsymbol{K}}=\eta f^{\mathrm{rel}}_{\xi\nu\boldsymbol{K}}\label{eq:frel}
\end{equation}
by comparing the theoretical
results for $f^{\mathrm{rel}}_{\xi\,1S\,\boldsymbol{K}}$ for the $1S$ exciton state
at the exciton-photon resonance $K=K_0$
with the experimentally obtained value of~\cite{GRE_26}
\begin{equation}
f_{\xi\,1S\,\boldsymbol{K}}=3.6\times 10^{-9}\label{eq:deteta}
\end{equation}
for $\boldsymbol{K}\parallel [110]$.
This yields $\eta = 825.9$. 
Knowing this scaling factor, we can give the 
absolute oscillator strengths of all exciton states in the following.
With this value the oscillator strengths
of, e.g., the $nP$ excitons are given by
\begin{equation}
f_{\xi\,nP\,\boldsymbol{K}}=3.75\times 10^{-4}\frac{n^2-1}{n^5}.\label{eq:deteta}
\end{equation}

For a complete description of the polariton problem, we also 
have to include the nonanalytic exchange interaction.
The splitting caused by this interaction
at $\boldsymbol{K}=\boldsymbol{0}$ must exactly equal the 
LT-splitting due to the polariton transformation.
As the rotating wave approximation does not hold for $\boldsymbol{K}\rightarrow\boldsymbol{0}$,
we determine at first the 
polariton energies $E_{\mu\xi\boldsymbol{K}}$ for the transverse exciton states 
via the conditional equation
\begin{equation}
\frac{\hbar^{2}c^{2}K^{2}}{\varepsilon_{\mathrm{b2}}E_{\mu\xi\boldsymbol{K}}^{2}}=1+\sum_{\nu}\frac{f_{\xi\nu\boldsymbol{K}}}{1-\left(E_{\mu\xi\boldsymbol{K}}/E_{\nu\boldsymbol{K}}\right)^{2}},\label{eq:polarener}
\end{equation}
which is obtained when using the complete exciton-photon 
interaction Hamiltonian~\cite{P1,TOE_206,P1_7}.
The conditional equation~(\ref{eq:polarener}) can be rewritten so that the polariton energies are the roots of the function
\begin{equation}
F\!\left(E_{\mu\xi\boldsymbol{K}}\right)=1-\frac{\hbar^2\omega_{\xi\boldsymbol{K}}^2}{E_{\mu\xi\boldsymbol{K}}^2}+\sum_{\nu}\frac{f_{\nu\xi\boldsymbol{K}}E_{\nu\boldsymbol{K}}^2}{E_{\nu\boldsymbol{K}}^2-E_{\mu\xi\boldsymbol{K}}^2},
\end{equation}
where the photon energy $\hbar\omega_{\xi\boldsymbol{K}}=\hbar Kc/\sqrt{\varepsilon_{\mathrm{b2}}}$
is used.
Note that a root of $F\!\left(E_{\mu\xi\boldsymbol{K}}\right)$ is always located between any neighboring pair
of the exciton energies $E_{\nu\boldsymbol{K}}$, $E_{\nu+1\,\boldsymbol{K}}$~\cite{P7,P7_4,P7_18}.

We calculate the effect of the nonanalytic exchange interaction
on the longitudinal states for $\boldsymbol{K}\rightarrow\boldsymbol{0}$
by diagonalizing the matrix $\boldsymbol{N}_{\{\nu\}\boldsymbol{K}}$ 
given in Eq.~(\ref{eq:zetamat}).
As the size of the nonanalytic exchange is a priori unknown, we
have scaled the matrix $\boldsymbol{N}_{\{\nu\}\boldsymbol{K}}$,
which describes this interaction, with a parameter $\zeta$.
We vary the parameter $\zeta$ in such a way that the energies of the 
longitudinal and transverse $nP$ exciton states are identical.
This yields $\zeta= 213.5\pm 2.0$.

Of prime interest are now the polariton dispersions
in the vicinity of the exciton phonon resonance.
In this range of $K$ the rotating wave approximation is valid
and we solve the eigenvalue problem~(\ref{eq:PNAeig}).
Note that the errors arising due to the use of the
rotating-wave approximation are smaller than
the error due to uncertainties in $\varepsilon_{\mathrm{b2}}$.

The numerical result for $\boldsymbol{K}\parallel [001]$ 
is shown in Fig.~\ref{fig:Figg1}.
In the panels (a) we show the exciton spectrum in dependence 
on $K=\left|\boldsymbol{K}\right|$.
Since the changes in the energy in dependence on $K$ are in the order of tens of $\mu\mathrm{eV}$,
the exciton states appear as straight horizontal lines. 

The spectrum is much more complicated
than when using the hydrogen-like model of excitons,
which was done in Refs.~\cite{74,BA_JaV}. 
Even for vanishing momentum of the center of mass, the complete valence band 
structure already leads to a complicated fine structure splitting
and to a mixing of the exciton states with even or odd parity.
This explains the observability of $F$ excitons in absorption spectra 
due to the admixture of $P$ excitons~\cite{28,100}.
Due to the finite momentum of the center of mass 
$S$ and $D$ excitons also obtain a small oscillator strength.

We state in the panels (a) of Fig.~\ref{fig:Figg1}
from which states at $K=0$ the exciton states originate
using the nomenclature $nL_{\mathrm{y/g}}(\Gamma_i^{\pm})$ of Ref.~\cite{200} 
with the abbreviations y and g for yellow and green.
In the case of the $P$ and $F$ excitons we do not give the symmetry of the complete 
exciton state but only the combined symmetry of envelope and hole~\cite{28}.

In the panels (b) we 
show the polariton dispersion. Only for the $P$ excitons, 
which have a comparatively large oscillator strength, significant deviations between the
polariton spectrum and the exciton spectrum can be observed.
The insets in the panels show that also for the other exciton states
avoided crossings appear due to their finite oscillator strength.
However, the panels (c), 
which show the photon-like part of the polariton states,
indicate that the mixing between excitons and photons is small in these cases.

Due to the proximity of the $nS$ and $nP$ states
as well as that of the $nF$ states, avoided crossings are not well-separated.
Hence, the multi-polariton 
concept has to be used and a single-polariton concept would lead to different results.
The large avoided crossings of the $P$ excitons clearly
affect the other avoided crossings as they 
shift them away from the position
where the dispersion of light and the dispersion 
of the other exciton states cross. 

As regards the other orientations 
$\boldsymbol{K}\parallel[110]$ or $\boldsymbol{K}\parallel[111]$
of the momentum of center of mass the differences in the
polariton spectra are slight.
Since the $K$-dependent shift of the exciton energies
is on the order of tens of $\mu\mathrm{eV}$, the exciton energies 
are almost the same for the three orientations of $\boldsymbol{K}$
considered. 
The main difference between the spectra for $\boldsymbol{K}\parallel[001]$, 
$\boldsymbol{K}\parallel[110]$ or $\boldsymbol{K}\parallel[111]$ is the 
values of the oscillator strengths due to the different 
symmetry breaking.
Therefore, also the polariton dispersions are different, however,
mainly in the vicinity of the avoided crossings~\cite{300_SF}.

Note that we do not show the polariton dispersion for exciton states with $n\geq 5$
since the number of states with a finite oscillator strength
increases rapidly. However, the oscillator strengths of the $G$, $H$, $\ldots$ exciton states
are very small so that polariton effects are likewise very small.
Hence, we do not expect to observe considerably new effects for the exciton-polaritons
with $n\geq 5$.

Using the criteria of Tait~\cite{DB_12} for temporal and spatial coherence 
presented in Sec.~\ref{sub:criteria}, we can estimate
the possible observability of the polariton effects.
If we set $n=2$, the criterion of temporal coherence
reads 
\begin{equation}
\hbar\Gamma < \frac{\hbar}{2}\Omega_{\mathrm{R}} \approx 6.4\,\mathrm{meV}.\label{eq:tc}
\end{equation} 
For the criterion of spatial coherence, we obtain
\begin{equation}
\hbar\Gamma < \frac{\hbar}{2}\Omega_{\mathrm{R}}\sqrt{\frac{16\varepsilon_{\mathrm{b2}}}{Mc^2}E_{\nu\boldsymbol{K}_0}}\approx 0.11\,\mathrm{meV}\label{eq:sc}
\end{equation}
with the isotropic mass $M\approx 1.57 m_0$ of the exciton
and $m_0$ denoting the free electron mass.
In Ref.~\cite{GRE} the spectrum of the giant Rydberg excitons has been investigated
in an absorption experiment and, thus, experimental values of the exciton linewidths 
$\hbar\Gamma$ for the criterion of spatial coherence are known.
For $n=2$ the experimental line width is $\hbar\Gamma=1.58\,\mathrm{meV}$~\cite{GRE,75}, which is
significantly larger than $0.11\,\mathrm{meV}$.
Therefore, we expect that the polariton effects for the yellow exciton states
with $n\geq 2$ in $\mathrm{Cu_{2}O}$ may only be observed using nonlinear
spectroscopy methods and high quality crystals~\cite{P3_46,P8}.
As regards the $1S$ ortho exciton state, the linewidth
is small enough to clearly observe polariton effects~\cite{GRE_26,P3_107_12,GRE_28}.

\section{Yellow $\boldsymbol{1S}$ ortho exciton polariton\label{sec:complete1S}}

We now come to the yellow $1S$ ortho exciton polariton, for which a pronounced
polariton effect has been proven in experiments.
In Sec.~\ref{sec:1S1} we will discuss
at first the effect of finite momentum of the center of mass $\hbar K\neq 0$
on the exciton spectrum. We will especially pay attention to the 
small quadrupole oscillator strength~\cite{GRE_26} 
and the $\boldsymbol{K}$ dependent splitting of this state~\cite{9_1,8,9,100}.
In Sec.~\ref{sec:1S} we set up a $5\times 5$ matrix model, which allows
us to calculate the anisotropic dispersion
of the $1S$ ortho-exciton polariton for any direction of $\boldsymbol{K}$.
We then present in Sec.~\ref{sec:1S2} the polariton dispersion,
determine the group velocity~\cite{P3_107_12,GRE_28}
as well as the spectra for rotations about the 
$[1\bar{1}0]$ and the $[111]$ axis
and compare them with experimental values~\cite{8,9_1}.

\subsection{$K$-dependent splitting\label{sec:1S1}}

\begin{figure*}[t]
\begin{centering}
\includegraphics[width=2.0\columnwidth]{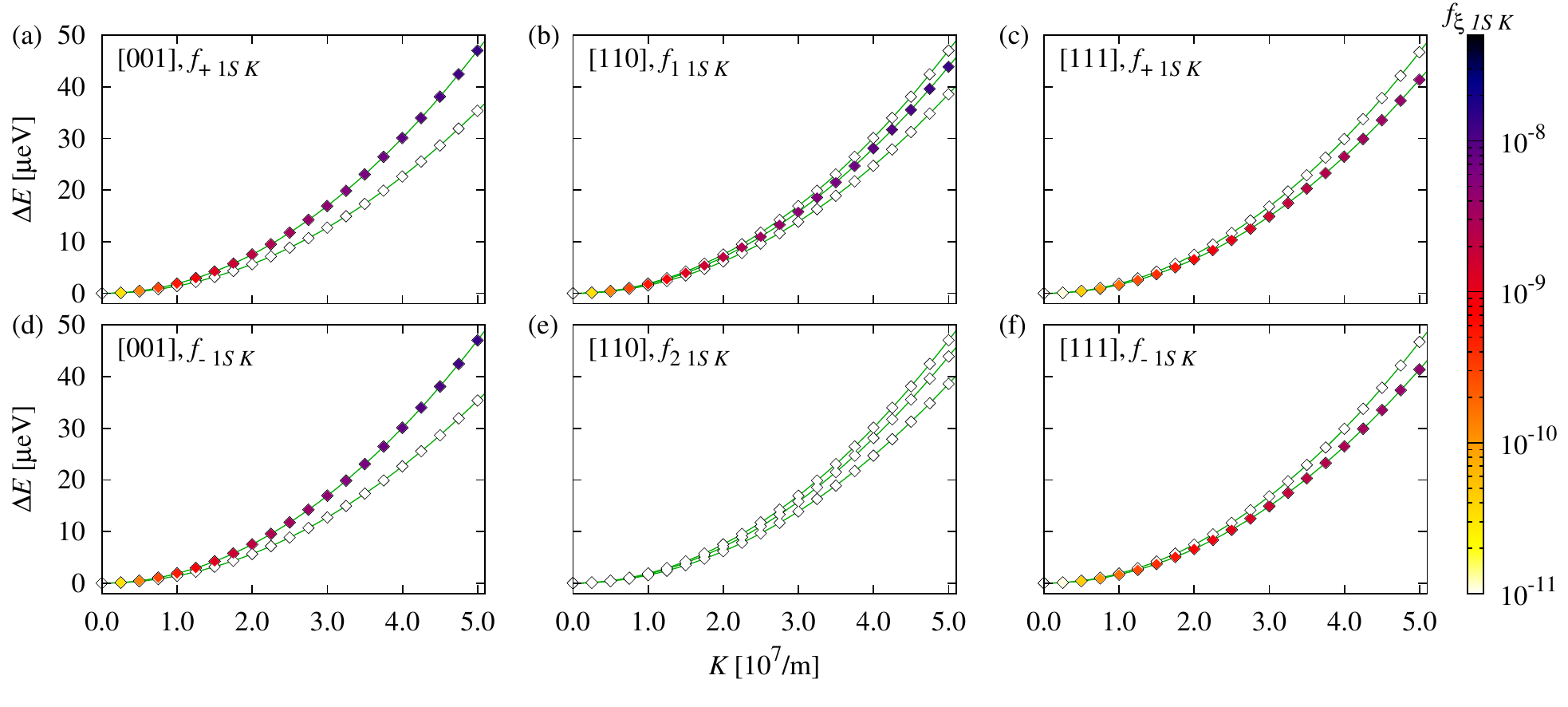}
\par\end{centering}

\protect\caption{Energy of the $1S_{\mathrm{y}}(\Gamma_5^+)$ ortho exciton state
in dependence of $K=\left|\boldsymbol{K}\right|$. We do not plot the
absolute energies but the energy difference 
$\Delta E=E_{1S,\boldsymbol{K}}^{\mathrm{o}}-E_{1S,\boldsymbol{0}}^{\mathrm{o}}$
to show the small increase of the energy for $K\neq 0$.
The $1S$ exciton obtains a finite oscillator strength for $K\neq 0$,
which increases quadratically with $K$.
Furthermore, a $K$-dependent splitting of the three ortho exciton states can be observed.
The green solid lines are the fits of the eigenvalues
of the Hamiltonian~(\ref{eq:TtK})
to the numerical results.
For further information see text.~\label{fig:Fig3}}
\end{figure*}

As has already been stated in Sec.~\ref{sec:complete}, we have to consider the 
reduction of the irreducible representations
of the cubic group $O_{\mathrm{h}}$ by the groups 
$C_{\mathrm{4v}}$, $C_{\mathrm{2v}}$, and $C_{\mathrm{3v}}$
for the three cases of $\boldsymbol{K}$ being oriented along the
$[001]$, $[110]$, or the $[111]$ direction, respectively.
In particular, for 
an exciton state having the symmetry 
$\Gamma_5^{\pm}$ at $K=0$ degeneracies are lifted for $K\neq 0$.

We therefore observe a splitting of the 
$1S_{\mathrm{y}}(\Gamma_5^+)$ ortho exciton state
depending on $K$ when solving the 
full $K$-dependent Hamiltonian of excitons in $\mathrm{Cu_{2}O}$.
This splitting is shown in Fig.~\ref{fig:Fig3}.
It was observed experimentally in Refs.~\cite{9_1,8,9}
and originally discussed in terms of a $K$-dependent exchange interaction.
However, a closer examination of this interaction revealed that it
is far too weak in $\mathrm{Cu_{2}O}$ to describe the observed splitting~\cite{150}.
Instead, it could be shown that the effects due to the cubic valence band structure
lead to a $K$-dependent effective mass and a $K$-dependent splitting
of the $1S$ ortho exciton~\cite{100}. Hence, the directional dispersion
is the true cause of the experimentally observed splitting.

Note that in Ref.~\cite{100} the splitting was treated within a perturbation approach and
it was already emphasized that the complete $K$-dependent Schr\"odinger equation
including the central cell corrections would have to be
solved to obtain correct results. This has now been done.

As the $1S$ ortho exciton state exciton state has the symmetry $\Gamma_5^+$ for $K=0$,
we expect for $\boldsymbol{K}\parallel[001]$ and 
$\boldsymbol{K}\parallel[111]$ a splitting into two degenerate and one
non-degenerate state. For $\boldsymbol{K}\parallel[110]$ all degeneracies 
are lifted.
This splitting can be described by the Hamiltonian
\begin{align}
\boldsymbol{H}_{\mathrm{disp}}\left(\boldsymbol{K}\right) = & \Delta_{1}\left(\begin{array}{ccc}
K^{2} & 0 & 0\\
0 & K^{2} & 0\\
0 & 0 & K^{2}
\end{array}\right)\nonumber \\
\displaybreak[1] + & \Delta_{3}\left(\begin{array}{ccc}
3K_{x}^{2}-K^{2} & 0 & 0\\
0 & 3K_{y}^{2}-K^{2} & 0\\
0 & 0 & 3K_{z}^{2}-K^{2}
\end{array}\right)\nonumber \\
\displaybreak[1] + & \Delta_{5}\left(\begin{array}{ccc}
0 & K_{x}K_{y} & K_{x}K_{z}\\
K_{x}K_{y} & 0 & K_{y}K_{z}\\
K_{x}K_{z} & K_{y}K_{z} & 0
\end{array}\right)\label{eq:TtK}.
\end{align}
We can prove the consistency with this formula
by diagonalizing the Hamiltonian~(\ref{eq:TtK})
for $\boldsymbol{K}\parallel[001]$, $\boldsymbol{K}\parallel[110]$, and 
$\boldsymbol{K}\parallel[111]$ 
and fitting the resulting eigenvalues to our numerical results obtained with the 
full exciton Hamiltonian~(\ref{eq:Hges}). This is shown in Fig.~\ref{fig:Fig3}.
The values of the fit parameters $\Delta_{1}$, $\Delta_{3}$, and $\Delta_{5}$
are given in Table~\ref{tab:3}.
Since we have only three independent parameters $\Delta_i$
but seven exciton states to be fitted, the consistency
is proven by the fact that we obtain the same values of the parameters 
$\Delta_{i}$ in all fits.

When performing the fit, it is not necessary to account for the
$K$-dependent nonanalytic exchange interaction.
In the next section~\ref{sec:1S} a formula for the size of the nonanalytic exchange
interaction of the $1S$ exciton will be derived:
\begin{equation}
\Delta_Q=2f_0 E_0/K_0^2.
\end{equation}
Here $f_0=3.6\times 10^{-9}$ and $E_0$ denote the oscillator strength and the 
energy of the transverse exciton for $\boldsymbol{K}\parallel[001]$
at the exciton-photon resonance $K_0$, respectively.
From the numerical results we obtain
\begin{equation}
K_0=2.614\times 10^{7}\,\frac{1}{\mathrm{m}},\,E_0=2.032\,\mathrm{eV}
\end{equation}
and, therefore,
\begin{equation}
\Delta_Q=2.135\times 10^{-17}\,\mathrm{\mu eV\,m^2}.
\end{equation}
Obviously, the nonanalytic exchange interaction
is two to three orders of magnitude smaller
than the anisotropic dispersion (cf.~Table~\ref{tab:3}).

\begin{table}[t]

\protect\caption{Results for the three coefficients
$\Delta_i$ when fitting the eigenvalues the Hamiltonian~(\ref{eq:TtK})
to the theoretical spectra of Fig.~\ref{fig:Fig3} for the different
orientations of $\boldsymbol{K}$.
For the $[001]$ and the $[111]$ directions the values
$\Delta_5$ and $\Delta_3$ cannot be determined, respectively.
All results are given in $10^{-14}\,\mathrm{\mu eV\,m^2}$.
For a comparison, we also list the experimentally determined
values of Refs.~\cite{9_1,8,9}. For further information see text.
\label{tab:3}}

\begin{centering}
\begin{tabular}{c|ccc|c}
\hline 
 & $[001]$ & $[110]$ & $[111]$ & expt\tabularnewline
\hline 
$\Delta_{1}$ & $\phantom{-}1.727$ & $\phantom{-}1.729$ & $\phantom{-}1.728$ & $\phantom{-}1.074$\tabularnewline
$\Delta_{3}$ & $-0.156$ & $-0.155$ & - & $-0.189$\tabularnewline
$\Delta_{5}$ & - & $\phantom{-}0.213$ & $\phantom{-}0.213$ & $\phantom{-}0.292$\tabularnewline
\hline 
\end{tabular}
\par\end{centering}

\end{table}

The experimentally observed splitting was
described by the ansatz~(\ref{eq:TtK}) as well~\cite{9_1,8,9}.
When comparing the results, we have to note that
the values of the $\Delta_i$ are given with respect to $K_0$
and that the factor $\Delta_1$ of Refs.~\cite{9_1,8,9} only describes the ``exchange interaction,'' i.e.,
the interaction without the spherically symmetric part of
the kinetic energy. 
Therefore, we have to compare $\Delta_1/K_0^2+\hbar^2/2M$,
$\Delta_3/K_0^2$, and $\Delta_5/K_0^2$ with our results of the parameters $\Delta_i$.
Here $M=m_{\mathrm{e}}+m_{\mathrm{e}}\approx 1.64m_0$ is the value of the isotropic exciton mass 
used in Refs.~\cite{9_1,8,9}.

Within our model the average mass of the $1S$ ortho exciton is
\begin{equation}
M=\frac{\hbar^2 }{2 K^2}\frac{3}{\mathrm{Tr}\left[\boldsymbol{H}_{\mathrm{disp}}\left(\boldsymbol{K}\right)\right]}=\frac{\hbar^2}{2\Delta_1}=2.2\,m_0,
\end{equation}
which is significantly larger than the 
sum $M=m_{\mathrm{e}}+m_{\mathrm{h}}=m_{\mathrm{e}}+m_0/\gamma_1 \approx 1.56m_0$ of the
isotropic quasi-particle masses. However, there is
a clear deviation from the experimentally determined
value of $M=(3.0\pm 0.2)m_0$~\cite{10a,Diss41,9_3}.
The disparity in the masses can be explained
by the fact that the electron and the hole are not bare particles
but polarize the surrounding lattice and are thus
accompanied by clouds of longitudinal optical phonons~\cite{SO}.
Hence, in the experiment always polaron masses are measured which are
larger than the bare particle masses~\cite{SO}.
Although we accounted for this effect via the Haken potential 
in the Hamiltonian, we already stated in Ref.~\cite{200}
that there are some difficulties in applying this theory
to $\mathrm{Cu_{2}O}$ due to the existence of two optical 
phonons contributing to the Fr\"ohlich interaction and the small
distance between the electron and the hole in the exciton ground state,
for which the Haken potential cannot describe the non-Coulombic electron-hole interaction.

For the other two coefficients $\Delta_3$ and $\Delta_5$
a good agreement is obtained (see Table~\ref{tab:3}),
in particular, as regards the sign of the parameters.
Note, however, that in Refs.~\cite{9_1,8,9} $\Delta_Q$
has been assumed to be of the same order of magnitude
as the parameters $\Delta_i$ and that it has been included
in the fit of the experimental data. This affects the values
for $\Delta_3$ and $\Delta_5$ and 
is, therefore, another reason for the
difference between the experimental and the theoretical values
in Table~\ref{tab:3}.

Finally, differences between the experimental and theoretical values
of the parameters $\Delta_i$ could furthermore be explained by
the neglection of polariton nature of the ortho exciton in Refs.~\cite{9_1,8,9}, in terms
of small strains in the crystal~\cite{8} and uncertainties in the experimental
values (cf.~the large error bars in the figures of Ref.~\cite{9_1}). In particular, the 
fact that the experimental
spectra are not identical when changing the angle of the laser beam
by the same amount in opposite directions in Ref.~\cite{9_1} shows the presence of strains.

We wish to note that also splittings in the same order of magnitude
are obtained, e.g., for the $2P$ exciton state.
However, these splittings cannot be observed experimentally 
due to the large linewidth of the $2P$ exciton.

\subsection{$\boldsymbol{5\times 5}$ matrix model\label{sec:1S}}

The yellow $1S$ ortho exciton is well separated from the other exciton
states regarding its energy. Hence, it can be treated separately 
from the other ones as regards polariton effects..
In this section we set up a model with a $5\times 5$ matrix, which allows calculating the
dispersion of the $1S$ ortho exciton polariton for any direction of $\boldsymbol{K}$
close to the resonance $\left(K\approx K_0\right)$. This model
includes the two photon states with the polarization vectors $\hat{\boldsymbol{e}}_{\xi\boldsymbol{K}}$
and the three ortho exciton states $\Psi_i$, which transform according to $yz$, $zx$ and $xy$.

First we will treat the oscillator strength and the Rabi frequency.
Let us consider the most simple case with $\boldsymbol{K}\parallel [001]$.
Due to group theoretical reasons, the three states $\Psi_{yz}$, $\Psi_{zx}$,
and $\Psi_{xy}$ are good eigenstates of the Hamiltonian.
The $\Psi_{zx}$ exciton interacts with the photon in $x$ polarization
and the $\Psi_{yz}$ exciton interacts with the photon in $y$ polarization.
Let us denote the oscillator strength of these exciton states at the
exciton-photon resonance by $f_0$.

For other orientations of $\boldsymbol{K}$ superpositions 
of the form $\sum_i a_i\Psi_i$ are eigenstates of the Hamiltonian.
From the expression~(\ref{eq:fgesres}) or especially from the form of the states 
$|T^Q_{\xi\boldsymbol{K}}\rangle$~(\ref{eq:TQ}) we can see that
the $K$-dependent oscillator strength of these exciton states is given by
\begin{equation}
f_{1S\;\xi\boldsymbol{K}}=f_{0}\left|\left(\begin{array}{c}
\hat{e}_{\xi\boldsymbol{K},y}K_{z}+\hat{e}_{\xi\boldsymbol{K},z}K_{y}\\
\hat{e}_{\xi\boldsymbol{K},z}K_{x}+\hat{e}_{\xi\boldsymbol{K},x}K_{z}\\
\hat{e}_{\xi\boldsymbol{K},x}K_{y}+\hat{e}_{\xi\boldsymbol{K},y}K_{x}
\end{array}\right)\cdot\left(\begin{array}{c}
a_{yz}\\
a_{zx}\\
a_{xy}
\end{array}\right)\right|^{2}\label{eq:symmcrossprod}
\end{equation}
with the components $\hat{e}_{\xi\boldsymbol{K},i}$ of the
polarization vector $\hat{\boldsymbol{e}}_{\xi\boldsymbol{K}}$.

For states of symmetry $\Gamma_5^+$ it has been shown in Ref.~\cite{150}
that the nonanalytic exchange interaction can be written as
\begin{eqnarray}
\boldsymbol{H}_{\mathrm{exch}}^{\mathrm{NA}}\left(\boldsymbol{K}\right) & = & \frac{\Delta_{Q}}{K^2}\left(\begin{array}{ccc}
K_{y}^{2}K_{z}^{2} & K_{z}^{2}K_{y}K_{x} & K_{y}^{2}K_{x}K_{z}\\
K_{z}^{2}K_{y}K_{x} & K_{z}^{2}K_{x}^{2} & K_{x}^{2}K_{y}K_{z}\\
K_{y}^{2}K_{x}K_{z} & K_{x}^{2}K_{y}K_{z} & K_{x}^{2}K_{y}^{2}
\end{array}\right).\nonumber\\
\label{eq:HexchKmatrix}
\end{eqnarray}
Contrary to dipole allowed excitons, the nonanalytic exchange
energy depends on the fourth power of the angular coordinates of $\boldsymbol{K}$.
The prefactor $\Delta_Q$ is connected to the oscillator strength and can be determined
for the $1S$ state in the following way:
For the special case of $\boldsymbol{K}$ being oriented in [111] direction, the
$\Gamma_5^+$ state splits into one longitudinal $\Gamma_1$ and two transverse $\Gamma_5$
states. The longitudinal state is an eigenstate of the operator~(\ref{eq:HexchKmatrix})
with the eigenvalue $\Delta_Q/3$.
An excitation of the longitudinal exciton leads to an oscillating
longitudinal polarization. Due to the Maxwell equation $\nabla\cdot\boldsymbol{D}=0$,
the dielectric function must be zero.
Hence, we have
\begin{equation}
\varepsilon(\omega,\,K_0)\left[\Gamma_1\right]=\varepsilon_{\mathrm{b}2}+\frac{\frac{4}{3}f_{0}\varepsilon_{\mathrm{b}2}}{1-\left(E_0+\frac{1}{3}\Delta_Q K_0^2\right)^2/\left(E_0\right)^2}=0
\end{equation}
[cf.~also Eq.~(\ref{eq:polarener})]. Here $E_0=\hbar c K_0/\sqrt{\varepsilon_{\mathrm{b}2}}$ 
is the energy of the $\Gamma_1$ exciton at $K=K_0$ without the nonanalytic exchange interaction .
Using $f_0\ll 1$~\cite{GRE_26,P3_107_12,GRE_28}, we obtain
\begin{equation}
\Delta_Q=2f_0 E_0/K_0^2=2f_0\hbar^2 c^2/\varepsilon_{\mathrm{b}2}.
\end{equation}

Combining all the $K$ dependent effects for the $1S$ ortho exciton, we arrive at the
Hamiltonian for the $1S$ ortho exciton-polariton in the rotating-wave approximation:
\begin{equation}
\boldsymbol{H}=\left(\begin{array}{cc}
\boldsymbol{H}_{\mathrm{ph}} & \boldsymbol{H}_{\mathrm{exc-ph}}\\
\boldsymbol{H}_{\mathrm{exc-ph}}^{\mathrm{T}} & \boldsymbol{H}_{\mathrm{exc}}
\end{array}\right)\label{eq:5by5}
\end{equation}
with the $2\times 2$ matrix $\boldsymbol{H}_{\mathrm{ph}}$ containing the photon dispersion,
\begin{equation}
\boldsymbol{H}_{\mathrm{ph}}=\frac{\hbar cK}{\sqrt{\varepsilon_{\mathrm{b2}}}}\left(\begin{array}{cc}
1 & 0\\
0 & 1
\end{array}\right)=E_{0}\frac{K}{K_{0}}\left(\begin{array}{cc}
1 & 0\\
0 & 1
\end{array}\right),
\end{equation}
a $2\times 3$ matrix $\boldsymbol{H}_{\mathrm{exc-ph}}$ with the Rabi energies $\hbar\Omega_{\mathrm{R}}=E_0\sqrt{f_0}$,
\begin{widetext}
\begin{equation}
\boldsymbol{H}_{\mathrm{exc-ph}}=\frac{1}{2}\hbar\Omega_{\mathrm{R}}\frac{1}{K_{0}}\left(\begin{array}{ccc}
\left(\hat{e}_{1\boldsymbol{K},y}K_{z}+\hat{e}_{1\boldsymbol{K},z}K_{y}\right) & \left(\hat{e}_{1\boldsymbol{K},z}K_{x}+\hat{e}_{1\boldsymbol{K},x}K_{z}\right) & \left(\hat{e}_{1\boldsymbol{K},x}K_{y}+\hat{e}_{1\boldsymbol{K},y}K_{x}\right)\\
\left(\hat{e}_{2\boldsymbol{K},y}K_{z}+\hat{e}_{2\boldsymbol{K},z}K_{y}\right) & \left(\hat{e}_{2\boldsymbol{K},z}K_{x}+\hat{e}_{2\boldsymbol{K},x}K_{z}\right) & \left(\hat{e}_{2\boldsymbol{K},x}K_{y}+\hat{e}_{2\boldsymbol{K},y}K_{x}\right)
\end{array}\right),\label{eq:Hexph}
\end{equation}
\end{widetext}
and 
\begin{equation}
\boldsymbol{H}_{\mathrm{exc}}=E_{0}\,\boldsymbol{1}+\boldsymbol{H}_{\mathrm{disp}}\left(\boldsymbol{K}\right)+\boldsymbol{H}_{\mathrm{exch}}^{\mathrm{NA}}\left(\boldsymbol{K}\right),
\end{equation}
where $\boldsymbol{1}$ is the $3\times 3$ identity matrix.

Note that the Rabi energy depends on the square root of the
oscillator strength [cf.~Eq.~(\ref{eq:Omegar})]. Hence, questions about
the sign of the terms $\left(\hat{e}_{1\boldsymbol{K},y}K_{z}+\hat{e}_{1\boldsymbol{K},z}K_{y}\right)$
in Eq.~(\ref{eq:Hexph}) may arise. However, for reasons of symmetry
the terms must be linear in $\boldsymbol{K}$. As the photon has
negative parity, i.e., since it transforms according to 
$\Gamma_4^-$ in $O_{\mathrm{h}}$, the terms have to change the sign
if the direction of $\boldsymbol{K}$ is reversed.

The eigenstates of the Hamiltonian~(\ref{eq:5by5}) can be calculated
using an appropriate LAPACK routine~\cite{Lapack}.

\begin{figure*}[t]
\begin{centering}
\includegraphics[width=2.0\columnwidth]{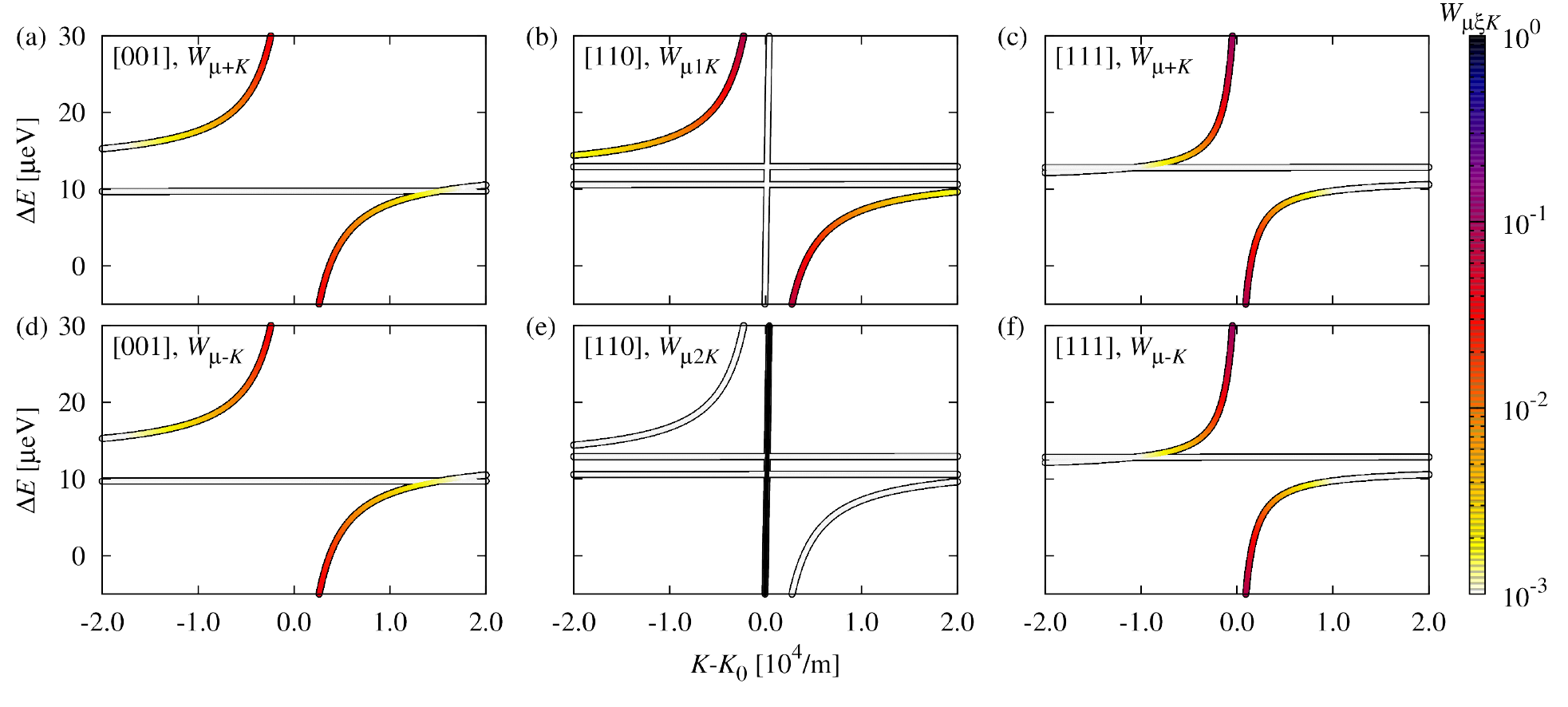}
\par\end{centering}

\protect\caption{Dispersion of the $1S$ ortho
exciton polariton calculated using the $5\times 5$ matrix
model. For the different orientations of
$\boldsymbol{K}$ we give the photon-like part $W_{\mu\xi\boldsymbol{K}}$ for the two polarizations $\xi$.
It can be seen that for $\boldsymbol{K}\parallel [110]$ and 
light being polarized along $\hat{\boldsymbol{e}}_{2\boldsymbol{K}}$
no exciton-photon coupling occurs.
We do not plot the
absolute energies but the energy difference 
$\Delta E=E-E_0$. For further information see text.
~\label{fig:FigD1S}}
\end{figure*}

\subsection{Polariton dispersion\label{sec:1S2}}

As the $1S$ ortho state shows a pronounced polariton effect, we
will now come to its dispersion, which can be calculated
using the $5\times 5$ matrix model.
For the subsequent calculations we will use the
parameters
\begin{equation}
f_0=3.6\times 10^{-9},\quad E_0=2.0239\,\mathrm{eV}
\end{equation}
of Ref.~\cite{GRE_26}, which yields
\begin{equation}
K_0=2.618\times 10^7\,\frac{1}{\mathrm{m}},\quad \Delta_Q=2.135\times 10^{-17}\,\mathrm{\mu eV\,m^2},
\end{equation}
and the average values of the theoretical 
results for the parameters $\Delta_i$ in Table~\ref{tab:3}.

In Fig.~\ref{fig:FigD1S} we first present the results for $\boldsymbol{K}$
being oriented along one of the axes of high symmetry.
The longitudinal exciton states do not couple to photons
and, therefore, their dispersion appears as almost horizontal lines.
In particular, it can be seen that for $\boldsymbol{K}\parallel [110]$ and 
light being polarized along $\hat{\boldsymbol{e}}_{2\boldsymbol{K}}$
no exciton-photon coupling occurs.
For the transverse states we present the photon-like 
part for the two polarizations.

\begin{figure*}[t]
\begin{centering}
\includegraphics[width=2.0\columnwidth]{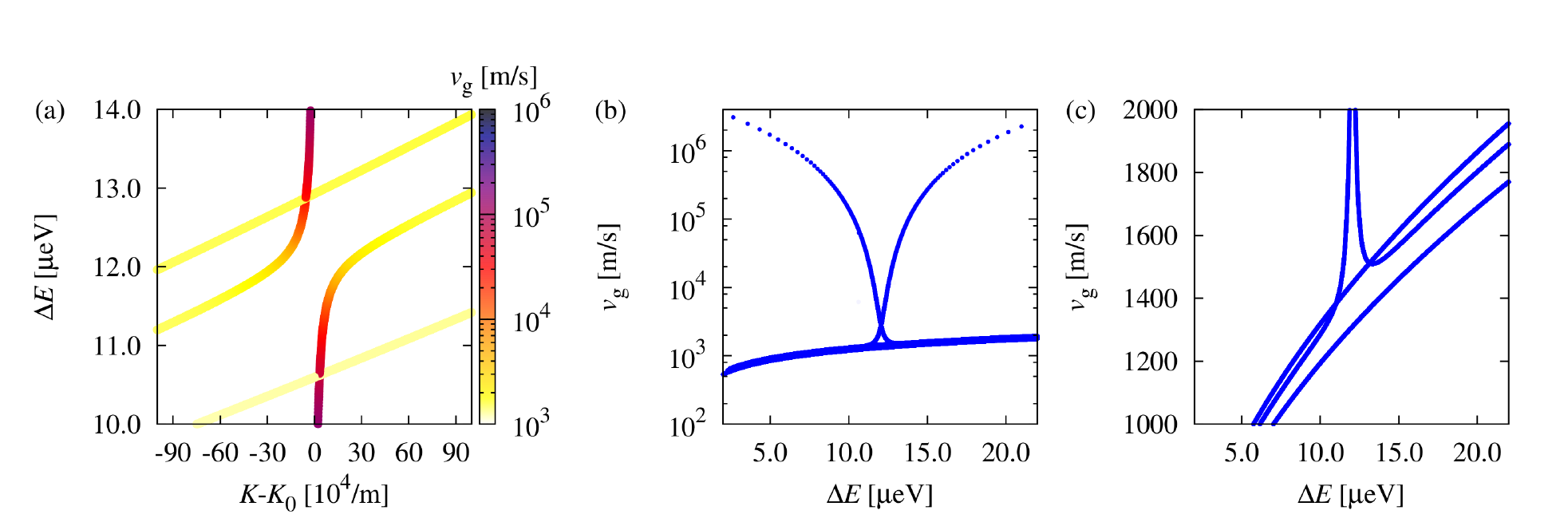}
\par\end{centering}

\protect\caption{(a) Group velocity $v_{\mathrm{g}}$ 
(colorscale) of the $1S$ ortho
exciton polariton with $\boldsymbol{K}\parallel[110]$
and $\hat{\boldsymbol{e}}_{1\boldsymbol{K}}$ polarization
in dependence on the energy difference
$\Delta E=E-E_0$ and the wave vector $K$.
(b) and (c) The group velocity $v_{\mathrm{g}}$ 
only in dependence on the energy difference
$\Delta E=E-E_0$ for a comparison with Fig.~3 of Ref.~\cite{P3_107_12}.
The minimum value of $v_{\mathrm{g}}$ depends
on $f_0$ and is given by $v_{\mathrm{g, min}}=1.5\,\mathrm{\frac{km}{s}}$.
~\label{fig:Figvg}}
\end{figure*}

In Ref.~\cite{P3_107_12} the group velocity of the
$1S$ ortho exciton-polariton has been measured
as a function of the photon energy for $\boldsymbol{K}\parallel[110]$.
From the dispersion shown in Fig.~\ref{fig:FigD1S}(b) and (e), we can directly
calculate the group velocity via
\begin{equation}
v_{\mathrm{g}}=\frac{\mathrm{d}\omega}{\mathrm{d}K}=\frac{1}{\hbar}\frac{\mathrm{d}E}{\mathrm{d}K}.\label{eq:vg}
\end{equation}
The result is shown in Fig.~\ref{fig:Figvg}.
It can be seen that the group velocity decreases
on the lower polariton branch close to the resonance
for increasing values of $K$.
However, for large values of $K$ the polariton dispersion
approaches the exciton dispersion, which increases
quadratically in $K$, so that $v_{\mathrm{g}}$
is then proportional to $K$.
Hence, there must be a minimum value of the group velocity.
In the experiment group velocities as low as $40\,\mathrm{\frac{km}{s}}$
could be measured~\cite{P3_107_12}. However, it was not possible
to measure the complete dispersion. In particular, the region
of very low group velocities is not experimentally accessible.
From Eq.~(\ref{eq:vg}) and the theoretical results it is possible to calculate all 
group velocities. The minimum value of $v_{\mathrm{g}}$
obtained in our calculations is
\begin{equation}
v_{\mathrm{g, min}}=1.5\,\mathrm{\frac{km}{s}}.
\end{equation}
at $K=2.758\,\frac{1}{\mathrm{m}}$.
When comparing our result for the group velocity, i.e., Fig.~\ref{fig:Figvg}(b),
with the experimental results of Ref.~\cite{P3_107_12},
a very good agreement can be observed.

From the results of the $5\times 5$ matrix model it
is also possible to calculate the polarization vector
of the photon-like part of the polariton
or the orientation of the electric field as 
$\sum_{i=1}^{2} W_{\mu\,i\boldsymbol{K}}\hat{\boldsymbol{e}}_{i\boldsymbol{K}}$
and the polarization vector connected with the exciton-like
part of the polariton via the symmetric cross product of
$\hat{\boldsymbol{e}}_{i\boldsymbol{K}}$ and 
$\boldsymbol{K}$ [cf.~Eq.~(\ref{eq:symmcrossprod})]
according to the group theoretical condition $\Gamma_5^+\otimes\Gamma_4^-\rightarrow\Gamma_4^-$.
One obtains for $\Delta_3=\Delta_5=0$ states
with purely longitudinal or transverse polarization.
However, in the general case, the states are mixed longitudinal-transverse states
and the polarization is not parallel to the applied electrical field.

Another interesting point is the polariton energies
if the crystal is rotated about the $[1\bar{1}0]$
or the $[111]$ axis (cf.~Refs.~\cite{9_1,8,9}).
In the case that the crystal is rotated about the $[1\bar{1}0]$ plane,
the $K$ vector passes all three orientations of high symmetry
from $[001]$ to $[111]$ and then to $[110]$:
\begin{equation}
\boldsymbol{K}=\frac{K}{\sqrt{2}}\left(\begin{array}{c}
\sin\varphi\\
\sin\varphi\\
\sqrt{2}\cos\varphi
\end{array}\right).\label{eq:K11m2}
\end{equation} 
The polarization vectors are given by
\begin{equation}
\hat{\boldsymbol{e}}_{1\boldsymbol{K}} =\frac{1}{\sqrt{2}}\left(\begin{array}{c}
\cos\varphi\\
\cos\varphi\\
-\sqrt{2}\sin\varphi
\end{array}\right),\quad 
\hat{\boldsymbol{e}}_{2\boldsymbol{K}} =\frac{1}{\sqrt{2}}\left(\begin{array}{c}
1\\
-1\\
0
\end{array}\right).\label{eq:e11m2}
\end{equation}
Independent of the angle $\varphi$ the symmetry of the problem is always $C_{\mathrm{s}}$~\cite{G3}.
This group contains the identity and a reflection at the plane with 
the normal vector $\hat{\boldsymbol{n}}=(1,\,-1,\,0)/\sqrt{2}$, which is
identical to one of the six $\sigma_{\mathrm{d}}$ reflections of $O_{\mathrm{h}}$.
The irreducible representations of $C_{\mathrm{s}}$ are either symmetric $(\Gamma_1)$
or antisymmetric $(\Gamma_2)$ under reflection.
Hence, the complete problem falls into two parts:
The linear combinations $\Psi_{xy}$ and $(\Psi_{zx}+\Psi_{yz})/\sqrt{2}$
of the ortho exciton states
transform according to $\Gamma_1$ while the linear combination
$(\Psi_{zx}-\Psi_{yz})/\sqrt{2}$ transforms according to $\Gamma_2$.
Furthermore, photons transform in $C_{\mathrm{s}}$ also according to
$\Gamma_1$ $(\hat{\boldsymbol{e}}_{1\boldsymbol{K}})$
and $\Gamma_2$ $(\hat{\boldsymbol{e}}_{2\boldsymbol{K}})$. 
Therefore, the problem decouples group theoretically
in a $\Gamma_1$ problem with three polariton branches
and a $\Gamma_2$ problem with only two polariton branches.

\begin{figure}[t]
\begin{centering}
\includegraphics[width=1.0\columnwidth]{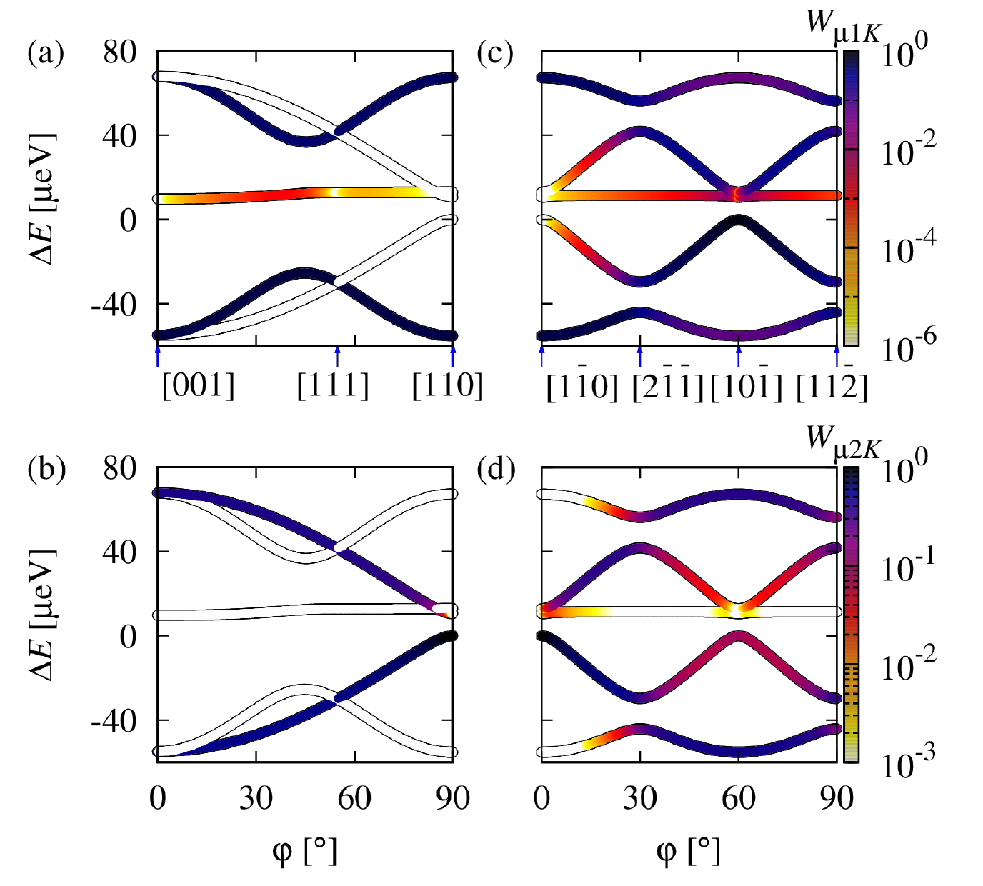}
\par\end{centering}

\protect\caption{(a), (b) Polariton energies at $K=K_0$ when rotating
the vector $\boldsymbol{K}$ in the plane with the normal vector
$\hat{\boldsymbol{n}}=(1,\,-1,\,0)/\sqrt{2}$. Again, we only plot the
energy difference
$\Delta E=E-E_0$.
The colorscale
denotes the photon-like part of the polariton states.
As is discussed in the text, the problem decouples in two problems
for the two irreducible representations $\Gamma_1$ [panel (a)]
and $\Gamma_2$ [panel (b)] of the group $C_{\mathrm{s}}$. 
(c), (d) Polariton energies at $K=K_0$ when rotating 
$\boldsymbol{K}$ in the plane with the normal vector
$\hat{\boldsymbol{n}}=(1,\,1,\,1)/\sqrt{3}$. 
Every $60\degree$ $\boldsymbol{K}$ is oriented in a direction
of high symmetry.
~\label{fig:Fig1S110}}
\end{figure}

We keep the amount of $\boldsymbol{K}$ fixed at $K=K_0$ and increase the
angle $\varphi$ from $0\degree$ to $90\degree$. The result is
shown in Fig.~\ref{fig:Fig1S110}(a) and (b). From the photon-like part
of the polaritons we can clearly see the decoupling of the problem.
Furthermore, the expected degeneracies of the polariton states occur
if $\boldsymbol{K}$ is oriented in $[001]$, $[111]$ or $[110]$ direction.

\begin{figure}[t]
\begin{centering}
\includegraphics[width=0.8\columnwidth]{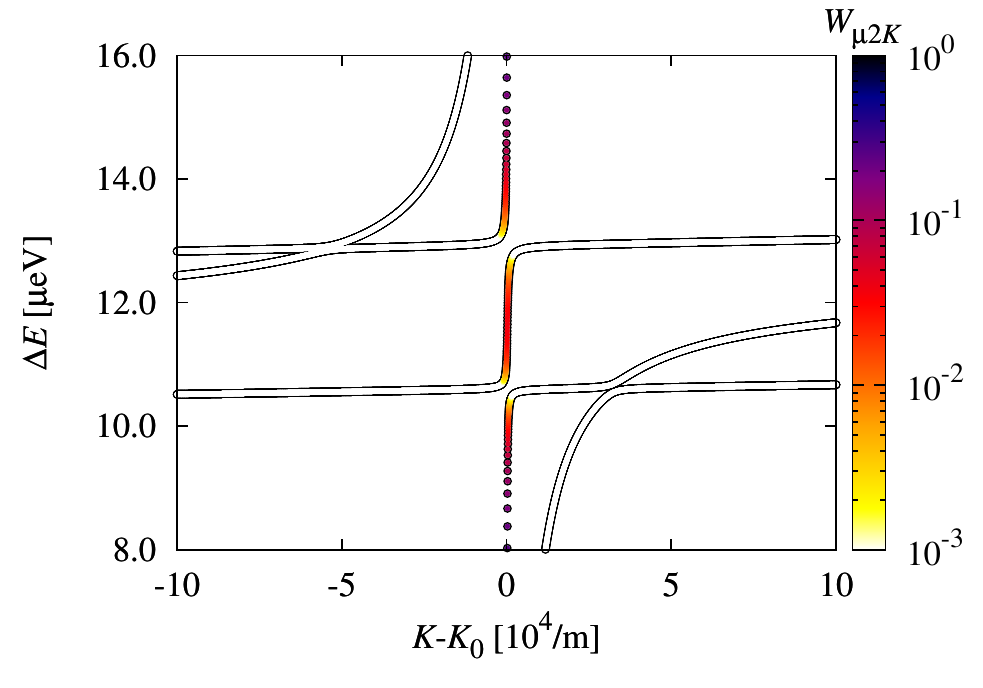}
\par\end{centering}

\protect\caption{Same as Fig.~\ref{fig:FigD1S} but
for $\boldsymbol{K}\parallel[11\bar{2}]$, i.e., for 
$\varphi=\arccos(-2/\sqrt{6})$ in Eqs.~(\ref{eq:K11m2})
and~(\ref{eq:e11m2}).
In panel (a) there are three hybrid polariton branches 
having a polarization within the $(1,\bar{1},0)$-plane.
This is in contrast to the directions of high symmetry, where 
only two polariton branches occur.
For further information see text.
~\label{fig:11m2}}
\end{figure}

Especially the polariton dispersion for $\boldsymbol{K}\parallel[11\bar{2}]$ 
$(\varphi=\arccos(-2/\sqrt{6}))$
is interesting as regards second harmonic generation~(SHG)
measurements. 
This dispersion is shown in Fig.~\ref{fig:11m2}.
SHG measurements are selective with respect to $\omega$ and $K$ creating bulk polaritons with 
$2\omega$ and $2K$. This allows the excitation of polaritons on the upper polariton branch. 
The lower polariton branch is inaccessible since the total $K$-vector is too small. 
Due to symmetry reasons it is not possible to observe SHG for $\boldsymbol{K}$ being parallel 
to $[001]$ or $[110]$. The SHG measurements in Ref.~\cite{93} were 
therefore performed in a configuration 
where $\boldsymbol{K}$ is parallel to $[111]$ or $[11\bar{2}]$. Especially SHG measurements 
with $\boldsymbol{K}\parallel[11\bar{2}]$ are interesting because 
there exist 3 hybrid polariton branches having a polarization within the 
$(1,\bar{1},0)$-plane. The polariton states on the upper and 
intermediate polariton branch are both accessible in SHG. Therefore, this is a situation
where two exciton polaritons can get coherently excited by means of the same nonlinear 
polarization $P(2\omega,\,2K)$. 
The interference of both exciton polaritons should give rise to a polariton 
beat showing up in time resolved SHG measurements.


Let us now discuss the rotation about the $[111]$ axis.
The $[111]$ axis is a threefold axis, for which reason after every $60\degree$
$\boldsymbol{K}$ is oriented in a direction of the form $[110]$.
Between each two of these cases $\boldsymbol{K}$ is oriented in 
a direction of the form $[112]$. 
We have
\begin{equation}
\boldsymbol{K}=\frac{K}{\sqrt{6}}\left(\begin{array}{c}
\sqrt{3}\cos\varphi+\sin\varphi\\
-\sqrt{3}\cos\varphi+\sin\varphi\\
-2\sin\varphi
\end{array}\right),\label{eq:K111}
\end{equation} 
and the polarization vectors
\begin{equation}
\hat{\boldsymbol{e}}_{j\boldsymbol{K}} =\frac{1}{3\sqrt{j}}\left(\begin{array}{c}
3j-4+\cos\varphi-\sqrt{3}\sin\varphi\\
3j-4+\cos\varphi+\sqrt{3}\sin\varphi\\
3j-4-2\cos\varphi
\end{array}\right),\quad j=1,2.
\end{equation}

The resulting spectrum when varying the angle $\varphi$
and keeping the amount of $\boldsymbol{K}$ fixed at $K=K_0$ is
shown in Fig.~\ref{fig:Fig1S110}(c) and (d).
In the cases with $\boldsymbol{K}$ being of the form $[110]$
only one exciton state is allowed for the polarization
vector of the form $[001]$ and all exciton states are forbidden
for the other polarization vector.
Note, however, that if we start with the configuration
$\boldsymbol{K}\parallel[0\bar{1}0]$, $\hat{\boldsymbol{e}}_{1\boldsymbol{K}}\parallel[00\bar{1}]$,
and $\hat{\boldsymbol{e}}_{1\boldsymbol{K}}\parallel[110]$ of Ref.~\cite{9_1}
and perform a rotation about the $[111]$ axis, none of the polarization vectors
is oriented in the $[010]$ direction at 
$\varphi=60\degree$ while $\boldsymbol{K}\parallel[10\bar{1}]$ holds.
Hence, only for $\varphi=0\degree$ all exciton states are
forbidden for the polarization $\hat{\boldsymbol{e}}_{2\boldsymbol{K}}$ in Fig.~\ref{fig:Fig1S110}(c) and (d).

In the cases with $\boldsymbol{K}$ being of the form $[112]$ 
all exciton states are allowed.
Since the difference between $\boldsymbol{K}\parallel [1\bar{1}0]$
and $\boldsymbol{K}\parallel [2\bar{1}\bar{1}]$ are only $\Delta\varphi=30\degree$,
a rotation about the $[111]$ axis with $\varphi=4\degree$
already shows a significant effect on the spectrum as can be seen
especially from Fig.~\ref{fig:Fig1S110}(d).

Exactly for this case with $\varphi=4\degree$ and the polarization
$\hat{\boldsymbol{e}}_{2\boldsymbol{K}}$ the transmission
spectrum of the ortho exciton-polariton is shown 
in Fig.~2(c) of Ref.~\cite{9_1}.
The shape of the transmission spectrum of
is clearly affected by polariton dispersion.
Photons with $\boldsymbol{K}\parallel[\bar{1}10]$ and 
$\hat{\boldsymbol{e}}_{\xi\boldsymbol{K}}\parallel[110]$
do not interact with the ortho exciton. 
Tilting the $K$-vector by $4\degree$ leads to a very weak exciton photon interaction 
so that the two of the excitons show up in this figure as two extremely narrow absorption 
peaks. The product of Rabi frequency and exciton lifetime is however so low 
that these states are excitons rather than polaritons. This explains 
the sharp absorption lines with a line width of less than $1\,\mathrm{\mu eV}$. 
The third quadrupole-allowed state is responsible for the comparatively broad 
absorption line observed in $[001]$
polarization. This state is in contrast to the other ones a perfect polariton. 
The much broader line width of $20\,\mathrm{\mu eV}$ is caused 
by polariton dispersion. If single photon spectroscopy was
$K$-selective, this exciton polariton would show up with a similar 
small line width as is in the pure exciton case. 

\begin{figure}[t]
\begin{centering}
\includegraphics[width=1.0\columnwidth]{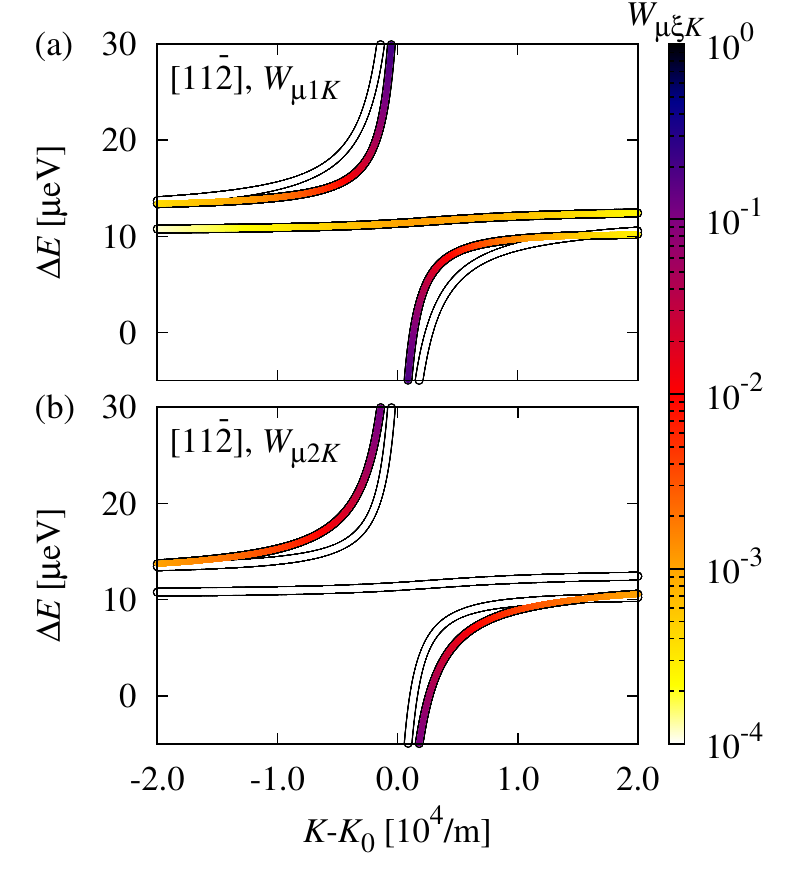}
\par\end{centering}

\protect\caption{Polariton energies for the vector $\boldsymbol{K}$ 
given by Eq.~(\ref{eq:K111}) with $\varphi=4\degree$
and its amount is varied. 
Again, we only plot the energy difference
$\Delta E=E-E_0$.
The colorscale denotes the photon-like part of the polariton states.
~\label{fig:Fig1T111}}
\end{figure}

For a comparison with the experiment, 
we calculate the polariton dispersion for different values
of $K$ and keeping $\varphi$ fixed to $\varphi=4\degree$.
The result is shown in Fig.~\ref{fig:Fig1T111}.
It can be seen that for the polarization $\hat{\boldsymbol{e}}_{2\boldsymbol{K}}$
two states have a significant photon amplitude. Hence,
these are the two states, which could be observed as very narrow
lines in the experimental transmission spectrum~\cite{9_1}.
However, while in the experiment the energy difference between the two states
is about $4\,\mathrm{\mu eV}$, we obtain only $2.4\,\mathrm{\mu eV}$.
We can exclude uncertainties in the parameter $\Delta_1$
or the angle $\varphi$ since a variation of these parameters
does not change the size of the splitting.
Only a variation of $\Delta_3$ and $\Delta_5$ changes the size
of the splitting. However, even if we set these parameters to the
experimentally determined values listed in Table~\ref{tab:3}, the splitting increases
only to $3.0\,\mathrm{\mu eV}$.
As we have already stated above, the presence of small strains
in the crystal cannot be excluded. These may be the major reason for the
observed discrepancy.

Of course, it is also possible that the positions of the transmission
maxima in the experiment are not exactly given by the positions
of the avoided crossings. In this case, one would have to calculate the
transmission spectrum for the polariton dispersion using 
Pekar's boundary conditions~\cite{P9}. This is likewise difficult 
and beyond the scope of this work.
Despite all that, we obtain a good qualitative agreement with the experiment.




\section{Summary and outlook\label{sec:Summary}}

We presented the theory of exciton-polaritons in $\mathrm{Cu_{2}O}$.
In the derivation of the formulas we accounted for all
relevant effects which are needed to describe the spectra
theoretically in an appropriate way, i.e., the complete valence-band structure,
the exchange interaction, and the central-cell corrections~\cite{200}.
This leads to a likewise complicated expression for the momentum-dependent
Hamiltonian of excitons. Our method of solving the corresponding
Schr\"odinger equation allows calculating dipole and quadrupole oscillator strengths,
for which general formulas have been derived. The subsequent polariton
transformation can be performed within the so-called rotating-wave approximation.
Within this approximation it is straightforward to additionally account for the nonanalytic
exchange interaction.

We have treated the dispersion of polaritons in $\mathrm{Cu_{2}O}$
using a multi-polariton concept. 
When considering the correct model of excitons in $\mathrm{Cu_{2}O}$,
all states which have the symmetry $\Gamma_5^+$ for $K=0$ obtain
a finite $K$-dependent oscillator strength, i.e., $S$, $D$, $G$, and higher
excitons.
The more complex spectrum of excitons when using the full Hamiltonian
leads to a more complex polariton dispersion than in previous works~\cite{74,BA_JaV}.
We also estimated that polariton effects
for the yellow exciton states with $n\geq 2$ may
only be observed using nonlinear spectroscopy methods
and high quality crystals.

When including the complete valence band structure, a
$K$-dependent splitting of the three
components of the yellow $1S$ ortho exciton appears.
In contrast to Ref.~\cite{100},
we have solved the complete $K$-dependent Schr\"odinger equation numerically
including also the correct values for the central-cell corrections of Ref.~\cite{200}.
The splittings calculated are on the same order of magnitude as 
the splittings observed experimentally~\cite{9_1,8,9}.

As the exciton ground state is well separated from the other exciton states, it
can be treated separately as regards polariton effects. 
Exploiting the symmetry properties of the $1S$ ortho exciton,
we were able to set up a $5\times 5$ matrix model, which allows for the calculation of the
corresponding polariton dispersion for any direction of $\boldsymbol{K}$.
Using the $5\times 5$ matrix model, we investigated the dispersion 
of the $1S$ ortho exciton polariton for different orientations of $\boldsymbol{K}$.
In a comparison of the results for the group velocity with experimental
results we obtained a very good agreement. 
The calculations allowed us to determine the minimum value of the group velocity,
which is not directly accessible in the experiment, to $v_{\mathrm{g, min}}=1.5\,\mathrm{\frac{km}{s}}$.
We also presented results for the two special cases that the $\boldsymbol{K}$-vector
is rotated in planes perpendicular to $[1\bar{1}0]$ and $[111]$.
Especially in the second case, another comparison with experimental results
was possible for a rotation angle of $4\degree$. The splitting between the two
allowed states obtained is smaller than in the experiment. This can, however, be
explained in terms of the presence of small strains and 
uncertainties in the parameters $\Delta_i$.

As a next step, it is possible to calculate transmission spectra
from the polariton dispersion obtained via the $5\times 5$ matrix model
by assuming Pekar's boundary conditions~\cite{P9} and 
applying the Fresnel equations. This would allow for an even better comparison
of theoretical results with the experiment and may also lead
to an understanding of the shape of the peaks in the spectra.

Furthermore, we plan to investigate exciton spectra in an external magnetic field
in Voigt configuration, where, in contrast to
the Faraday configuration of Ref.~\cite{125}, a finite value 
of the wave vector $K$ must be considered.

\acknowledgments
F.S. is  grateful  for  support  from  the
Landesgraduiertenf\"orderung of the Land Baden-W\"urttemberg.



\section*{SUPPLEMENTARY INFORMATION}

\section{Hamiltonian and matrix elements\label{sub:Hamiltonianrm}}

In the paper the Hamiltonian of excitons
has been written in the form
\begin{equation}
H=H_{0}+\left(\hbar K\right)H_{1}+\left(\hbar K\right)^{2}H_{2}.\label{eq:Hges}
\end{equation}
We now give in Secs.~\ref{sub:A001}-\ref{sub:A111} the expressions for $H_0$, $H_1$ and $H_2$
in terms of irreducible tensors
for the case that the wave vector $\boldsymbol{K}$ is oriented in $[001]$, $[110]$ or $[111]$ direction~\cite{EDsub,7_11sub,44sub,100sub}.
Note that we rotate the coordinate system to make the quantization axis or $z$ axis
coincide with the direction of $\boldsymbol{K}$.
Hence, we rotate the coordinate system by the Euler angles
$\left(\alpha,\,\beta,\,\gamma\right)=\left(\pi,\,\pi/2,\,\pi/4\right)$ for
$\boldsymbol{K}\parallel [110]$ and by
$\left(\alpha,\,\beta,\,\gamma\right)=\left(0,\,\arccos(1/\sqrt{3}),\,\pi/4\right)$ for
$\boldsymbol{K}\parallel [111]$.

The first-order and second-order tensor operators
used in the following
correspond, as in Ref.~\cite{44sub}, to the vector operators 
$\boldsymbol{r}$,~$\boldsymbol{p}$,~$\boldsymbol{I}$,~$\boldsymbol{S}_{\mathrm{h}}$
and to the second-rank Cartesian operators
\begin{subequations}
\begin{eqnarray}
I_{mn} & = & 3\left\{ I_{m},\, I_{n}\right\} -\delta_{mn}I^{2},\\
P_{mn} & = & 3\left\{ p_{m},\, p_{n}\right\} -\delta_{mn}p^{2},
\end{eqnarray}
\end{subequations}
respectively. We also use the abbreviation
\begin{equation}
D_{k}^{(2)}=\left[I^{(1)}\times S_{\mathrm{h}}^{(1)}\right]_{k}^{(2)}.
\end{equation}
The coefficients $\gamma'_{1}$, $\mu'$ and $\delta'$
are given by~\cite{7_11sub,7sub}
\begin{equation}
\gamma'_{1}=\gamma_{1}+\frac{m_{0}}{m_{\mathrm{e}}},\quad\mu'=\frac{6\gamma_{3}+4\gamma_{2}}{5\gamma'_{1}},\quad\delta'=\frac{\gamma_{3}-\gamma_{2}}{\gamma'_{1}},
\end{equation}
and we define by analogy~\cite{100sub}
\begin{equation}
\nu=\frac{6\eta_{3}+4\eta_{2}}{5\eta_{1}},\quad\tau=\frac{\eta_{3}-\eta_{2}}{\eta_{1}}.
\end{equation}

In Sec.~\ref{sub:Matrix-elements} we then give the matrix elements 
of the terms of the Hamiltonian in the basis $\left|\Pi\right\rangle  = \left|N,\, L;\,\left(I,\, S_{\mathrm{h}}\right)\, J;\, F,\, S_{\mathrm{e}};\, F_{t},\, M_{F_{t}}\right\rangle$
in Hartree units using the formalism of irreducible tensors~\cite{EDsub}.
The matrix elements of the first part $H_{0}$ of the Hamiltonian
are given in the Appendices of Refs.~\cite{100sub,200sub}. 
The functions of the form $\left(R_{1}\right)_{NL}^{j}$ and the quantities
$I_{N'\, L';N\, L}$ are taken from the Appendix of Ref.~\cite{100sub}.

\begin{widetext}

\subsection{Wave vector in [001] direction \label{sub:A001}}

\begin{subequations}
\begin{align}
H_{0} = &\:  E_{\mathrm{g}}-\frac{e^{2}}{4\pi\varepsilon_{0}\varepsilon}\frac{1}{r}-\frac{e^{2}}{4\pi\varepsilon_{0}r}\left[\frac{1}{2\varepsilon_{1}^{*}}\left(e^{-r/\rho_{\mathrm{h}1}}+e^{-r/\rho_{\mathrm{e}1}}\right)+\frac{1}{2\varepsilon_{2}^{*}}\left(e^{-r/\rho_{\mathrm{h}2}}+e^{-r/\rho_{\mathrm{e}2}}\right)\right]\displaybreak[3]\nonumber\\
\nonumber\\
 &\: +\frac{2}{3}\Delta\left(1+\frac{1}{\hbar^{2}}I^{(1)}\cdot S_{\mathrm{h}}^{(1)}\right)+\left[-V_0 +J_{0}\left(\frac{1}{4}-\frac{1}{\hbar^{2}}S_{\mathrm{e}}^{(1)}\cdot S_{\mathrm{h}}^{(1)}\right)\right]V_{\mathrm{uc}}\delta\!\left(\boldsymbol{r}\right)\displaybreak[3]\nonumber \\
\nonumber \\
 &\: +\frac{\gamma_{1}'}{2\hbar^{2}m_{0}}\left\{ \hbar^{2}p^{2}-\frac{\mu'}{3}\left(P^{(2)}\cdot I^{(2)}\right)\right.\displaybreak[3]\nonumber \\
\nonumber \\
 &\: \qquad\qquad\qquad\left.+\frac{\delta'}{3}\left(\sum_{k=\pm4}\left[P^{(2)}\times I^{(2)}\right]_{k}^{(4)}+\frac{\sqrt{70}}{5}\left[P^{(2)}\times I^{(2)}\right]_{0}^{(4)}\right)\right\} \displaybreak[3]\nonumber \\
\nonumber \\
 &\: +\frac{3\eta_{1}}{\hbar^{2}m_{0}}\left\{ \frac{1}{3}p^{2}\left(I^{(1)}\cdot S_{\mathrm{h}}^{(1)}\right)-\frac{\nu}{3}\left(P^{(2)}\cdot D^{(2)}\right)\right.\displaybreak[3]\nonumber \\
\nonumber \\
 &\: \qquad\qquad\qquad\left.+\frac{\tau}{3}\left(\sum_{k=\pm4}\left[P^{(2)}\times D^{(2)}\right]_{k}^{(4)}+\frac{\sqrt{70}}{5}\left[P^{(2)}\times D^{(2)}\right]_{0}^{(4)}\right)\right\} ,
\\
\displaybreak[1]
\nonumber \\
H_{1} = &\:  \frac{1}{2\hbar^{2}m_{\mathrm{e}}}\left\{ -2\sqrt{\frac{5}{3}}\mu'\left[P^{(1)}\times I^{(2)}\right]_{0}^{(1)}-4\sqrt{\frac{2}{5}}\delta'\left[P^{(1)}\times I^{(2)}\right]_{0}^{(3)}\right\}\displaybreak[3] \nonumber \\
\nonumber \\
 &\: +\frac{3\eta_{1}}{\gamma{}_{1}'\hbar^{2}m_{\mathrm{e}}}\left\{ -\frac{2}{3}P_{0}^{(1)}\left(I^{(1)}\cdot S_{\mathrm{h}}^{(1)}\right)-2\sqrt{\frac{5}{3}}\nu\left[P^{(1)}\times D^{(2)}\right]_{0}^{(1)}-4\sqrt{\frac{2}{5}}\tau\left[P^{(1)}\times D^{(2)}\right]_{0}^{(3)}\right\}, 
\\
\displaybreak[1]
\nonumber \\
H_{2} = &\:  \frac{\gamma_{1}'m_{\mathrm{e}}-m_{0}}{2\gamma_{1}'m_{\mathrm{e}}^{2}}+\frac{m_{0}}{2\gamma_{1}'\hbar^{2}m_{\mathrm{e}}^{2}}\left\{ \left(-\sqrt{\frac{2}{3}}\mu'+\frac{2}{5}\sqrt{6}\delta'\right)I_{0}^{(2)}\right\}\displaybreak[3] \nonumber \\
\nonumber \\
 &\: +\frac{3\eta_{1}m_{0}}{\gamma_{1}^{'2}\hbar^{2}m_{\mathrm{e}}^{2}}\left\{ \frac{1}{3}\left(I^{(1)}\cdot S_{\mathrm{h}}^{(1)}\right)+\left(-\sqrt{\frac{2}{3}}\nu+\frac{2}{5}\sqrt{6}\tau\right)D_{0}^{(2)}\right\}.
\end{align}
\end{subequations}

\subsection{Wave vector in [110] direction \label{sub:A110}}

\begin{subequations}
\begin{align}
H_{0} = &\:  E_{\mathrm{g}}-\frac{e^{2}}{4\pi\varepsilon_{0}\varepsilon}\frac{1}{r}-\frac{e^{2}}{4\pi\varepsilon_{0}r}\left[\frac{1}{2\varepsilon_{1}^{*}}\left(e^{-r/\rho_{\mathrm{h}1}}+e^{-r/\rho_{\mathrm{e}1}}\right)+\frac{1}{2\varepsilon_{2}^{*}}\left(e^{-r/\rho_{\mathrm{h}2}}+e^{-r/\rho_{\mathrm{e}2}}\right)\right]\displaybreak[3]\nonumber\\
\nonumber\\
 &\: +\frac{2}{3}\Delta\left(1+\frac{1}{\hbar^{2}}I^{(1)}\cdot S_{\mathrm{h}}^{(1)}\right)+\left[-V_0 +J_{0}\left(\frac{1}{4}-\frac{1}{\hbar^{2}}S_{\mathrm{e}}^{(1)}\cdot S_{\mathrm{h}}^{(1)}\right)\right]V_{\mathrm{uc}}\delta\!\left(\boldsymbol{r}\right)\displaybreak[3]\nonumber \\
\nonumber \\
 &\: +\frac{\gamma_{1}'}{2\hbar^{2}m_{0}}\left\{ \hbar^{2}p^{2}-\frac{\mu'}{3}\left(P^{(2)}\cdot I^{(2)}\right)+\frac{\delta'}{4}\left(\sum_{k=\pm4}\left[P^{(2)}\times I^{(2)}\right]_{k}^{(4)}\right)\right.\displaybreak[3]\nonumber \\
\nonumber \\
 &\: \qquad\qquad\qquad\left.-\frac{\sqrt{7}}{6}\delta'\left(\sum_{k=\pm2}\left[P^{(2)}\times I^{(2)}\right]_{2}^{(4)}+\sqrt{\frac{1}{10}}\left[P^{(2)}\times I^{(2)}\right]_{0}^{(4)}\right)\right\} \displaybreak[3]\nonumber \\
\nonumber \\
 &\: +\frac{3\eta_{1}}{\hbar^{2}m_{0}}\left\{ \frac{1}{3}p^{2}\left(I^{(1)}\cdot S_{\mathrm{h}}^{(1)}\right)-\frac{\nu}{3}\left(P^{(2)}\cdot D^{(2)}\right)+\frac{\tau}{4}\left(\sum_{k=\pm4}\left[P^{(2)}\times D^{(2)}\right]_{k}^{(4)}\right)\right.\displaybreak[3]\nonumber \\
\nonumber \\
 &\: \qquad\qquad\qquad\left.-\frac{\sqrt{7}}{6}\tau\left(\sum_{k=\pm2}\left[P^{(2)}\times D^{(2)}\right]_{k}^{(4)}+\sqrt{\frac{1}{10}}\left[P^{(2)}\times D^{(2)}\right]_{0}^{(4)}\right)\right\}, 
\\
\displaybreak[1]
\nonumber \\
H_{1} = &\:  \frac{1}{2\hbar^{2}m_{\mathrm{e}}}\left\{ -2\sqrt{\frac{5}{3}}\mu'\left[P^{(1)}\times I^{(2)}\right]_{0}^{(1)}+\sqrt{3}\delta'\left(\sum_{k=\pm2}\left[P^{(1)}\times I^{(2)}\right]_{k}^{(3)}+\sqrt{\frac{2}{15}}\left[P^{(1)}\times I^{(2)}\right]_{0}^{(3)}\right)\right\}\displaybreak[3] \nonumber \\
\nonumber \\
 &\: +\frac{3\eta_{1}}{\gamma{}_{1}'\hbar^{2}m_{\mathrm{e}}}\left\{ -\frac{2}{3}P_{0}^{(1)}\left(I^{(1)}\cdot S_{\mathrm{h}}^{(1)}\right)-2\sqrt{\frac{5}{3}}\nu\left[P^{(1)}\times D^{(2)}\right]_{0}^{(1)}\right.\displaybreak[3]\nonumber \\
\nonumber \\
 &\: \qquad\qquad\qquad\left.+\sqrt{3}\tau\left(\sum_{k=\pm2}\left[P^{(1)}\times D^{(2)}\right]_{k}^{(3)}+\sqrt{\frac{2}{15}}\left[P^{(1)}\times D^{(2)}\right]_{0}^{(3)}\right)\right\}, 
\\
\displaybreak[1]
\nonumber \\
H_{2} = &\:  \frac{\gamma_{1}'m_{\mathrm{e}}-m_{0}}{2\gamma_{1}'m_{\mathrm{e}}^{2}}+\frac{m_{0}}{2\gamma_{1}'\hbar^{2}m_{\mathrm{e}}^{2}}\left\{ -\frac{1}{2}\delta'\left(I_{-2}^{(2)}+I_{2}^{(2)}\right)+\left(-\sqrt{\frac{2}{3}}\mu'-\frac{1}{5}\sqrt{\frac{3}{2}}\delta'\right)I_{0}^{(2)}\right\}\displaybreak[3] \nonumber \\
\nonumber \\
 &\: +\frac{3\eta_{1}m_{0}}{\gamma_{1}^{'2}\hbar^{2}m_{\mathrm{e}}^{2}}\left\{ \frac{1}{3}\left(I^{(1)}\cdot S_{\mathrm{h}}^{(1)}\right)-\frac{1}{2}\tau\left(D_{-2}^{(2)}+D_{2}^{(2)}\right)+\left(-\sqrt{\frac{2}{3}}\nu-\frac{1}{5}\sqrt{\frac{3}{2}}\tau\right)D_{0}^{(2)}\right\}. 
\end{align}
\end{subequations}

\subsection{Wave vector in [111] direction \label{sub:A111}}

\begin{subequations}
\begin{align}
H_{0} = &\:  E_{\mathrm{g}}-\frac{e^{2}}{4\pi\varepsilon_{0}\varepsilon}\frac{1}{r}-\frac{e^{2}}{4\pi\varepsilon_{0}r}\left[\frac{1}{2\varepsilon_{1}^{*}}\left(e^{-r/\rho_{\mathrm{h}1}}+e^{-r/\rho_{\mathrm{e}1}}\right)+\frac{1}{2\varepsilon_{2}^{*}}\left(e^{-r/\rho_{\mathrm{h}2}}+e^{-r/\rho_{\mathrm{e}2}}\right)\right]\displaybreak[3]\nonumber\\
\nonumber\\
 &\: +\frac{2}{3}\Delta\left(1+\frac{1}{\hbar^{2}}I^{(1)}\cdot S_{\mathrm{h}}^{(1)}\right)+\left[-V_0 +J_{0}\left(\frac{1}{4}-\frac{1}{\hbar^{2}}S_{\mathrm{e}}^{(1)}\cdot S_{\mathrm{h}}^{(1)}\right)\right]V_{\mathrm{uc}}\delta\!\left(\boldsymbol{r}\right)\displaybreak[3]\nonumber \\
\nonumber \\
 &\: +\frac{\gamma_{1}'}{2\hbar^{2}m_{0}}\left\{ \hbar^{2}p^{2}-\frac{\mu'}{3}\left(P^{(2)}\cdot I^{(2)}\right)\right.\displaybreak[3]\nonumber \\
\nonumber \\
 &\: \qquad\qquad\qquad\left.+\frac{4}{27}\delta'\left(\sum_{k=\pm3}k\left[P^{(2)}\times I^{(2)}\right]_{k}^{(4)}-3\sqrt{\frac{7}{10}}\left[P^{(2)}\times I^{(2)}\right]_{0}^{(4)}\right)\right\} \displaybreak[3]\nonumber \\
\nonumber \\
 &\: +\frac{3\eta_{1}}{\hbar^{2}m_{0}}\left\{ \frac{1}{3}p^{2}\left(I^{(1)}\cdot S_{\mathrm{h}}^{(1)}\right)-\frac{\nu}{3}\left(P^{(2)}\cdot D^{(2)}\right)\right.\displaybreak[3]\nonumber \\
\nonumber \\
 &\: \qquad\qquad\qquad\left.+\frac{4}{27}\tau\left(\sum_{k=\pm3}k\left[P^{(2)}\times D^{(2)}\right]_{3}^{(4)}-3\sqrt{\frac{7}{10}}\left[P^{(2)}\times D^{(2)}\right]_{0}^{(4)}\right)\right\}, 
\\
\displaybreak[1]
\nonumber \\
H_{1} = &\:  \frac{1}{2\hbar^{2}m_{\mathrm{e}}}\left\{ -2\sqrt{\frac{5}{3}}\mu'\left[P^{(1)}\times I^{(2)}\right]_{0}^{(1)}-\frac{4}{9}\delta'\left(\sum_{k=\pm3}k\left[P^{(1)}\times I^{(2)}\right]_{k}^{(3)}-6\sqrt{\frac{2}{5}}\left[P^{(1)}\times I^{(2)}\right]_{0}^{(3)}\right)\right\} \displaybreak[3]\nonumber \\
\nonumber \\
 &\: +\frac{3\eta_{1}}{\gamma{}_{1}'\hbar^{2}m_{\mathrm{e}}}\left\{ -\frac{2}{3}P_{0}^{(1)}\left(I^{(1)}\cdot S_{\mathrm{h}}^{(1)}\right)-2\sqrt{\frac{5}{3}}\nu\left[P^{(1)}\times D^{(2)}\right]_{0}^{(1)}\right.\displaybreak[3]\nonumber \\
\nonumber \\
 &\: \qquad\qquad\qquad\left.-\frac{4}{9}\tau\left(\sum_{k=\pm3}k\left[P^{(1)}\times D^{(2)}\right]_{k}^{(3)}-6\sqrt{\frac{2}{5}}\left[P^{(1)}\times D^{(2)}\right]_{0}^{(3)}\right)\right\}, 
\\
\displaybreak[1]
\nonumber \\
H_{2} = &\:  \frac{\gamma_{1}'m_{\mathrm{e}}-m_{0}}{2\gamma_{1}'m_{\mathrm{e}}^{2}}+\frac{m_{0}}{2\gamma_{1}'\hbar^{2}m_{\mathrm{e}}^{2}}\left\{ -\sqrt{\frac{2}{3}}\left(\mu'+\frac{4}{5}\delta'\right)I_{0}^{(2)}\right\}\displaybreak[3] \nonumber \\
\nonumber \\
 &\: +\frac{3\eta_{1}m_{0}}{\gamma_{1}^{'2}\hbar^{2}m_{\mathrm{e}}^{2}}\left\{ \frac{1}{3}\left(I^{(1)}\cdot S_{\mathrm{h}}^{(1)}\right)-\sqrt{\frac{2}{3}}\left(\nu+\frac{4}{5}\tau\right)D_{0}^{(2)}\right\}. 
\end{align}
\end{subequations}

\subsection{Matrix elements and reduced matrix elements\label{sub:Matrix-elements}}

\begin{align}
\left\langle \Pi'\left|I_{q}^{(2)}\right|\Pi\right\rangle = & \:\delta_{LL'}3\sqrt{5}\left(-1\right)^{F'_{t}+F_{t}-M'_{F_{t}}+2F'+2J+L}\displaybreak[3]\nonumber \\
\nonumber \\
\times & \:\left[\left(2F_{t}+1\right)\left(2F'_{t}+1\right)\left(2F+1\right)\left(2F'+1\right)\left(2J+1\right)\left(2J'+1\right)\right]^{\frac{1}{2}}\displaybreak[3]\nonumber \\
\nonumber \\
\times & \:\left(\begin{array}{ccc}
F'_{t} & 2 & F_{t}\\
-M'_{F_{t}} & q & M_{F_{t}}
\end{array}\right)\left\{ \begin{array}{ccc}
F' & F'_{t} & \frac{1}{2}\\
F_{t} & F & 2
\end{array}\right\} \left\{ \begin{array}{ccc}
J' & F' & L\\
F & J & 2
\end{array}\right\} \left\{ \begin{array}{ccc}
1 & J' & \frac{1}{2}\\
J & 1 & 2
\end{array}\right\}\displaybreak[3] \nonumber \\
\nonumber \\
\times & \:\sum_{j=-1}^{1}\left(R_{1}\right)_{NL}^{j}\left[N+L+j+1\right]^{-1}\delta_{N',N+j},\\
\displaybreak[1]\nonumber \\
\left\langle \Pi'\left|D_{q}^{(2)}\right|\Pi\right\rangle = & \:\delta_{LL'}3\sqrt{5}\left(-1\right)^{F'_{t}+F_{t}-M'_{F_{t}}+2F'+L+J+\frac{1}{2}}\displaybreak[3]\nonumber \\
\nonumber \\
\times & \:\left[\left(2F_{t}+1\right)\left(2F'_{t}+1\right)\left(2F+1\right)\left(2F'+1\right)\left(2J+1\right)\left(2J'+1\right)\right]^{\frac{1}{2}}\displaybreak[3]\nonumber \\
\nonumber \\
\times & \:\left(\begin{array}{ccc}
F'_{t} & 2 & F_{t}\\
-M'_{F_{t}} & q & M_{F_{t}}
\end{array}\right)\left\{ \begin{array}{ccc}
F' & F'_{t} & \frac{1}{2}\\
F_{t} & F & 2
\end{array}\right\} \left\{ \begin{array}{ccc}
J' & F' & L\\
F & J & 2
\end{array}\right\} \left\{ \begin{array}{ccc}
1 & 1 & 1\\
\frac{1}{2} & \frac{1}{2} & 1\\
J' & J & 2
\end{array}\right\}\displaybreak[3] \nonumber \\
\nonumber \\
\times & \:\sum_{j=-1}^{1}\left(R_{1}\right)_{NL}^{j}\left[N+L+j+1\right]^{-1}\delta_{N',N+j},\\
\displaybreak[1]\nonumber \\
\left\langle \Pi'\left|\left(I^{(1)}\cdot S_{\mathrm{h}}^{(1)}\right)\right|\Pi\right\rangle = & \:\delta_{LL'}\delta_{M_{F_{t}}M'_{F_{t}}}\delta_{F_{t}F'_{t}}3\sqrt{3}\left(-1\right)^{F_{t}+2F'+L+J+\frac{3}{2}}\displaybreak[3]\nonumber \\
\nonumber \\
\times & \:\left[\left(2F_{t}+1\right)\left(2F+1\right)\left(2F'+1\right)\left(2J+1\right)\left(2J'+1\right)\right]^{\frac{1}{2}}\displaybreak[3]\nonumber \\
\nonumber \\
\times & \:\left\{ \begin{array}{ccc}
F' & F{}_{t} & \frac{1}{2}\\
F_{t} & F & 0
\end{array}\right\} \left\{ \begin{array}{ccc}
J' & F' & L\\
F & J & 0
\end{array}\right\} \left\{ \begin{array}{ccc}
1 & 1 & 1\\
\frac{1}{2} & \frac{1}{2} & 1\\
J' & J & 0
\end{array}\right\}\displaybreak[3] \nonumber \\
\nonumber \\
\times & \:\sum_{j=-1}^{1}\left(R_{1}\right)_{NL}^{j}\left[N+L+j+1\right]^{-1}\delta_{N',N+j},\\
\displaybreak[1]\nonumber \\
\left\langle \Pi'\left|\left[P^{(1)}\times I^{(2)}\right]_{q}^{(K)}\right|\Pi\right\rangle = & \:3\sqrt{5}\left(-1\right)^{F'_{t}+F_{t}-M'_{F_{t}}+F'+J+K}\left\langle N'\, L'\left\Vert P^{(1)}\right\Vert N\, L\right\rangle\displaybreak[3] \nonumber \\
\nonumber \\
\times & \:\left[\left(2F_{t}+1\right)\left(2F'_{t}+1\right)\left(2F+1\right)\left(2F'+1\right)\left(2J+1\right)\left(2J'+1\right)\left(2K+1\right)\right]^{\frac{1}{2}}\displaybreak[3]\nonumber \\
\nonumber \\
\times & \:\left(\begin{array}{ccc}
F'_{t} & K & F_{t}\\
-M'_{F_{t}} & q & M_{F_{t}}
\end{array}\right)\left\{ \begin{array}{ccc}
F' & F'_{t} & \frac{1}{2}\\
F_{t} & F & K
\end{array}\right\} \left\{ \begin{array}{ccc}
L' & L & 1\\
J' & J & 2\\
F' & F & K
\end{array}\right\} \left\{ \begin{array}{ccc}
1 & J' & \frac{1}{2}\\
J & 1 & 2
\end{array}\right\}, \\
\displaybreak[1]\nonumber \\
\left\langle \Pi'\left|\left[P^{(1)}\times D^{(2)}\right]_{q}^{(K)}\right|\Pi\right\rangle = & \:3\sqrt{5}\left(-1\right)^{F'_{t}+F_{t}-M'_{F_{t}}+F'+\frac{1}{2}+K}\left\langle N'\, L'\left\Vert P^{(1)}\right\Vert N\, L\right\rangle\displaybreak[3] \nonumber \\
\nonumber \\
\times & \:\left[\left(2F_{t}+1\right)\left(2F'_{t}+1\right)\left(2F+1\right)\left(2F'+1\right)\left(2J+1\right)\left(2J'+1\right)\left(2K+1\right)\right]^{\frac{1}{2}}\displaybreak[3]\nonumber \\
\nonumber \\
\times & \:\left(\begin{array}{ccc}
F'_{t} & K & F_{t}\\
-M'_{F_{t}} & q & M_{F_{t}}
\end{array}\right)\left\{ \begin{array}{ccc}
F' & F'_{t} & \frac{1}{2}\\
F_{t} & F & K
\end{array}\right\} \left\{ \begin{array}{ccc}
L' & L & 1\\
J' & J & 2\\
F' & F & K
\end{array}\right\} \left\{ \begin{array}{ccc}
1 & 1 & 1\\
\frac{1}{2} & \frac{1}{2} & 1\\
J' & J & 2
\end{array}\right\}, \\
\displaybreak[1]\nonumber \\
\left\langle \Pi'\left|P_{0}^{(1)}\left(I^{(1)}\cdot S_{\mathrm{h}}^{(1)}\right)\right|\Pi\right\rangle = & \:\delta_{M_{F_{t}}M'_{F_{t}}}9\left(-1\right)^{F'_{t}+F_{t}-M{}_{F_{t}}+F'+\frac{1}{2}}\left\langle N'\, L'\left\Vert P^{(1)}\right\Vert N\, L\right\rangle\displaybreak[3] \nonumber \\
\nonumber \\
\times & \:\left[\left(2F_{t}+1\right)\left(2F'_{t}+1\right)\left(2F+1\right)\left(2F'+1\right)\left(2J+1\right)\left(2J'+1\right)\right]^{\frac{1}{2}}\displaybreak[3]\nonumber \\
\nonumber \\
\times & \:\left(\begin{array}{ccc}
F'_{t} & 1 & F_{t}\\
-M{}_{F_{t}} & 0 & M_{F_{t}}
\end{array}\right)\left\{ \begin{array}{ccc}
F' & F'_{t} & \frac{1}{2}\\
F_{t} & F & 1
\end{array}\right\} \left\{ \begin{array}{ccc}
L' & L & 1\\
J' & J & 0\\
F' & F & 1
\end{array}\right\} \left\{ \begin{array}{ccc}
1 & 1 & 1\\
\frac{1}{2} & \frac{1}{2} & 1\\
J' & J & 0
\end{array}\right\}. 
\end{align}

\begin{align}
\left\langle N'\, L'\left\Vert P^{(1)}\right\Vert N\, L\right\rangle = & \:\delta_{L',L+1}\,\left(-i\right)\left[\frac{\left(2L+3\right)\left(2L+1\right)}{L+1}\right]^{\frac{1}{2}}\displaybreak[3]\nonumber \\
\nonumber \\
 & \:\times\left[\sum_{j=-1}^{1}\sum_{k=-2}^{0}\left(L_{1}\right)_{N',L+1}^{k,1}\left\{ \left(RP_{1}\right)_{N,L}^{j}\left(N_{1}\right)_{L,0}^{1}+\delta_{0j}\left(D_{1}\right)_{L,0}^{1}\right\} I_{N'+k\, L+2;\, N+j\, L}\right]\displaybreak[3]\nonumber \\
\nonumber \\
+ & \:\delta_{L',L-1}\,\left(+i\right)\left[\frac{\left(2L+1\right)\left(2L-1\right)}{L}\right]^{\frac{1}{2}}\displaybreak[3]\nonumber \\
\nonumber \\
 & \:\times\left[\sum_{j=-1}^{1}\sum_{k=-2}^{0}\left(L_{1}\right)_{N+j,L}^{k,1}\left\{ \left(RP_{1}\right)_{N,L}^{j}\left(N_{1}\right)_{L,0}^{-1}+\delta_{0j}\left(D_{1}\right)_{L,0}^{-1}\right\} I_{N'\, L-1;\, N+j+k\, L+1}\right].
\end{align}

\end{widetext}

\section{Oscillator strength \label{sec:Oscrm}}

In this section we give the formulas for the oscillator
strength 
\begin{equation}
f_{\xi\nu\boldsymbol{K}}=\eta\left| \lim_{r\rightarrow 0}\left[-i\frac{\partial}{\partial r}\langle T^D_{\xi\boldsymbol{K}} |\Psi_{\nu\boldsymbol{K}}\rangle +\frac{\alpha K}{\sqrt{6}}\langle T^Q_{\xi\boldsymbol{K}} |\Psi_{\nu\boldsymbol{K}}\rangle\right]\right|^2
\end{equation}
for the three different orientations of 
$\boldsymbol{K}$ in
$[001]$, $[110]$, or $[111]$ direction.
For these directions we have to consider the
reduction of the irreducible representations
of the cubic group $O_{\mathrm{h}}$ by the
corresponding subgroups.
Furthermore, as in Ref.~\cite{125sub},
we also rotate the coordinate system to make the
$z$ axis of the new coordinate system coincide 
with the direction of $\boldsymbol{K}$.

\begin{table}[t]

\protect\caption{Reduction of the irreducible representations
of the cubic group $O_{\mathrm{h}}$ by the groups 
$C_{\mathrm{4v}}\,\left(\boldsymbol{K}\parallel[001]\right)$, 
$C_{\mathrm{2v}}\,\left(\boldsymbol{K}\parallel[110]\right)$, and 
$C_{\mathrm{3v}}\,\left(\boldsymbol{K}\parallel[111]\right)$~\cite{G3sub}.
Note that only the irreducible representation $\Gamma_5$ of $C_{\mathrm{4v}}$
and the irreducible representation $\Gamma_3$ of $C_{\mathrm{3v}}$ are two-dimensional.
Hence, almost all degeneracies in the exciton spectrum are lifted for $K\neq 0$.
\label{tab:2}}

\begin{centering}
\begin{tabular}{c|ccc}
$O_{\mathrm{h}}$ \hspace{0.2cm}& $C_{\mathrm{4v}}$ & $C_{\mathrm{2v}}$ & $C_{\mathrm{3v}}$\tabularnewline
\hline 
\hline 
$\Gamma_{1}^{+}$\hspace{0.2cm} & $\Gamma_{1}$ & $\Gamma_{1}$ & $\Gamma_{1}$\tabularnewline
$\Gamma_{2}^{+}$\hspace{0.2cm} & $\Gamma_{3}$ & $\Gamma_{2}$ & $\Gamma_{2}$\tabularnewline
$\Gamma_{3}^{+}$\hspace{0.2cm} & $\Gamma_{1}\oplus\Gamma_{3}$ & $\Gamma_{1}\oplus\Gamma_{2}$ & $\Gamma_{3}$\tabularnewline
$\Gamma_{4}^{+}$\hspace{0.2cm} & $\Gamma_{2}\oplus\Gamma_{5}$ & $\Gamma_{2}\oplus\Gamma_{3}\oplus\Gamma_{4}$ & \hspace{0.2cm}$\Gamma_{2}\oplus\Gamma_{3}$\tabularnewline
$\Gamma_{5}^{+}$\hspace{0.2cm} & \hspace{0.2cm}$\Gamma_{4}\oplus\Gamma_{5}$\hspace{0.2cm} & \hspace{0.2cm}$\Gamma_{1}\oplus\Gamma_{3}\oplus\Gamma_{4}$\hspace{0.2cm} & \hspace{0.2cm}$\Gamma_{1}\oplus\Gamma_{3}$\hspace{0.2cm}\tabularnewline
\hline 
$\Gamma_{1}^{-}$\hspace{0.2cm} & $\Gamma_{2}$ & $\Gamma_{3}$ & $\Gamma_{2}$\tabularnewline
$\Gamma_{2}^{-}$\hspace{0.2cm} & $\Gamma_{4}$ & $\Gamma_{4}$ & $\Gamma_{1}$\tabularnewline
$\Gamma_{3}^{-}$\hspace{0.2cm} & $\Gamma_{2}\oplus\Gamma_{4}$ & $\Gamma_{3}\oplus\Gamma_{4}$ & $\Gamma_{3}$\tabularnewline
$\Gamma_{4}^{-}$\hspace{0.2cm} & $\Gamma_{1}\oplus\Gamma_{5}$ & $\Gamma_{1}\oplus\Gamma_{2}\oplus\Gamma_{4}$ & \hspace{0.2cm}$\Gamma_{1}\oplus\Gamma_{3}$\tabularnewline
$\Gamma_{5}^{-}$\hspace{0.2cm} & \hspace{0.2cm}$\Gamma_{3}\oplus\Gamma_{5}$\hspace{0.2cm} & \hspace{0.2cm}$\Gamma_{1}\oplus\Gamma_{2}\oplus\Gamma_{3}$\hspace{0.2cm} & \hspace{0.2cm}$\Gamma_{2}\oplus\Gamma_{3}$\hspace{0.2cm}\tabularnewline
\hline 
\end{tabular}
\par\end{centering}

\end{table}

\subsection{Wave vector $\boldsymbol{K}\parallel [001]$}

For $\boldsymbol{K}\parallel [001]$, the
symmetry $O_{\mathrm{h}}$ of the system is reduced to $C_{\mathrm{4v}}$ 
and we have to consider the reduction of the 
irreducible representations of $O_{\mathrm{h}}$ by the group $C_{\mathrm{4v}}$
(see Table~\ref{tab:2}).
Especially for $\Gamma_4^-$ it is
\begin{equation}
\Gamma_{4}^{-} \rightarrow \Gamma_{1}\oplus\Gamma_{5}.
\end{equation}

Using the method of projection operators~\cite{G1sub},
we can determine the correct linear combinations of the states
\begin{subequations}
\begin{align}
|\pi_x^D\rangle= &\; \frac{i}{\sqrt{2}}\left[\left|2,\,-1\right\rangle_D+\left|2,\,1\right\rangle_D\right],\\
\displaybreak[1]
|\pi_y^D\rangle= &\; \frac{1}{\sqrt{2}}\left[\left|2,\,-1\right\rangle_D-\left|2,\,1\right\rangle_D\right],\\
\displaybreak[1]
|\pi_z^D\rangle= &\; \frac{i}{\sqrt{2}}\left[\left|2,\,-2\right\rangle_D-\left|2,\,2\right\rangle_D\right].
\end{align}
\label{eq:Dxyz}%
\end{subequations}
and
\begin{subequations}
\begin{align}
\left|\pi_x^Q\right\rangle= &\; \hat{K}_y \left|1,\,0\right\rangle_Q\nonumber\\
&\; \qquad+\hat{K}_z \frac{i}{\sqrt{2}}\left[\left|1,\,-1\right\rangle_Q+\left|1,\,1\right\rangle_Q\right],\\
\displaybreak[1]
\left|\pi_y^Q\right\rangle= &\; \hat{K}_x \left|1,\,0\right\rangle_Q\nonumber\\
&\; \qquad+\hat{K}_z \frac{1}{\sqrt{2}}\left[\left|1,\,-1\right\rangle_Q-\left|1,\,1\right\rangle_Q\right],\\
\displaybreak[1]
\left|\pi_z^Q\right\rangle= &\; \hat{K}_y \frac{1}{\sqrt{2}}\left[\left|1,\,-1\right\rangle_Q-\left|1,\,1\right\rangle_Q\right]\nonumber\\
&\; \qquad+\hat{K}_x \frac{i}{\sqrt{2}}\left[\left|1,\,-1\right\rangle_Q+\left|1,\,1\right\rangle_Q\right].
\end{align}
\label{eq:Qxyz}%
\end{subequations}
which transform according to the irreducible
representations of $C_{\mathrm{4v}}$.
On the other hand, it is instructive that the correct linear combinations
are given according to the direction of $\boldsymbol{K}$ and the polarization vectors 
$\hat{\boldsymbol{e}}_{\xi\boldsymbol{K}}$ transverse to $\boldsymbol{K}$:
\begin{subequations}
\begin{alignat}{2}
\hat{\boldsymbol{K}} &\; = (0,\,0,\,1)^{\mathrm{T}},\\
\hat{\boldsymbol{e}}_{1\boldsymbol{K}} &\; = (1,\,0,\,0)^{\mathrm{T}},\\
\hat{\boldsymbol{e}}_{2\boldsymbol{K}} &\; = (0,\,1,\,0)^{\mathrm{T}}.
\end{alignat}
\end{subequations}
Since light is always transversely polarized, only states of symmetry $\Gamma_5$ are allowed.
The correct linear combinations of the states~(\ref{eq:Dxyz}) are 
\begin{subequations}
\begin{alignat}{2}
\Gamma_{1}: &\; |L^D_{\boldsymbol{K}}\rangle=|\pi_z^D\rangle,\\
\nonumber \\
\Gamma_{5}: &\; |T^D_{1\boldsymbol{K}}\rangle=|\pi_x^D\rangle\;\mathrm{and}\;\\
&\; |T^D_{2\boldsymbol{K}}\rangle=|\pi_y^D\rangle,
\end{alignat}
\label{eq:Dpi001}%
\end{subequations}
and likewise for the states of Eq.~(\ref{eq:Qxyz}).
If we assume the incident light to be circularly
polarized, the oscillator strength is given by
\begin{equation}
f_{\xi\nu\boldsymbol{K}}=\eta\left| \lim_{r\rightarrow 0}\left[-i\frac{\partial}{\partial r}\langle T^D_{\pm\boldsymbol{K}}|\Psi_{\nu\boldsymbol{K}}\rangle +\frac{\alpha K}{\sqrt{6}}\langle T^Q_{\pm\boldsymbol{K}}|\Psi_{\nu\boldsymbol{K}}\rangle\right]\right|^2\label{eq:frel001}
\end{equation}
with
\begin{subequations}
\begin{alignat}{2}
|T^D_{+\boldsymbol{K}}\rangle = & \frac{-i}{\sqrt{2}}\left[|T^D_{1\boldsymbol{K}}\rangle+i|T^D_{2\boldsymbol{K}}\rangle\right] = \left|2,\,-1\right\rangle_D,\\
\nonumber \\
|T^D_{-\boldsymbol{K}}\rangle = & \frac{i}{\sqrt{2}}\left[|T^D_{1\boldsymbol{K}}\rangle-i|T^D_{2\boldsymbol{K}}\rangle\right] = -\left|2,\,1\right\rangle_D
\end{alignat}
\end{subequations}
and
\begin{subequations}
\begin{alignat}{2}
|T^Q_{+\boldsymbol{K}}\rangle = & \frac{-i}{\sqrt{2}}\left[|T^Q_{1\boldsymbol{K}}\rangle+i|T^Q_{2\boldsymbol{K}}\rangle\right] = \left|1,\,-1\right\rangle_Q,\\
\nonumber \\
|T^Q_{-\boldsymbol{K}}\rangle = & \frac{i}{\sqrt{2}}\left[|T^Q_{1\boldsymbol{K}}\rangle-i|T^Q_{2\boldsymbol{K}}\rangle\right] = -\left|1,\,1\right\rangle_Q.
\end{alignat}
\end{subequations}
Note that the sign $\pm$ is defined by the direction of rotation of the polarization
with respect to $\boldsymbol{K}$.

\begin{figure*}[t]
\begin{centering}
\includegraphics[width=1.95\columnwidth]{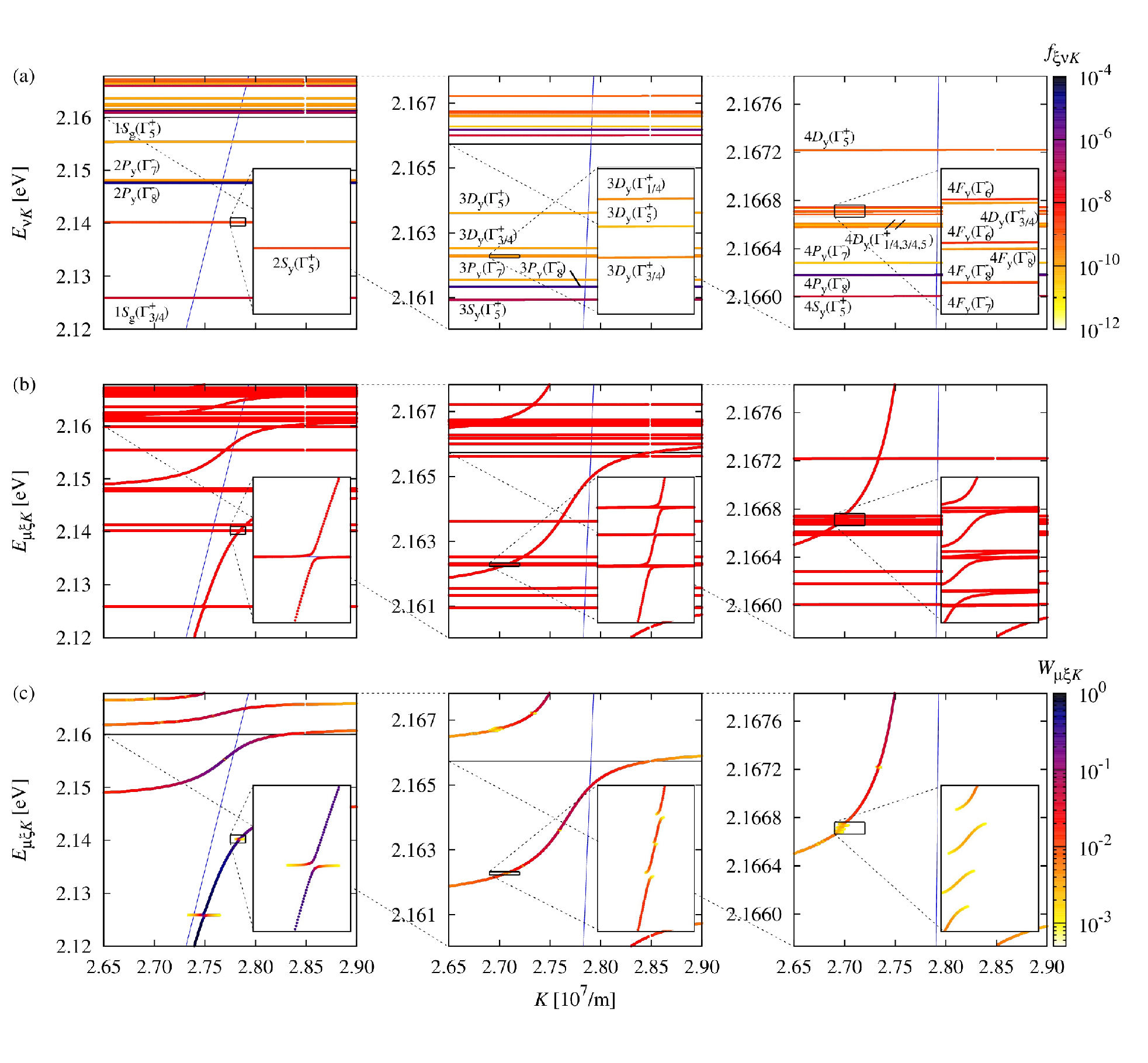}
\par\end{centering}

\protect\caption{(a) The exciton energies $E_{\nu\boldsymbol{K}}$ in dependence on
$K=\left|\boldsymbol{K}\right|$ for $\boldsymbol{K}\parallel[110]$. 
The colorbar shows the oscillator strengths for $\xi=\pi_{x'}$ polarized light.
We denote from which states at $K=0$ the exciton states originate (cf.~Ref.~\cite{200sub}).
For reasons of space we introduce the abbreviated notation $\Gamma_{i/j}^{\pm}$
to replace $\Gamma_{i}^{\pm},\,\Gamma_{j}^{\pm}$.
The blue solid line gives the photon dispersion $\hbar\omega_{\xi\boldsymbol{K}}=\hbar Kc/\sqrt{\varepsilon_{\mathrm{b2}}}$.
(b) Polariton dispersion obtained by solving the eigenvalue problem in Sec.~IVC. 
(c) Photon-like part $W_{\mu\xi\boldsymbol{K}}$ of the polariton states. 
For further information see text.~\label{fig:Figg2}}
\end{figure*}

\begin{figure*}[t]
\begin{centering}
\includegraphics[width=1.95\columnwidth]{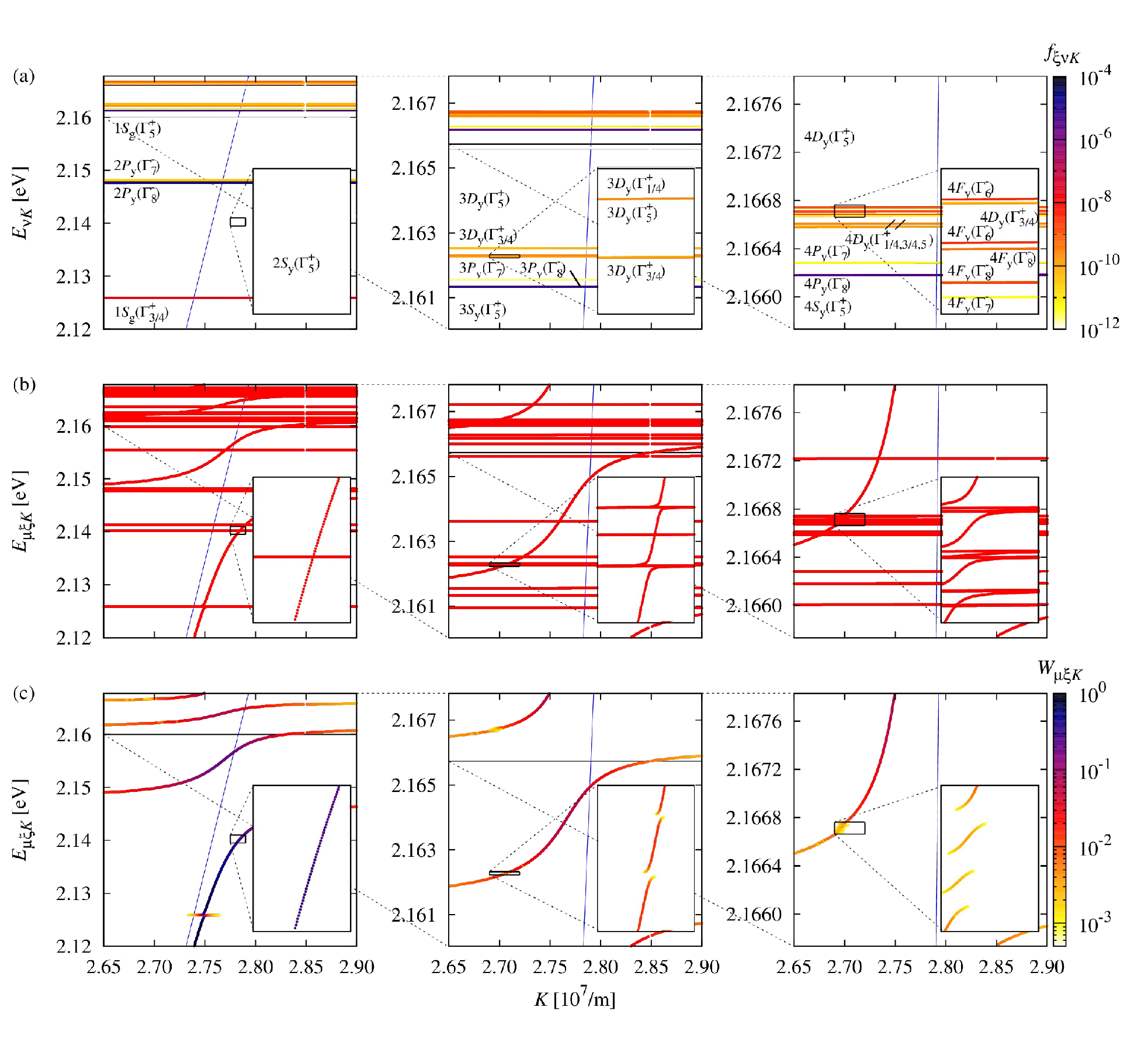}
\par\end{centering}

\protect\caption{Same spectra as in Fig.~\ref{fig:Figg2} but for 
$\boldsymbol{K}\parallel[110]$ and $\xi=\pi_{y'}$ polarized light.
Note that the states which have the symmetry $\Gamma_5^+$ for $K=0$
are dipole-forbidden in this case. 
For further information see text.~\label{fig:Figg3}}
\end{figure*}

\begin{figure*}[t]
\begin{centering}
\includegraphics[width=1.95\columnwidth]{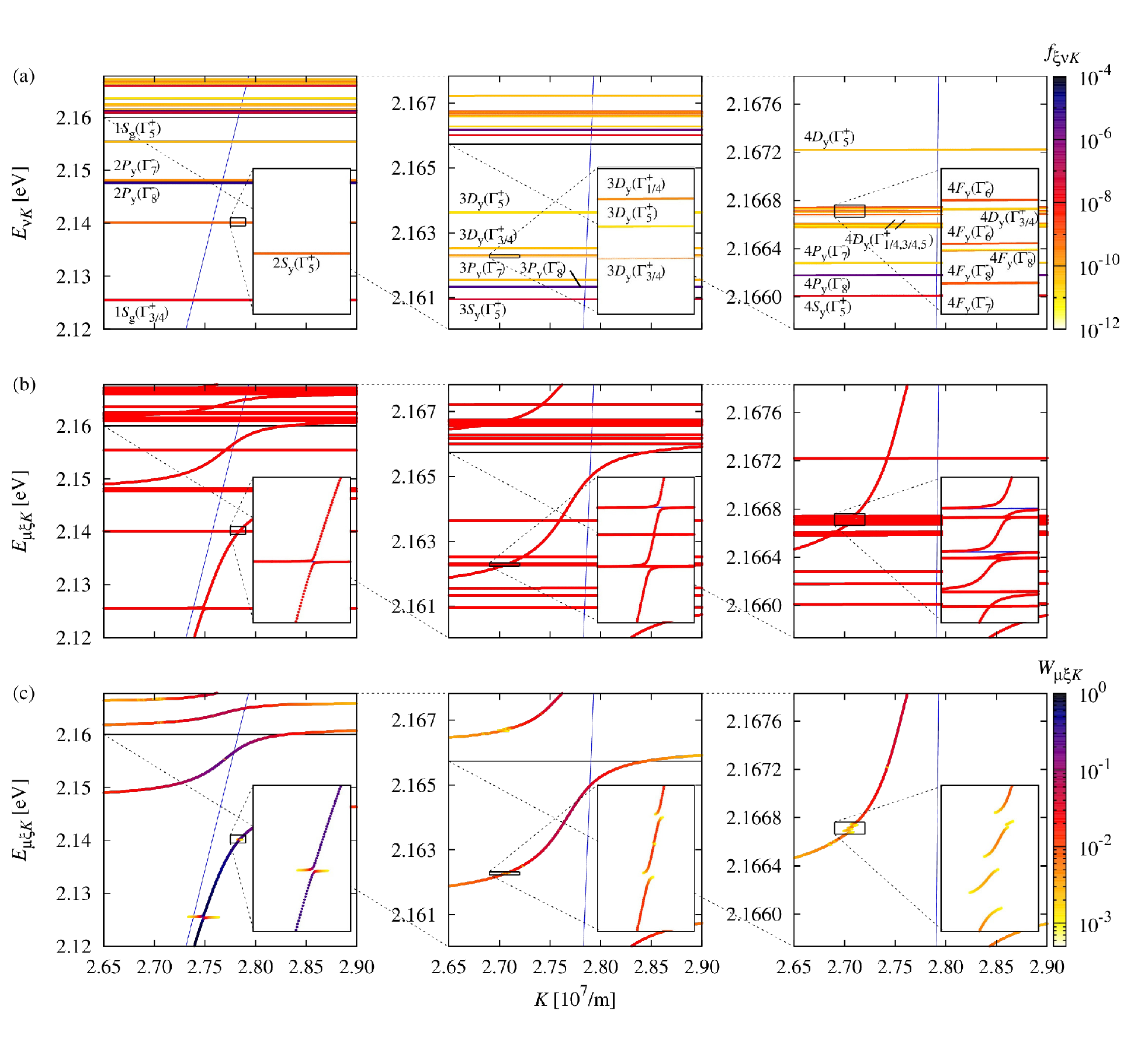}
\par\end{centering}

\protect\caption{Same spectra as in Fig.~\ref{fig:Figg2} but for 
$\boldsymbol{K}\parallel[111]$ and $\xi=\sigma_{z''}^{\pm}$ polarized light.
For both polarizations the spectrum is identical.
The oscillator strength of the $4F(\Gamma_7^-)$ state
is too weak to be seen here.
For further information see text.~\label{fig:Figg4}}
\end{figure*}

\subsection{Wave vector $\boldsymbol{K}\parallel [110]$}

For $\boldsymbol{K}\parallel [110]$ the symmetry is reduced to $C_{\mathrm{2v}}$ 
and it is
\begin{equation}
\Gamma_{4}^{-} \rightarrow \Gamma_{1}\oplus\Gamma_{2}\oplus\Gamma_{4}.
\end{equation}
The corresponding vectors are
\begin{subequations}
\begin{alignat}{2}
\hat{\boldsymbol{K}} &\; = (1,\,1,\,0)^{\mathrm{T}}/\sqrt{2},\\
\hat{\boldsymbol{e}}_{1\boldsymbol{K}} &\; = (1,\,-1,\,0)^{\mathrm{T}}/\sqrt{2},\\
\hat{\boldsymbol{e}}_{2\boldsymbol{K}} &\; = (0,\,0,\,1)^{\mathrm{T}},
\end{alignat}
\end{subequations}
so that the correct linear combinations of the states~(\ref{eq:Dxyz}) read
\begin{subequations}
\begin{alignat}{2}
\Gamma_{1}: &\; |L^D_{\boldsymbol{K}}\rangle=\frac{1}{\sqrt{2}}\left[\left|\pi_x^D\right\rangle+\left|\pi_y^D\right\rangle\right],\\
\nonumber\\
\Gamma_{2}: &\; |T^D_{1\boldsymbol{K}}\rangle=\frac{1}{\sqrt{2}}\left[\left|\pi_x^D\right\rangle-\left|\pi_y^D\right\rangle\right],\\
\nonumber\\
\Gamma_{4}: &\; |T^D_{2\boldsymbol{K}}\rangle=\left|\pi_z^D\right\rangle.
\end{alignat}
\label{eq:Dpi110}%
\end{subequations}

We now choose the quantization axis parallel to $\boldsymbol{K}$, i.e.,
we rotate the coordinate system by the Euler angles
$\left(\alpha,\,\beta,\,\gamma\right)=\left(\pi,\,\pi/2,\,\pi/4\right)$.
Rotating the states $|L^i_{\boldsymbol{K}}\rangle$
and $|T^i_{\xi\boldsymbol{K}}\rangle$ as well yields
\begin{subequations}
\begin{alignat}{2}
|L'^D_{\boldsymbol{K}}\rangle=|\pi_{z'}^D\rangle & = \frac{1}{\sqrt{2}}\left[\left|2,\,-1\right\rangle_D-\left|2,\,1\right\rangle_D\right],\\
\nonumber\\
|T'^D_{1\boldsymbol{K}}\rangle=|\pi_{y'}^D\rangle & = \frac{i}{\sqrt{2}}\left[\left|2,\,2\right\rangle_D-\left|2,\,-2\right\rangle_D\right],\\
\nonumber\\
|T'^D_{2\boldsymbol{K}}\rangle=|\pi_{x'}^D\rangle & =\frac{\sqrt{3}}{2}\left|2,\,0\right\rangle_D\nonumber\\
& +\frac{1}{\sqrt{8}}\left[\left|2,\,-2\right\rangle_D+\left|2,\,2\right\rangle_D\right],
\end{alignat}
\label{eq:Dpi110r}%
\end{subequations}
and
\begin{subequations}
\begin{alignat}{2}
|L'^Q_{\boldsymbol{K}}\rangle=|\pi_{z'}^Q\rangle & = \frac{1}{\sqrt{2}}\left[\left|1,\,-1\right\rangle_Q-\left|1,\,1\right\rangle_Q\right],\\
\nonumber\\
|T'^Q_{1\boldsymbol{K}}\rangle=|\pi_{y'}^Q\rangle & = 0,\\
\nonumber\\
|T'^Q_{2\boldsymbol{K}}\rangle=|\pi_{x'}^Q\rangle & = \left|1,\,0\right\rangle_Q.
\end{alignat}
\label{eq:Qpi110r}%
\end{subequations}
The labels $x'$, $y'$ and $z'$ are meant to indicate that the states are given in the rotated coordinate system.
Finally, we calculate the
oscillator strengths by evaluating
\begin{equation}
f_{\nu\xi\boldsymbol{K}}=\eta\left| \lim_{r\rightarrow 0}\left[-i\frac{\partial}{\partial r}\langle T'^D_{\xi\boldsymbol{K}}|\Psi_{\nu\boldsymbol{K}}\rangle +\frac{\alpha K}{\sqrt{6}}\langle T'^Q_{\xi\boldsymbol{K}}|\Psi_{\nu\boldsymbol{K}}\rangle\right]\right|^2\label{eq:frel110}
\end{equation}
for light which is polarized in [001] or in $[1\bar{1}0]$ direction.

\subsection{Wave vector $\boldsymbol{K}\parallel [111]$}

For $\boldsymbol{K}\parallel [111]$ the
symmetry is reduced to $C_{\mathrm{3v}}$ and we have
\begin{equation}
\Gamma_{4}^{-} \rightarrow \Gamma_{1}\oplus\Gamma_{3}
\end{equation}
with the vectors
\begin{subequations}
\begin{alignat}{2}
\hat{\boldsymbol{K}} &\; = (1,\,1,\,1)^{\mathrm{T}}/\sqrt{3},\\
\hat{\boldsymbol{e}}_{1\boldsymbol{K}} &\; = (1,\,1,\,-2)^{\mathrm{T}}/\sqrt{6},\\
\hat{\boldsymbol{e}}_{2\boldsymbol{K}} &\; = (-1,\,1,\,0)^{\mathrm{T}}/\sqrt{2}.
\end{alignat}
\end{subequations}
The correct linear combinations of the states
in Eqs.~(\ref{eq:Dxyz}) are therefore
\begin{subequations}
\begin{alignat}{2}
\Gamma_{1}: &\; |L^D_{\boldsymbol{K}}\rangle=\frac{1}{\sqrt{3}}\left[\left|\pi_x^D\right\rangle+\left|\pi_y^D\right\rangle+\left|\pi_z^D\right\rangle\right],\\
\nonumber\\
\Gamma_{3}: &\; |T^D_{1\boldsymbol{K}}\rangle=\frac{1}{\sqrt{6}}\left[\left|\pi_x^D\right\rangle+\left|\pi_y^D\right\rangle-2\left|\pi_z^D\right\rangle\right]\;\mathrm{and}\;\\
            &\; |T^D_{2\boldsymbol{K}}\rangle=\frac{1}{\sqrt{2}}\left[-\left|\pi_x^D\right\rangle+\left|\pi_y^D\right\rangle\right]
\end{alignat}
\label{eq:Dpi111}%
\end{subequations}
Here we rotate the coordinate system by the Euler angles
$\left(\alpha,\,\beta,\,\gamma\right)=\left(0,\,\arccos(1/\sqrt{3}),\,\pi/4\right)$.
Rotating the states of Eq.~(\ref{eq:Dpi111}) and assuming circularly polarized light yields
\begin{subequations}
\begin{alignat}{2}
|L''^D_{\boldsymbol{K}}\rangle & =\left|2,\,0\right\rangle_D,\\
\nonumber\\
|T''^D_{+\boldsymbol{K}}\rangle & = \frac{i}{\sqrt{3}}\left[\sqrt{2}\left|2,\,-2\right\rangle_D-\left|2,\,1\right\rangle_D\right]\;\mathrm{and}\;\\
|T''^D_{-\boldsymbol{K}}\rangle & = \frac{-i}{\sqrt{3}}\left[\sqrt{2}\left|2,\,2\right\rangle_D+\left|2,\,-1\right\rangle_D\right],
\end{alignat}
\label{eq:Dpi111r}%
\end{subequations}
and
\begin{subequations}
\begin{alignat}{2}
|L''^Q_{\boldsymbol{K}}\rangle & = \frac{2}{\sqrt{3}}\left|1,\,0\right\rangle_Q,\\
\nonumber\\
|T''^Q_{+\boldsymbol{K}}\rangle & = \frac{-i}{\sqrt{3}}\left|1,\,1\right\rangle_Q\;\mathrm{and}\;\\
|T''^Q_{-\boldsymbol{K}}\rangle & =  \frac{-i}{\sqrt{3}}\left|1,\,-1\right\rangle_Q.
\end{alignat}
\label{eq:Qpi111r}%
\end{subequations}
We finally calculate the 
oscillator strengths by evaluating
\begin{equation}
f_{\xi\nu\boldsymbol{K}}=\eta\left| \lim_{r\rightarrow 0}\left[-i\frac{\partial}{\partial r}\langle T''^D_{\pm\boldsymbol{K}}|\Psi_{\nu\boldsymbol{K}}\rangle +\frac{\alpha K}{\sqrt{6}}\langle T''^Q_{\pm\boldsymbol{K}}|\Psi_{\nu\boldsymbol{K}}\rangle\right]\right|^2.\label{eq:frel110}
\end{equation}

\section{Polariton dispersions for $\boldsymbol{K}\parallel[110]$ and $\boldsymbol{K}\parallel[111]$\label{sec:Polaritonrm}}

In this section we present the dispersions 
of the exciton-polaritons with $\boldsymbol{K}\parallel[110]$ and $\boldsymbol{K}\parallel[111]$ using the Hamiltonian which accounts for
the valence band structure, the exchange interaction, the central-cell corrections,
and the finite momentum $\hbar K$ of the center of mass. 

As we already stated in the manuscript, the differences in the dispersion
relations for the different orientations $\boldsymbol{K}\parallel[001]$
$\boldsymbol{K}\parallel[110]$ or $\boldsymbol{K}\parallel[111]$
are only slight.
A change in the energies of the exciton states cannot be observed in Figs.~\ref{fig:Figg2},
\ref{fig:Figg3}, and~\ref{fig:Figg4}, since it is
on the order of tens of $\mu\mathrm{eV}$.
However, there is an important difference between the spectra:
The values of the oscillator strengths significantly 
change due to the different symmetry breaking.
For example, for $\boldsymbol{K}\parallel[110]$ 
the states which have the symmetry $\Gamma_5^+$ at $K=0$
are dipole-forbidden.
Especially the change in the oscialltor strength
of the $P$ excitons affects the position of the avoided crossings 
of the other excitons as can be seen when comparing, e.g., the
insets in the lower right panels of Figs.~\ref{fig:Figg3}
and~\ref{fig:Figg4}.
Therefore, the polariton dispersions differ, however,
mainly in the vicinity of the avoided crossings.



\end{document}